\documentclass[oldversion]{aa}  
\usepackage{graphicx}
\usepackage{txfonts}
\usepackage{natbib}
\usepackage{enumitem}
\usepackage{subfig}
\usepackage{url}
\bibpunct{(}{)}{;}{a}{}{,}

\def\xmm{{\it XMM-Newton~}}
\def\chandra{{\it Chandra~}}

\def\lunits{$\rm erg\,s^{-1}$~}
\def\funits{$\rm erg\,cm^{-2}\,s^{-1}$~}
\def\cunits{$\rm cm^{-2}~$}

\begin{document}

\title{X-ray observations of dust obscured galaxies in the Chandra Deep Field South}


   \author{A. Corral
     \inst{1}
     \and
     I. Georgantopoulos\inst{1}
\and
A. Comastri\inst{2}
\and
P. Ranalli\inst{1} 
\and
A. Akylas\inst{1}
\and
M. Salvato\inst{3}
\and
G. Lanzuisi\inst{4}
\and
C. Vignali\inst{2,4}
\and
L. Koutoulidis\inst{1} 
}

  \titlerunning{X-ray DOGs in the CDF-S}
  \authorrunning{A. Corral et al.}

   \institute{Institute for Astronomy, Astrophysics, Space Applications, and Remote Sensing (IAASARS), National Observatory of Athens, 15236 Penteli, Greece    
\and
INAF – Osservatorio Astronomico di Bologna, via Ranzani 1, I–40127 Bologna, Italy
\and 
Max-Planck-Institut f\"ur extraterrestrische Physik (MPE), Giessenbachstrasse 1, D-85748, Garching bei M\"unchen, Germany
\and
Dipartimento di Fisica e Astronomia, Universit\'a di Bologna, Viale Berti Pichat 6/2, I-40127 Bologna, Italy
}

   \date{Received ; accepted }

\abstract{We present the properties of X-ray detected dust obscured galaxies (DOGs) in the {\it Chandra} Deep Field South. In recent years, it has been proposed that a significant percentage of the elusive Compton-thick (CT) active galactic nuclei (AGN) could be hidden among DOGs. This type of galaxy is characterized by a very high infrared (IR) to optical flux ratio ($f_{24\mu m}/f_{R}>1000$), which in the case of CT AGN could be due to the suppression of AGN emission by absorption and its subsequent re-emission in the IR. The most reliable way of confirming the CT nature of an AGN is by X-ray spectroscopy. In a previous work, we presented the properties of X-ray detected DOGs by making use of the deepest X-ray observations available at that time, the 2Ms observations of the {\it Chandra} deep fields, the Chandra Deep Field North (CDF-N), and the Chandra Deep Field South (CDF-S). In that work, we only found a moderate percentage ($<$ 50\%) of CT AGN among the DOGs sample. However, we pointed out that the limited photon statistics for most of the sources in the sample did not allow us to strongly constrain this number. In this paper, we further explore the properties of the sample of DOGs in the CDF-S presented in that work by using not only a deeper 6Ms {\it Chandra} survey of the CDF-S, but also by combining these data with the 3Ms {\it XMM-Newton} survey of the CDF-S. We also take advantage of the great coverage of the CDF-S region from the UV to the far-IR to fit the spectral energy distributions (SEDs) of our sources. Out of the 14 AGN composing our sample, 9 are highly absorbed ( N$_{H}$ $>$ 10$^{23}$ cm$^{-2}$), whereas 2 look unabsorbed, and the other 3 are only moderately absorbed. Among the highly absorbed AGN, we find that only three could be considered CT AGN. In only one of these three cases, we detect a strong Fe K$\alpha$ emission line; the source is already classified as a CT AGN with {\it Chandra} data in a previous work. Here we confirm its CT nature by combining {\it Chandra} and {\it XMM-Newton} data. For the other two CT candidates, the non-detection of the line could be because of the low number of counts in their X-ray spectra, but their location in the L$_{\rm 2-10\,keV}$/L$_{12\mu m}$ plot supports their CT classification. Although a higher number of CT sources could be hidden among the X-ray undetected DOGs, our results indicate that DOGs could be as well composed of only a fraction of CT AGN plus a number of moderate to highly absorbed AGN, as previously suggested. From our study of the X-ray undetected DOGs in the CDF-S, we estimate a percentage between 13 and 44\% of CT AGN among the whole population of DOGs. }

   \maketitle
%

\section{Introduction} 
There is mounting evidence that the growth of galaxies and the
super-massive black holes (SMBHs) at their centres must be strongly
connected
(\citealt{magorrian98}, \citealt{ferrarese00}, \citealt{geb00}, \citealt{tremaine02}, \citealt{tremaine02}, \citealt{marconi04}, \citealt{ferrarese05}, \citealt{kormendy13}).
Moreover, star formation history seems to follow that of SMBH growth
via accretion and traced by active galactic nuclei (AGN) emission
\citep{lafranca05}. Therefore, to obtain a complete census of the
AGN population is vital to understand cosmic evolution. 

X-rays surveys are extremely efficient in finding AGN since X-rays are
able to penetrate high amounts of gas and dust. Nevertheless, X-ray
surveys, even the deepest, such as the {\it Chandra} Deep Field
Surveys, and the hardest (i.e., those carried out above 10 keV),
are still biased against the most heavily absorbed AGN, the so-called
Compton-thick (CT) AGN (column densities N$_{H}$ $>$ 10$^{24}$
cm$^{-2}$). The actual percentage of CT AGN among the AGN population is
still unknown and it is usually inferred in an indirect way from the
modelling of the cosmic X-ray background (CXB; see
\citealt{gilli07}, \citealt{treister09a}, \citealt{akylas12}).

In the past few years, mid-infrared (IR) surveys have been proposed as
an alternative way of finding and studying CT AGN. The AGN emission
that is absorbed is then re-emitted by the heated dust, so CT AGN,
given their extremely obscured nature, should emit strongly in the
mid-IR while they are fainter at other wavelengths. This has been the
basic selection argument used in several recent works. For example,
\citet{sansigre05} argued that the missing AGN obscured population at
high redshifts displays bright 24 $\mu m$ emission with no 3.6 $\mu m$
detection. \citet{daddi07b} used ultraviolet (UV) selected sources
instead and found a very high percentage of CT AGN among those showing
mid-IR excess.  

In this work, we focus on mid-IR bright optically faint sources that
have been associated with dust obscured galaxies (DOGs). DOGs were
first discovered using {\it Spitzer} data \citep{houck05}, as a
population of 24 $\mu m$ bright, R-band faint sources at redshifts
$z\approx2$ (with a small scatter $\sigma_z=0.5$; \citealt{dey08},
\citealt{pope08}); this implies luminosities $\rm L_{IR} > 10^{12-14}
L_{\odot}$ that are comparable or in excess of low redshift
ultra-luminous infrared galaxies (ULIRGs; \citealt{sanders96}). At
those redshifts, the 24 $\mu m$ emission corresponds to the peak of
the torus IR emission in AGN. Therefore, this selection technique
should be very efficient in detecting heavily obscured AGN.

\citet{fiore08} proposed that most of the DOGs may be CT AGN. The
stacked X-ray signal of the undetected DOGs in X-ray surveys appears
to be flat, which is indicative of absorbed sources (\citealt{fiore08},
\citealt{georg08}, \citealt{fiore09}, \citealt{treister09b}). However,
\citet{georg08} pointed out that a flat stacked spectrum could also be
produced by a combination of low-luminosity AGN with moderate
absorption. Moreover, \citet{pope08} showed that the normal galaxy
(non-AGN) content of DOG samples may still be significant. In any
case, the most reliable way to confirm their CT AGN nature is to
obtain their intrinsic absorption, or a strong Fe K$\alpha$
emission line detection, directly from X-ray spectroscopy (see for
example \citealt{fuka11}).

\citet{lanzuisi09} performed an X-ray study of 44 luminous DOGs
($F_{24 \mu m}/F_R>2000$ and $F_{24 \mu m} >1.3$ mJy ) in the Spitzer
Wide-Area Infrared Extragalactic (SWIRE) survey, of which 23 are
detected in X-rays. Among the DOGs detected in X-rays, half of these
have column densities $\rm N_H>10^{23}$ \cunits, but only one could be
classified as a CT AGN. \citet{fiore09} investigated the X-ray
properties of 73 DOGs in the Cosmic Evolution Survey (COSMOS)
\citep{elvis09}. They derived their column densities from hardness
ratios, directly for the 31 detected in X-rays and from stacked images
from the undetected DOGs, and found that the fraction of CT AGN among
DOGs seems to increase as their IR luminosity increases. This is
consistent with previous and more recent results in which the percentage
of AGN has been found to increase as the IR luminosity increases
\citep{sacchi09,lee10,rigu15}. \citet{lanzuisi09} also found that the
hardness ratios of X-ray undetected DOGs were consistent with those of
the detected DOGs, and these authors argued that the very flat photon
index in both samples indicates a high percentage of CT AGN among
DOGs.
     
\citet{georgakakis10} compiled a sample of ``low redshift DOGs
analogues'' from the AEGIS and CDF-N surveys. These are sources for
which their spectral energy distributions (SEDs) would be similar to
those of DOGs if placed at redshift 2. Nine of their sources have
X-ray counterparts, and only three of these sources show tentative
evidence of CT obscuration. The SEDs of the X-ray undetected DOGs are
consistent with starburst activity showing no evidence for a hot dust
component. \citet{georgakakis10} concluded that there is little
evidence for the presence of a high percentage of luminous CT sources
in either the X-ray detected or undetected population of DOGs
analogues.

Finally, \citet{treister09b} examine the properties of 211 heavily
obscured AGN candidates in the extended CDF-S, selecting objects with
$f_{24\mu m}/f_R > 1000$ and $R-K >4.5$. Eighteen sources are detected
in X-rays, they have moderate column densities $\sim 10^{22-23}$
\cunits, and only two of these sources appear to be CT. The X-ray
undetected sources show a hard average spectrum that could be
interpreted as a mixture of 90\% CT objects and 10\% star-forming
galaxies.

In our previous work \citep{georg11}, we presented the properties of
26 X-ray detected DOGs from the 2Ms surveys in the CDF-N and CDF-S and
we found only a moderate percentage of CT AGN, although at least half
of the full sample show signs of heavy (but Compton-thin)
obscuration. It has to be noted that, because of poor photon
statistics, in many cases a heavily absorbed nature was inferred from
a very flat spectrum. We also found that the average spectrum of X-ray
detected and undetected DOGs are very similar with a very hard photon
index. This could indicate a high percentage of CT sources, but also a
combination of a moderate percentage of CTs plus a higher number of
only moderately absorbed AGN.

Here we further explore the properties of the X-ray detected DOGs in
the CDF-S presented in \citet{georg11} by combining 6 Ms {\it Chandra}
and 3 Ms {\it XMM-Newton} observations of this region, and thus, by
taking advantage of the improved signal-to-noise ratio of the new
X-ray spectra. We adopt H$_{o}$ =75 km s$^{-1}$, $\Omega_{M}$ = 0.3,
and $\Omega_{\Lambda}$ = 0.7 throughout this paper. Errors are
reported at the 90\% confidence level.

\section{Chandra Deep Field South}
Chandra Deep Field South (CDF-S) is the deepest {\it Chandra} survey
to date, covering an area of 465 arcmin$^{2}$. The most recent
catalogue of X-ray sources within the CDF-S was produced by using 52
observations with a total exposure time of $\sim$ 4 Ms
\citep{xue11}. A further approved 3 Ms is due to be added to this
field, which will bring the total exposure to 7 Ms by the end of
2015. This area has also been observed by {\it XMM-Newton} with a
total exposure time of $\sim$ 3 Ms \citep{ranalli13}. The flux limits
in the hard ({\it Chandra}: 2-8 KeV, {\it XMM-Newton}: 2-10 keV) band
for these surveys are 5.5$\times$10$^{-17}$ erg cm$^{-2}$ s$^{-1}$ and
6.6$\times$10$^{-16}$ erg cm$^{-2}$ s$^{-1}$ for the {\it Chandra} and
{\it XMM-Newton} observations, respectively. To maximize the spectral
quality of our sample, we extracted spectral data from all the {\it
  Chandra} observations that were publicly available by June 2015,
which resulted in a maximum exposure time of $\sim$ 6 Ms.

This area has also extensive multi-wavelength coverage from the UV to
the far-IR. Near-UV to near-IR data are available from the GOODS-MUSIC
catalogue \citep{grazian06}, including U photometry from ESO-La Silla
and ESO-VLT-VIMOS; B, V, i, and z photometry from HST-ACS; J, H, and K
photometry from VLT-ISAAC; and {\it Spitzer} photometry at 3.5,
4.5, 5.8, 8 $\mu m$ (IRAC), and 24 $\mu m$ (MIPS). We also used more
recent {\it Spitzer}-IRAC observations from the SIMPLE survey
\citep{damen11} and optical data from the MUSYC catalogue
\citep{gawa06}. Far-IR data in the 100 and 160 $\mu m$ bands are
available from the GOODS-Herschel survey \citep{elbaz07}, and the PACS
Evolutionary Probes programme \citep{lutz11}.

\section{Sample of dust obsured galaxies }

Our sample of X-ray detected DOGs is composed of the 14 sources in the
CDF-S studied in \citet{georg11}. For seven of these sources, there
are data available from both {\it XMM-Newton} and {\it Chandra}
observations. The selection of the sample is described in detail in
\citet{georg11}. We combined the catalogue in \citet{grazian06} with
the X-ray catalogue of \citet{luo08,luo10}, which is based on the 2 Ms
{\it Chandra} catalogue, and selected those sources with log($f_{24\mu
  m}$/f$_R$) $>$ 3 with a lower limit for optical non-detections of R
= 26.5(AB). There are 56 additional DOGs within the CDF-S footprint
with no X-ray counterpart in the 2 Ms Chandra catalogue, four of these
now detected using the 4 Ms Chandra data \citep{xue11}. However, the
X-ray data quality of these new four X-ray detected DOGs is too poor
to perform a reliable spectral analysis, which is the purpose of this
work, so we refer to these sources as X-ray undetected DOGs in this
paper.

Only {\it Chandra} spectral data from the 2 Ms survey was used in
\citet{georg11}. Here we combine 6 Ms of {\it Chandra} data with the
3 Ms of {\it XMM-Newton} data to better constrain the absorption
properties of the X-ray detected DOGs. The X-ray observations of the
DOGs in our sample are listed in Table \ref{sample}. Redshifts were
extracted from \citet{hsu14}, spectroscopic redshifts being available
only for four of the sources.

\begin{table*}[!htbp]
\caption{X-ray observations of the DOGs sample}
\label{sample}
\centering
\begin{tabular}{cccccccc}
\hline\hline
 LID   &  PID & RA & DEC & z& $f_{2-10\,{\rm keV}}$   & Cts(Chandra) & Cts (XMM)\\
(1)     &  (2)    & (3) &  (3)   & (4)  & (5)                           &  (6)                 &  (7) \\
  \hline
95  &  325 & 53.0349 & -27.6796 &  5.22   &   4.3 & 257 & 385  \\
117 &  (225)  & 53.0491 & -27.7745 & 1.51 &  2.5 &  964 & - \\
170 &  -   & 53.0720 & -27.8189 &  1.22 &  0.1 & 20 & - \\
197 &  140 & 53.0916 & -27.8532 &  1.81 &  0.6 &  125  &  126 \\
199 &  (193) & 53.0923 & -27.8032 &  2.45  &  0.6 & 213 & - \\   
230 &  -     & 53.1052 & -27.8752 &  2.61  & 0.1 & 48 & - \\
232 &  283 & 53.1070 & -27.7183 &  2.291  & 4.6 & 1791 & 2279 \\
293 &  -  & 53.1394 & -27.8744 &  3.88  & 0.2 & 81 & - \\
307 &  102 & 53.1467 & -27.8883 &  1.90  & 1.6  & 697 & 332 \\
309 &  172 & 53.1488 & -27.8211 &  2.579  & 1.2 & 129  & 131 \\
321 &  118 & 53.1573 & -27.8700 & 1.603 & 13.0 & 7661 & 9253 \\
326 &  - &  53.1597 & -27.9313 &  3.10  & 0.8 &  254 & - \\
346 &  64 & 53.1703 & -27.9297 &  1.221  & 5.8 &  1549 & 1581 \\
397 &  (74)  & 53.2049 & -27.9180 &  2.28  & 2.1 & 510 & - \\
\hline\hline
\end{tabular}
\begin{list}{}{}
\item The columns  are: (1) \chandra ID from the \citet{luo10} catalogue. (2) XMM ID from the \citet{ranalli13} catalogue; numbers in brackets denote sources that are detected by \xmm but with a limited number of counts (so no spectral fit has been carried out). (3) X-ray coordinates (J2000). 
(4) Redshift: two decimal and three decimal digits denote
photometric and spectroscopic redshifts, respectively. 
 (5) 2-10 keV \chandra flux in units of $10^{-15}$ \funits from the \citet{luo10} catalogue. 
(6) Background subtracted \chandra counts in the total band 0.3-8 keV
(7) Background subtracted  \xmm  EPIC(MOS+pn) counts in the 0.3-8 keV band.
\end{list}
\label{xrayobs}
\end{table*}

\section{X-ray Spectroscopy} 

To maximize the spectral quality of our sample, we extracted spectral
data from all the {\it Chandra} observations publicly available by
June 2015. In particular, we reduced all the {\it Chandra} survey data
in a uniform manner, screening for hot pixels and cosmic afterglows as
described in \citet{laird09} with the {\tt CIAO} data analysis
software version 4.8. We used the {\tt SPECEXTRACT} script in the {\tt
  CIAO} package to extract spectral information from the individual
CDF-S observations. The extraction radius increases with off-axis
angle to enclose 90\% of the PSF at 1.5 keV.  The same script extracts
response and auxiliary files.  The background spectrum was estimated
from source free regions of the image for each observation.  The
spectra from each observation were then merged to create a single
source spectrum, background spectrum, response, and auxiliary matrices
for each source using the {\tt FTOOL} tasks {\tt MATHPHA}, {\tt
  ADDRMF}, and {\tt ADDARF,} respectively. The addition of the spectra
from all observations results in a maximum exposure of $\sim$6
Ms. However, sources near the edges of the field may not be present in
all individual observations because the aim points and roll angles
vary between observations. In these cases, the total exposure is
significantly smaller.

We used {\tt Xspec} v12.8 \citep{arnaud} to carry out the spectral
analysis. We selected Cash statistics instead of the most commonly
used $\chi^{2}$ statistic to obtain reliable spectral results even for
very low count data.

The initial model was a power law modified by photoelectric absorption
at the source redshift. We modelled this absorption using the {\tt
  Xspec} model {\tt plcabs}, which properly takes not only absorption
but also Compton scattering into account, and can be applied up
to column densities $\sim$ 5$\times$10$^{24}$ cm$^{-2}$. If column
densities were found to be higher than this value, we attempted to use
the torus absorption model described in \citet{brightman11}, which can
be applied for higher column densities. We also added a second power
law, when neccessary, to account for soft scattered emission. We fixed
the photon index to 1.8 in all cases \citep{tozzi06} to better
constrain the intrinsic absorption.

The spectral fitting results are listed in Tables \ref{jointfits} and
\ref{chandrafits}. Upper limits correspond to components not
statistically neccesary in the spectral fits. The spectral fits are
plotted in the left panel of Fig.~\ref{ctsxrayfit}. Given the modest
quality of many X-ray spectra in our sample, we did not attempt to fit
more complex models, although some sources actually display signatures
of thermal emission; see for example the fitting residuals for source
346. To evaluate the improvement with respect to the 4 Ms {\it
  Chandra} spectral data, we computed the confidence limits on the
measured column density for both datasets. The results are plotted in
the middle panel of Fig.~\ref{ctsxrayfit}, which shows that the
additional 2 Ms provided stronger constraints for the column density
values in most cases.

Significant differences between the fluxes reported in Tables
\ref{jointfits} and \ref{chandrafits}, and those in Table
\ref{xrayobs}, can be attributed to significant deviations of the
spectral shape from a simple power law, which was the model assumed to
derive the fluxes in the \citet{luo08} catalogue. Variability could
also explain the flux differences, given the large time span covered
by the 6 Ms observations. To further explore this possibility, we
attempted to study the variability properties in our sample. Although
the CDF-S field has been observed many times during more than ten
years, the rather short exposure time of the individual observations,
given our sources average fluxes, does not allow us to derive the
individual fluxes for each individual observation. Therefore, we
divided the $\sim$ 10 years of observations into 4 epochs: up to 2000,
from 2000 to 2007, from 2007 to 2011, and from 2011 to 2015. We were
able to extract fluxes and errors for our entire sample except for
source 170, which was only detected in two epochs, and source 230,
which was only detected in three epochs. To quantify the variability
of our sample, we fitted a constant model to the available data. The
results are listed in Table~\ref{variability}. The sources showing the
most significant differences in the fluxes reported in Tables
\ref{jointfits}, \ref{chandrafits}, and \ref{xrayobs}, are also those
showing more variability, except for source 95, whose differences can
easily be attributed to the very different models used.

\begin{figure}[h]
\centering
   \includegraphics[angle=-90,width=9cm]{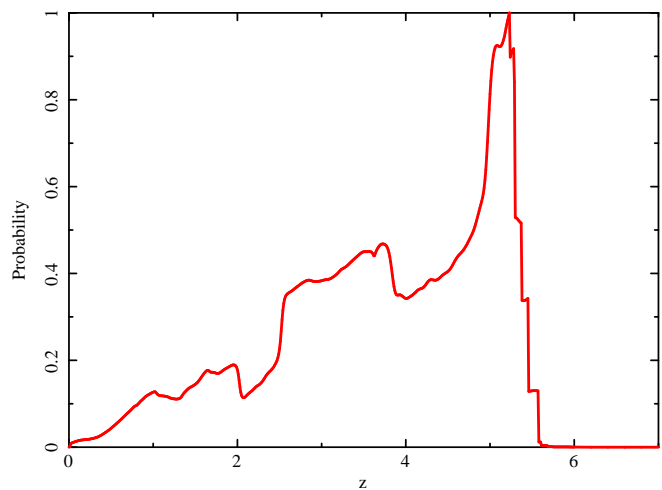} 
 \caption{Probability density function corresponding to the photometric redshift
   determination of source 95 \citep{hsu14}.}
\label{95_photz}
\end{figure}

We find that nine sources are highly absorbed with column densities
above 10$^{23}$ cm$^{-2}$, eight of which are at the 90\% confidence
level. However, only three of these can be considered as CT AGN:
sources 95, 230, and 309 ({\it Chandra} ids), although source 95 is only
marginally CT. In the case of source 95, it is very surprising to be
able to detect a near-CT AGN at redshift higher than 5. We examined
the probability density function (PDF) for the photometric redshift in
this case \citep{hsu14}. Although the PDF is somewhat flat towards
lower redshifts, there is a clear peak at z = 5.22 (see
Fig.~\ref{95_photz}). We also attempted to constrain the redshift
value directly from X-ray spectral fitting. There seems to be an
emission feature around 1.8 keV, although it is not very
significant. Assuming that this feature is real and corresponds to the
Fe K$\alpha$ emission line, we derived a z $\sim$ 2.8 and a resulting
column density of a few times 10$^{23}$ cm$^{-2}$. Source 309 was
already classified as a CT source in \citet{feruglio11}. Here we also
confirm its CT nature using {\it XMM-Newton} data. Although most CT
AGN display a very strong Fe K$\alpha$ emission line, we were unable
to detetect it in the case of source 230. However, CT AGN with very
high column densities and modest Fe K$\alpha$ emission lines are not
extremely rare, especially at high luminosities (L$_{X}>$ 10$^{44}$
erg s$^{-1}$; see \citealt{fuka11,iwa12}). Finally, although most of
the sources are obscured (N$_{H}$ $>$ 10$^{22}$ cm$^{-2}$, and $>$
10$^{23}$ cm$^{-2}$ in most cases), we also found that two sources
seem to be unabsorbed (sources 170 and 326).

Variability, and especially the lack of variability, has also been
proposed as a method to pinpoint CT sources. As can be seen in
Table~\ref{variability}, our CT candidates are among the less variable
sources, however, our observations span many years and CT AGN have
been shown to display variability in such long timescales.
 
\begin{table*}[!htbp]
\caption{\xmm and \chandra joint spectral fits} 
\label{jointfits}
\centering
\begin{tabular}{cccccccccc}
\hline\hline
LID &  z     & $\rm N_H$           &  P1/P2 & cstat/dof & EW     &  Flux             & log L$_{\rm X}$       & log L$_{\rm Xunabs}$ \\
    &        & 10$^{22}$ cm$^{-2}$  &        &           & keV    & $10^{-15}$ \funits &  \lunits &  \lunits  \\ 
(1) & (2) & (3) & (4) & (5) & (6) & (7) & (8 & (9)\\
 \hline
95  & 5.22  & $81^{+20}_{-15}$     & $<0.006$  & 2566/2427 & $<0.4$  &  $1.9^{+2.3}_{-1.4}$     & $43.6^{+0.4}_{-0.6}$  & 44.7\\
197 &  1.81 & $44^{+30}_{-12}$    & 0.005  & 2103/3031 & $0.20^{+0.07}_{-0.07}$ & $0.74^{+0.27}_{-0.28}$   & $42.63^{+0.13}_{-0.22}$  & 43.34 \\
232 &  2.291 & $15^{+2}_{-2}$       & 0.001  & 2917/3031 & $0.13^{+0.08}_{-0.10}$ & $5.3^{+0.3}_{-0.3}$ & $43.97^{+0.02}_{-0.03}$ & 44.30 \\
307  & 1.90  & $19^{+6}_{-3}$       & 0.01  & 2018/2173 & $0.34^{+0.19}_{-0.20}$  & $1.44^{+0.14}_{-0.20}$ & $43.20^{+0.05}_{-0.06}$ & 43.57 \\
309  & 2.579 & $580^{+550}_{-240}$ & 0.001 & 1548/1760 & $1.0^{+0.25}_{-0.29}$ & $1.5^{+0.3}_{-0.2}$  & $42.75^{+0.05}_{-0.05}$ & 44.18 \\
321  & 1.603 & $1.70^{+0.14}_{-0.13}$ & $<$0.0001 & 2509/2365 & $<0.10$ & $12.1^{+0.03}_{-0.03}$  & $44.21^{+0.01}_{-0.01}$ &  44.25 \\
346 & 1.221 & $13^{+3}_{-2}$ & 0.02 & 2773/2667 & $0.070^{+0.002}_{-0.002}$ & $6.0^{+0.4}_{-0.5}$  & $43.44^{+0.03}_{-0.03}$ & 43.69 \\
\hline \hline
\end{tabular}
\begin{list}{}{}
\item The columns are: (1) \chandra ID from the \citet{luo10}
  catalogue. (2) Redshift. (3) Intrinsic column density. (4) Ratio between the
  power-law normalizations in case of a double power-law model. (5)
  C-stat value to degrees of freedom ratio. (6) Fe K$\alpha$
  rest-frame equivalent width. (7) Observed flux in the 2-10 keV band. (8)
  Observed luminosity in the 2-10 keV band. (9) Unabsorbed luminosity
  in the 2-10 keV band.
\end{list}
\end{table*}

\begin{table*}[!htbp]
\caption{\chandra spectral fits} 
\label{chandrafits}
\centering
$$
\begin{tabular}{ccccccccc}
\hline\hline
LID  & z  & $\rm N_H$  &     P1/P2 &   cstat/dof   & EW  &  Flux  &         log L$_{\rm X}$       & log L$_{\rm Xunabs}$ \\
 &    &      10$^{22}$ cm$^{-2}$   &    &       & keV & $10^{-15}$ \funits & \lunits  & \lunits \\ 
(1) & (2) & (3) & (4) & (5) & (6) & (7) & (8) & (9)\\
 \hline
117 & 1.51 & $3.5^{+0.6}_{-0.6}$ & 0.02 & 392/390 & $0.025^{+0.200}_{-0.025}$ & $1.46^{+0.05}_{-0.05}$  & $43.22^{+0.02}_{-0.04}$ & 43.41 \\
170 & 1.22  & $<0.8$  & $<0.2$ & 93/117 &$<1.2$ & $0.10^{+0.02}_{-0.02}$ & $41.50^{+0.01}_{-41}$ & 41.50 \\
199 & 2.45  & $28^{+10}_{-6}$  & $<$0.02 & 162/205 & $0.52^{+0.43}_{-0.47}$  & $0.64^{+0.20}_{-0.20}$  & $42.99^{+0.04}_{-0.04}$ & 43.52 \\
230  & 2.61  & $900^{+500}_{-160}$  & 0.001 & 119/135 & $<0.43$ & $0.70^{+0.3}_{-0.3}$  & $42.37^{+0.02}_{-0.02}$ & 44.82 \\ 
293 & 3.88 & $20^{+32}_{16}$   & $<$0.002   & 195/228 & $<$1.0 & $0.20^{+0.07}_{-0.07}$ & $42.92^{+0.01}_{-0.01}$ & 43.31 \\ 
326 &  3.10  &$<7$ & $<0.01$ & 211/287 & $<0.3$ & $0.43^{+0.01}_{-0.01}$ & $43.36^{+0.01}_{-0.01}$ & 43.45  \\
397  &  2.28 & $1.3^{+2.1}_{-1.1}$ & $<$0.04 & 309/324 & $<0.2$ & $1.06^{+0.3}_{-0.3}$ & $43.44^{+0.02}_{-0.02}$ & 43.56 \\
\hline \hline
\end{tabular}
$$
\begin{list}{}{}
\item The columns are: (1) \chandra ID from the \citet{luo10}
  catalogue. (2) Redshift. (3) Intrinsic column density. (4) Ratio
  between the power-law normalizations in case of a double power-law
  model. (5) C-stat value to degrees of freedom ratio. (6) Fe
  K$\alpha$ rest-frame equivalent width. (7) Observed flux in the 2-10 keV
  band. (8) Observed luminosity in the 2-10 keV band. (9) Unabsorbed
  luminosity in the 2-10 keV band.
\end{list}
\end{table*}

\begin{table}[!htbp]
\caption{Variability of X-ray detected DOGS.} 
\centering
$$
\begin{tabular}{ccc}
\hline\hline
LID  & Epochs  & Reduced $\chi^{2}$\\
 \hline
95 & 4 & 0.38 \\
117 & 4 & 3.31 \\ 
170 & 2 & - \\
197 & 4 & 1.13 \\
199 & 4 & 0.99 \\
230 & 3 & 0.15 \\
232 & 4 & 9.35  \\
293 & 4 & 0.68 \\
307 & 4 & 0.31 \\
309 & 4 & 0.64 \\
321 & 4 & 25.72 \\
326 & 4 & 1.09 \\
346 & 4 & 3.36 \\
397 & 4 & 1.87 \\
\hline \hline
\end{tabular}
$$
\begin{list}{}{}
\item Results from $\chi^{2}$ fitting assuming a constant model.
\end{list}
\label{variability}
\end{table}

\begin{figure*}[!htbp]
\centering
$$
\begin{array}{ccc}
    \includegraphics[angle=0,width=0.3\textwidth]{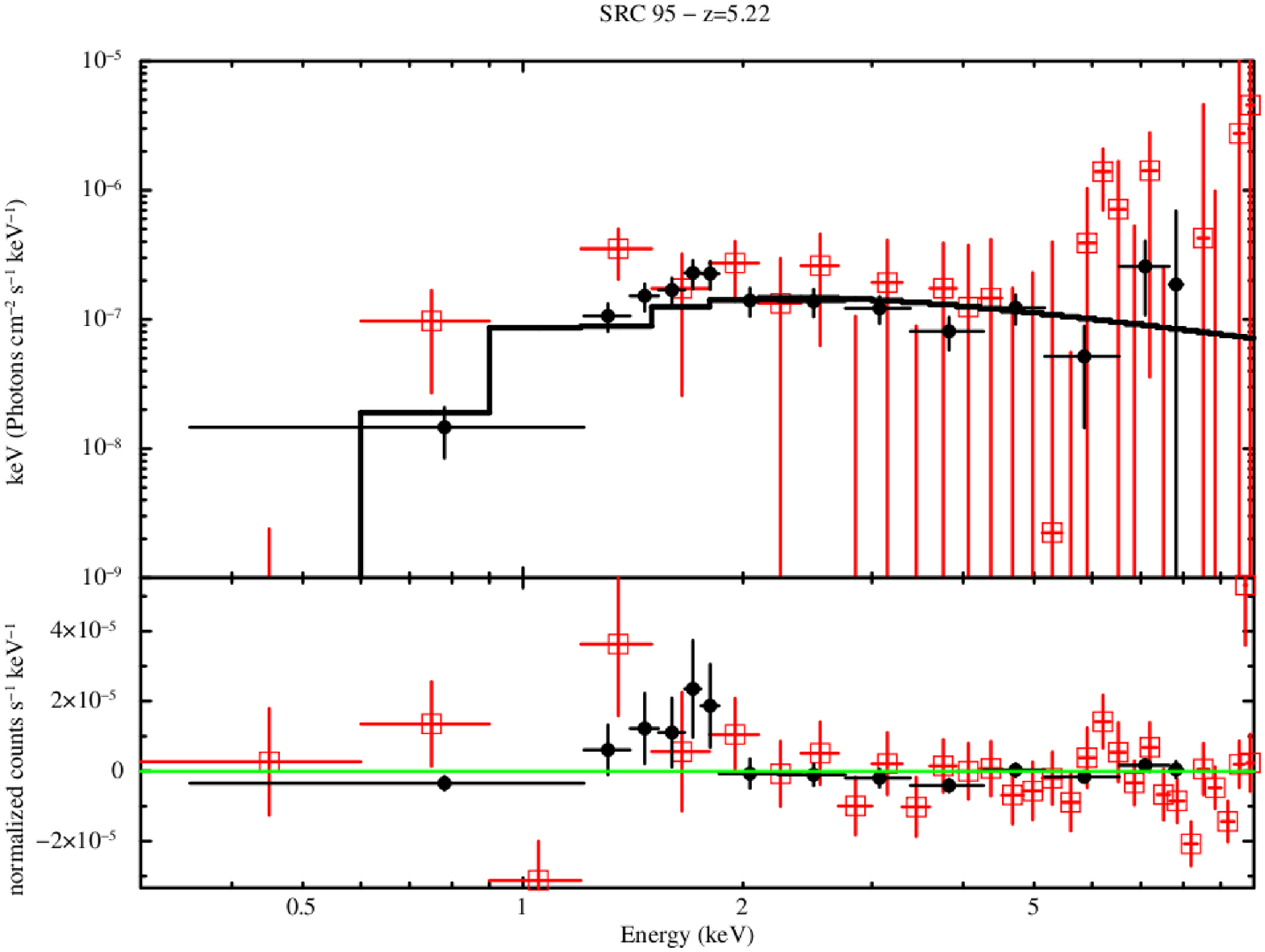}    &    \includegraphics[angle=0,width=0.3\textwidth]{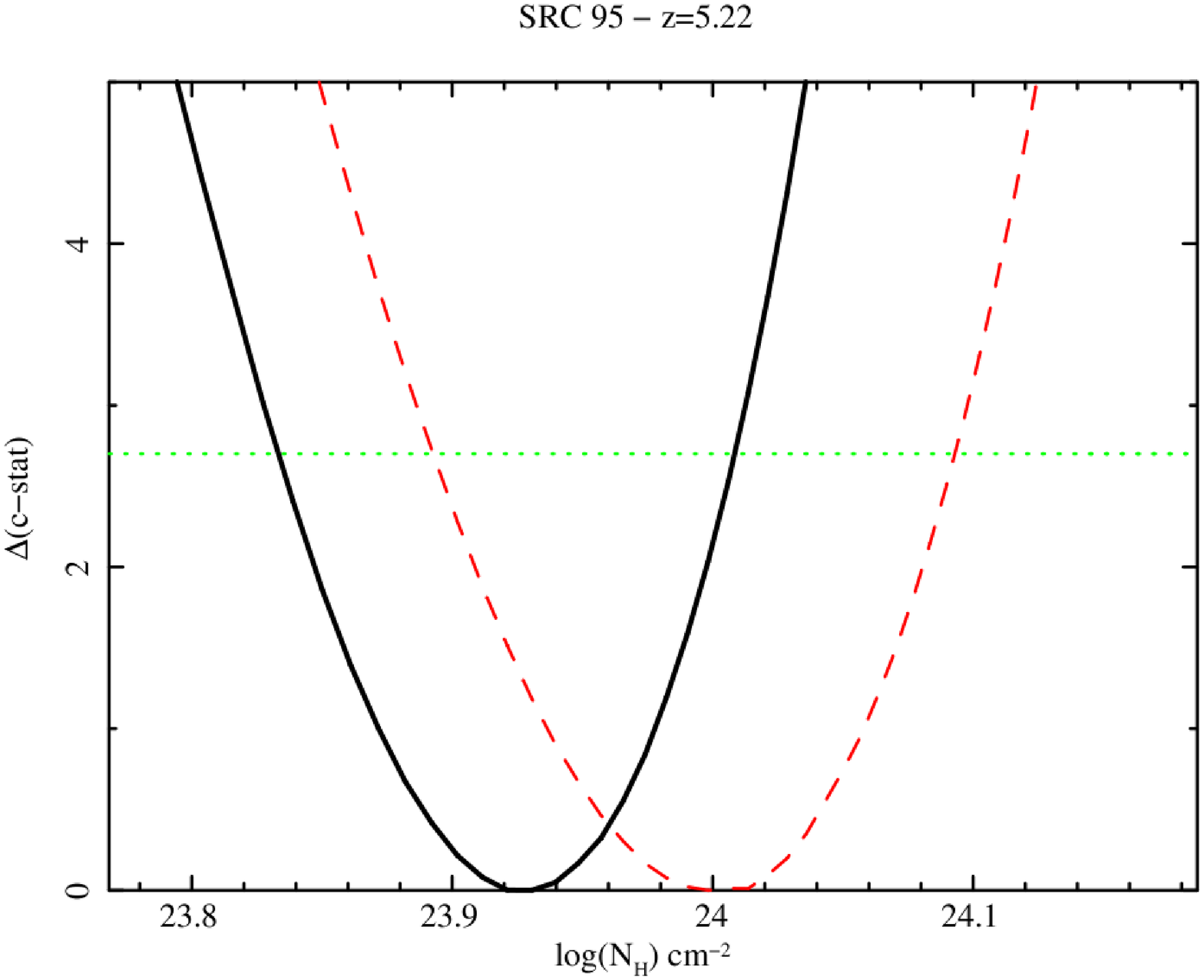}    &    \includegraphics[angle=0,width=0.3\textwidth]{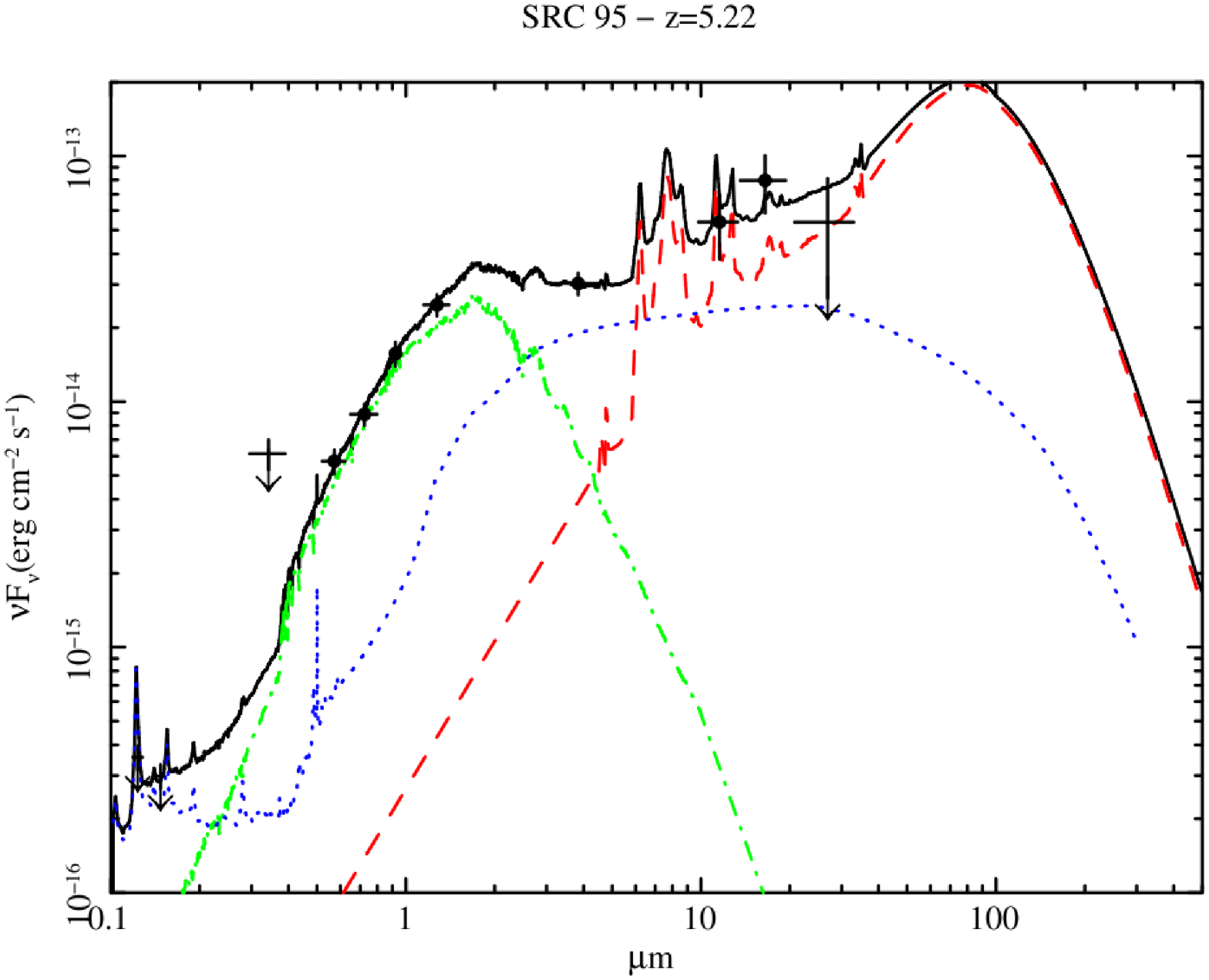}\\
    \includegraphics[angle=0,width=0.3\textwidth]{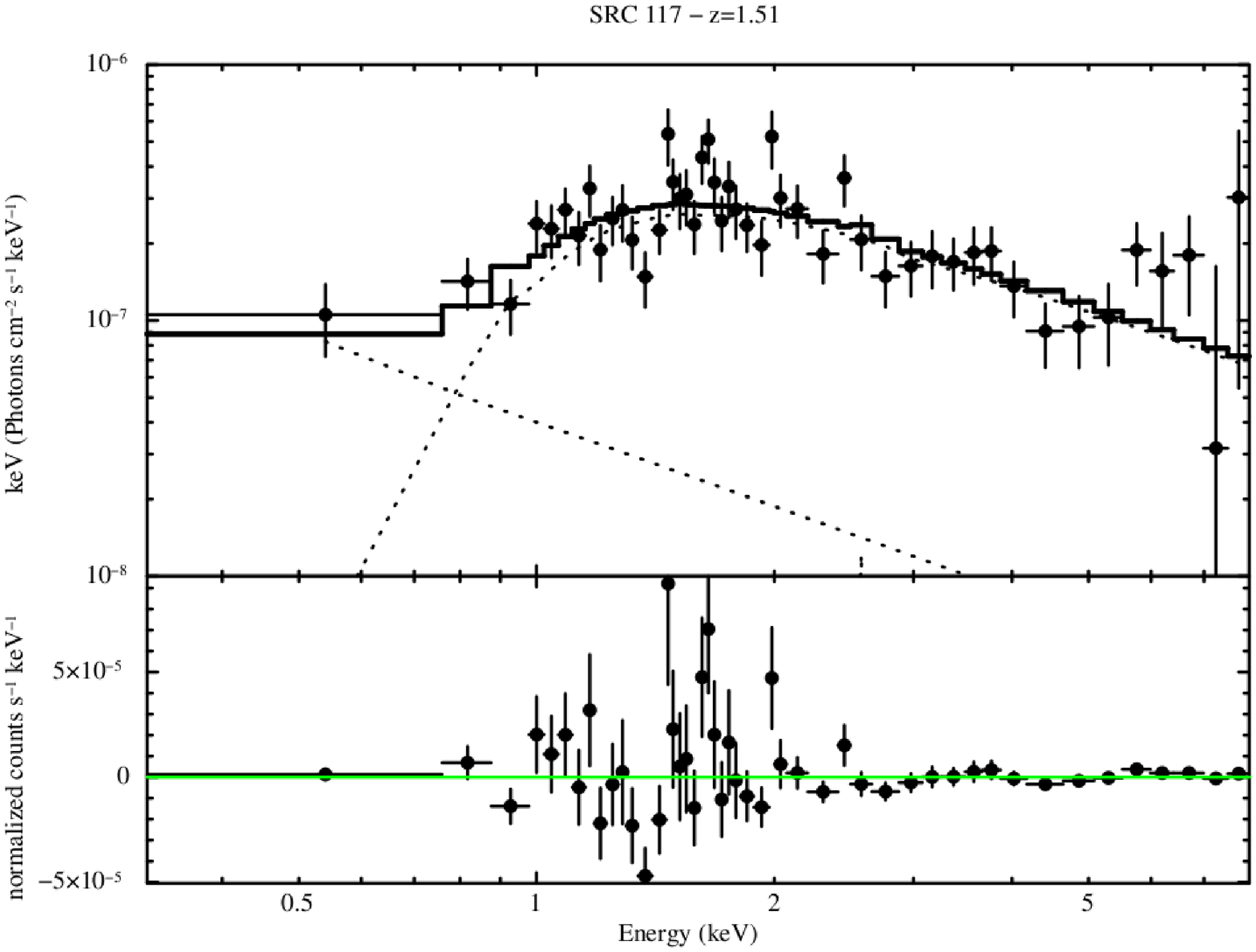}    &    \includegraphics[angle=0,width=0.3\textwidth]{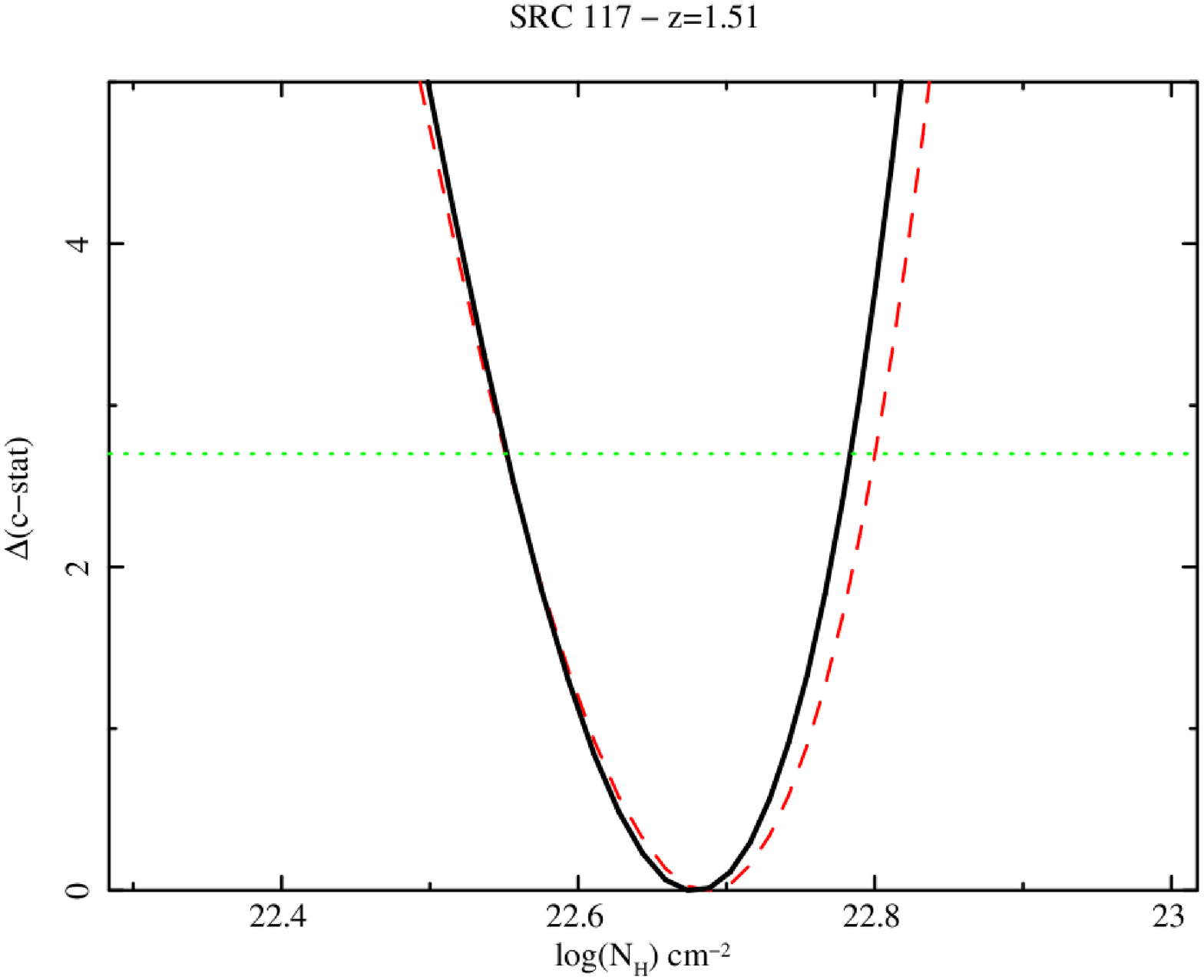}    &    \includegraphics[angle=0,width=0.3\textwidth]{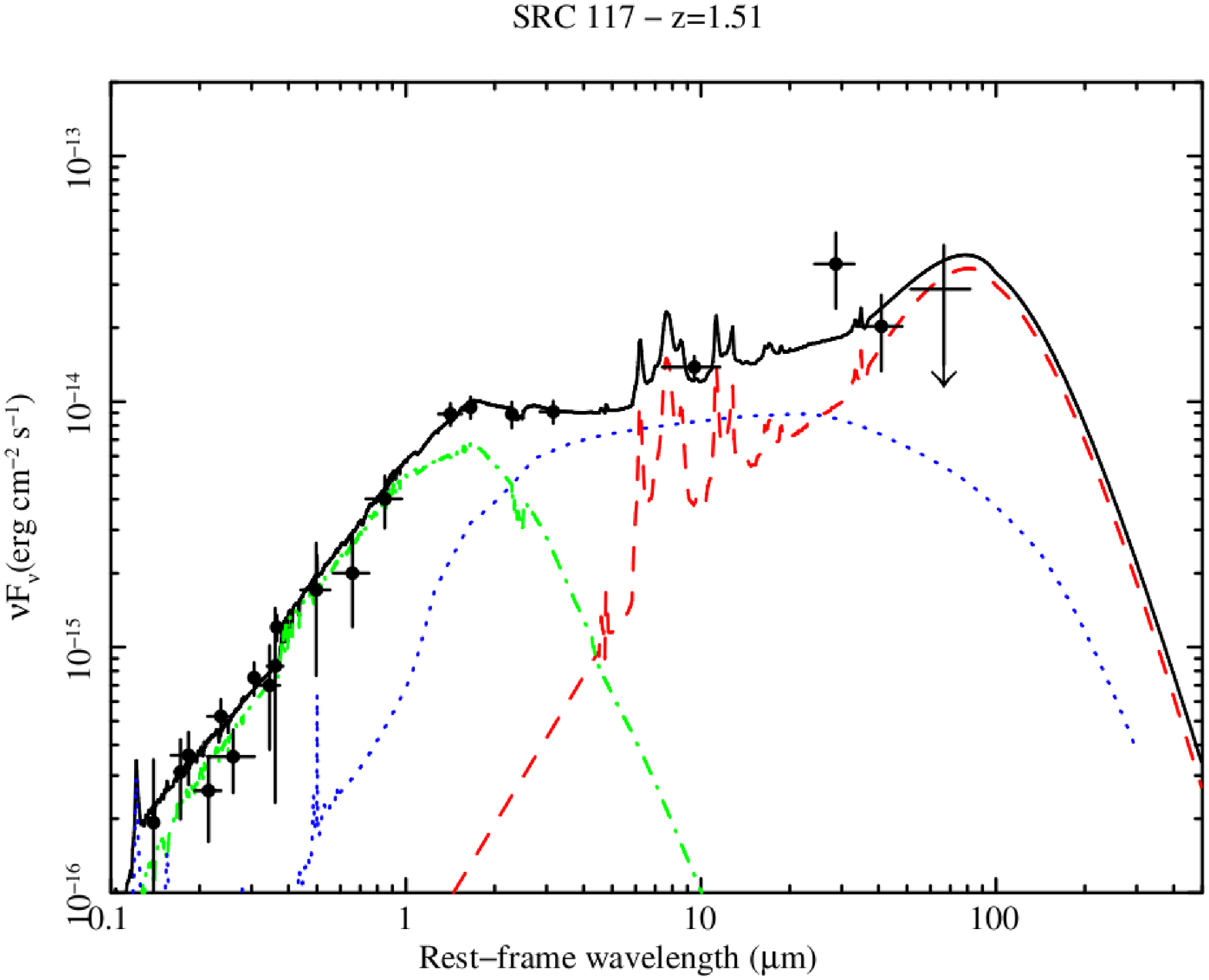}\\
    \includegraphics[angle=0,width=0.3\textwidth]{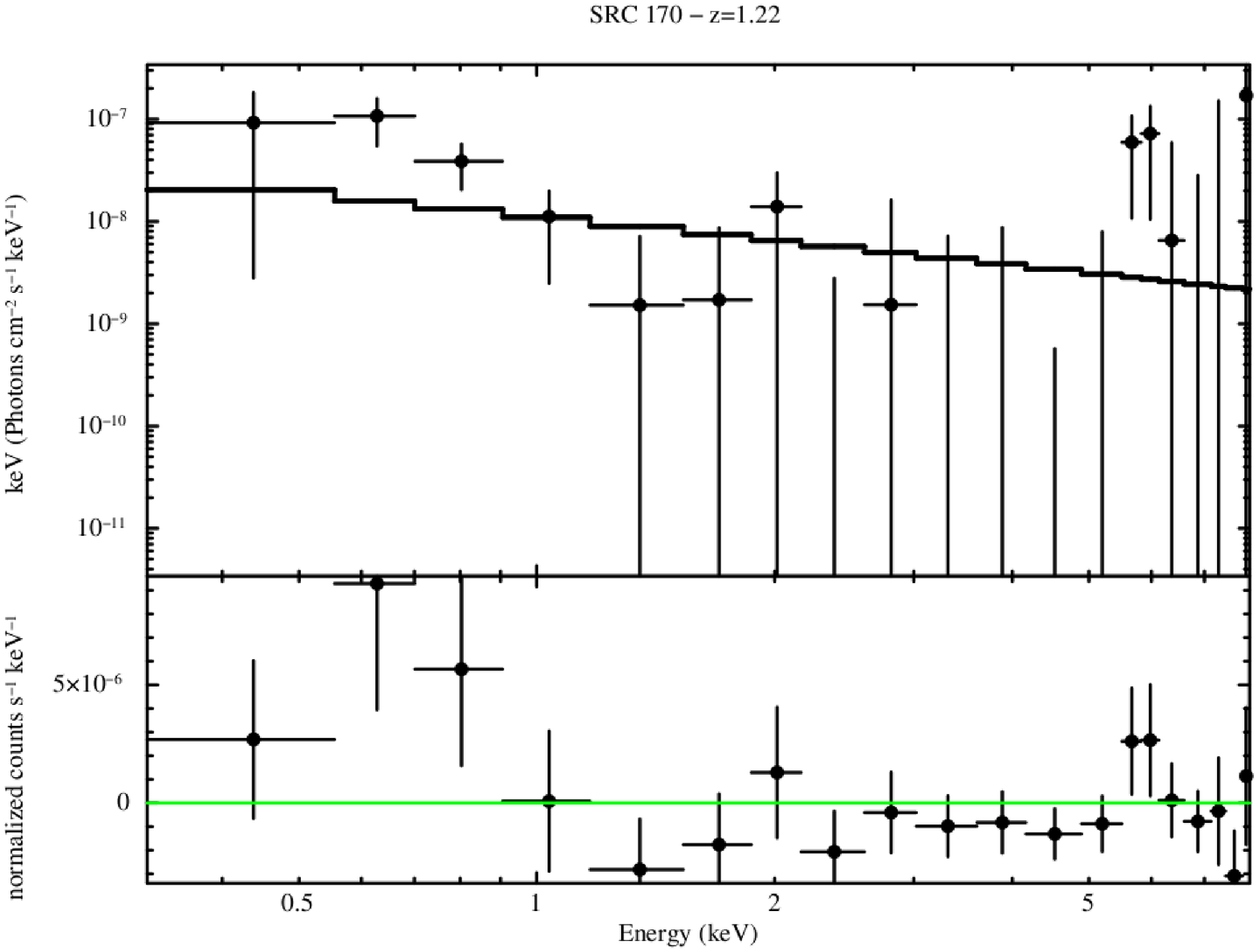}    &    \includegraphics[angle=0,width=0.3\textwidth]{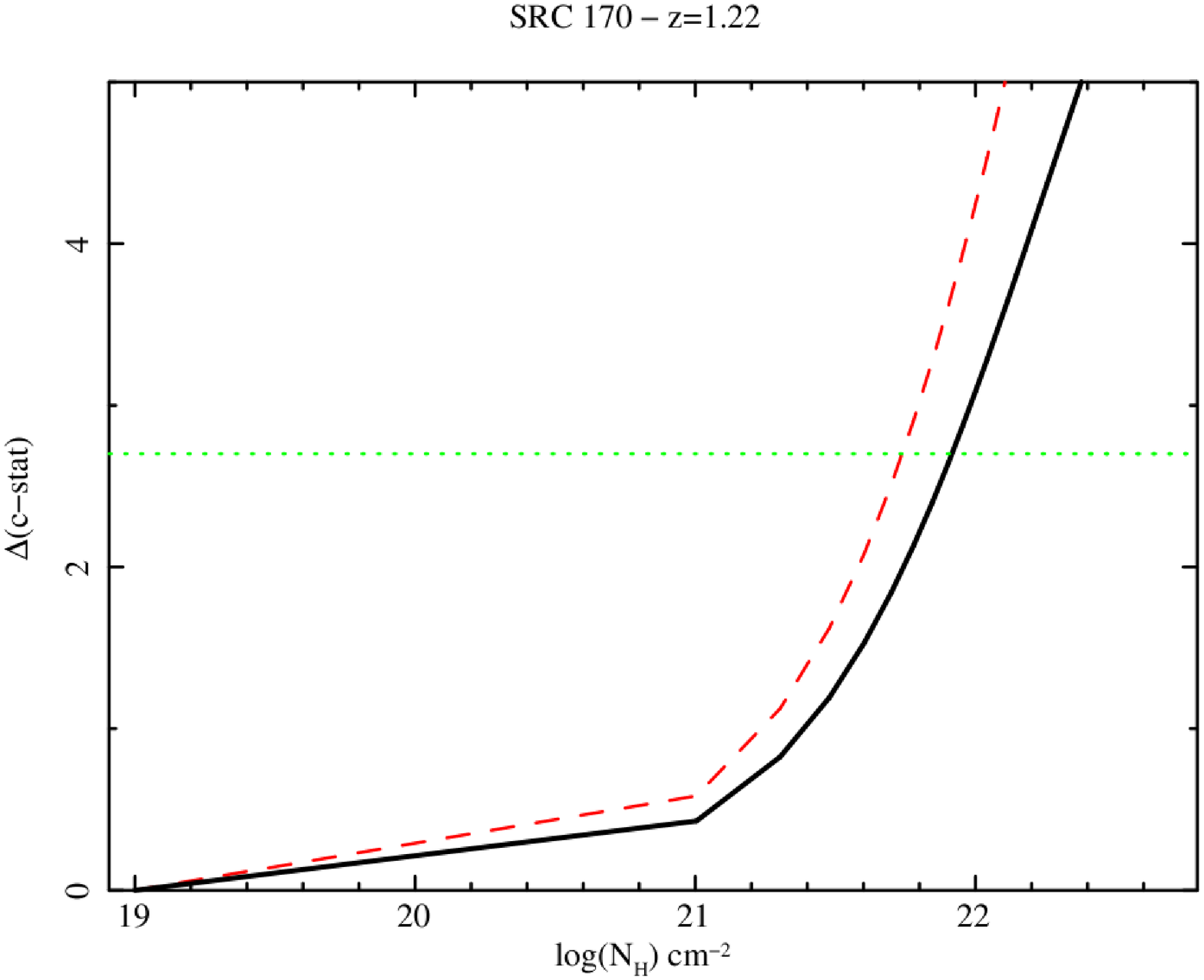}    &    \includegraphics[angle=0,width=0.3\textwidth]{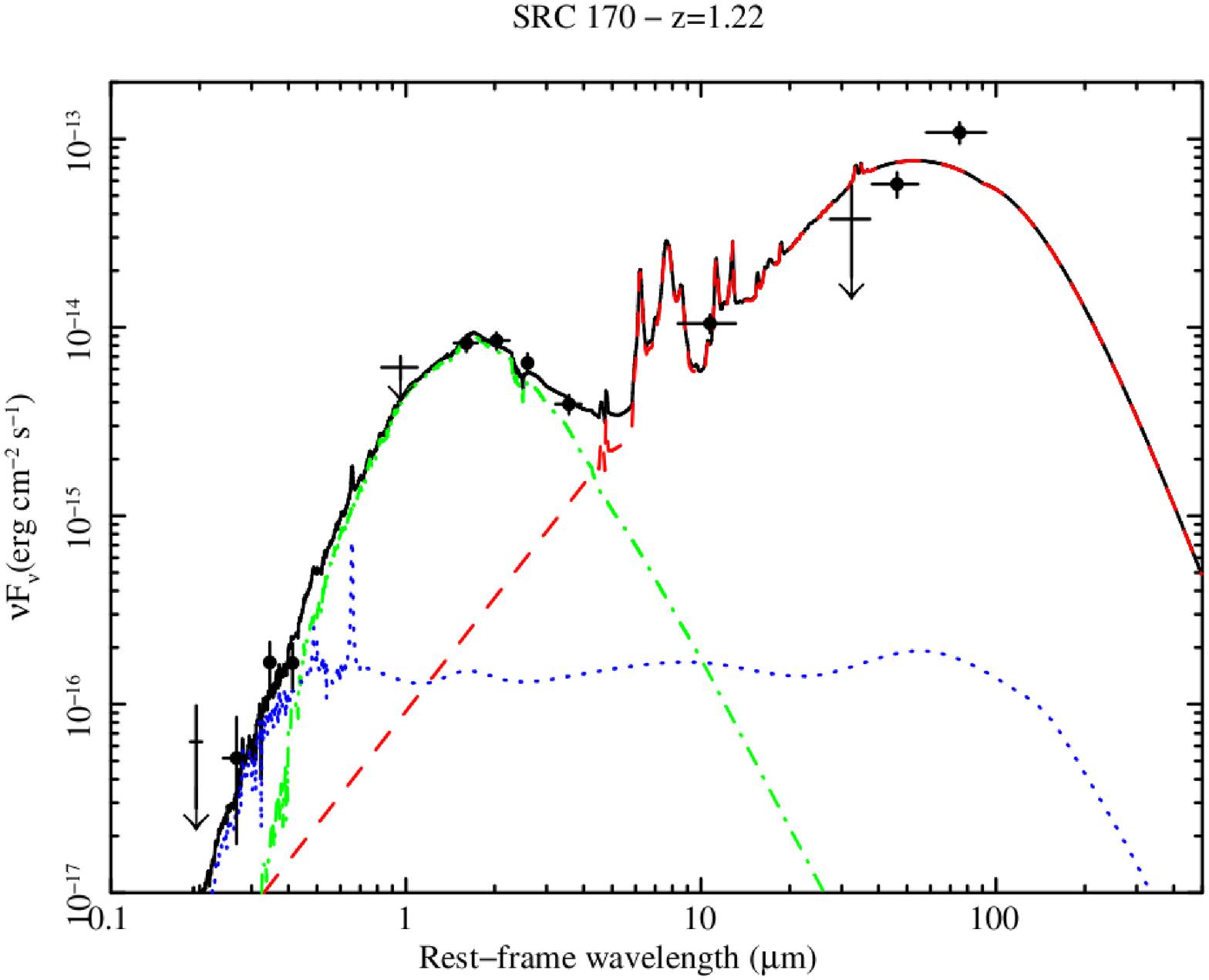}\\
    \includegraphics[angle=0,width=0.3\textwidth]{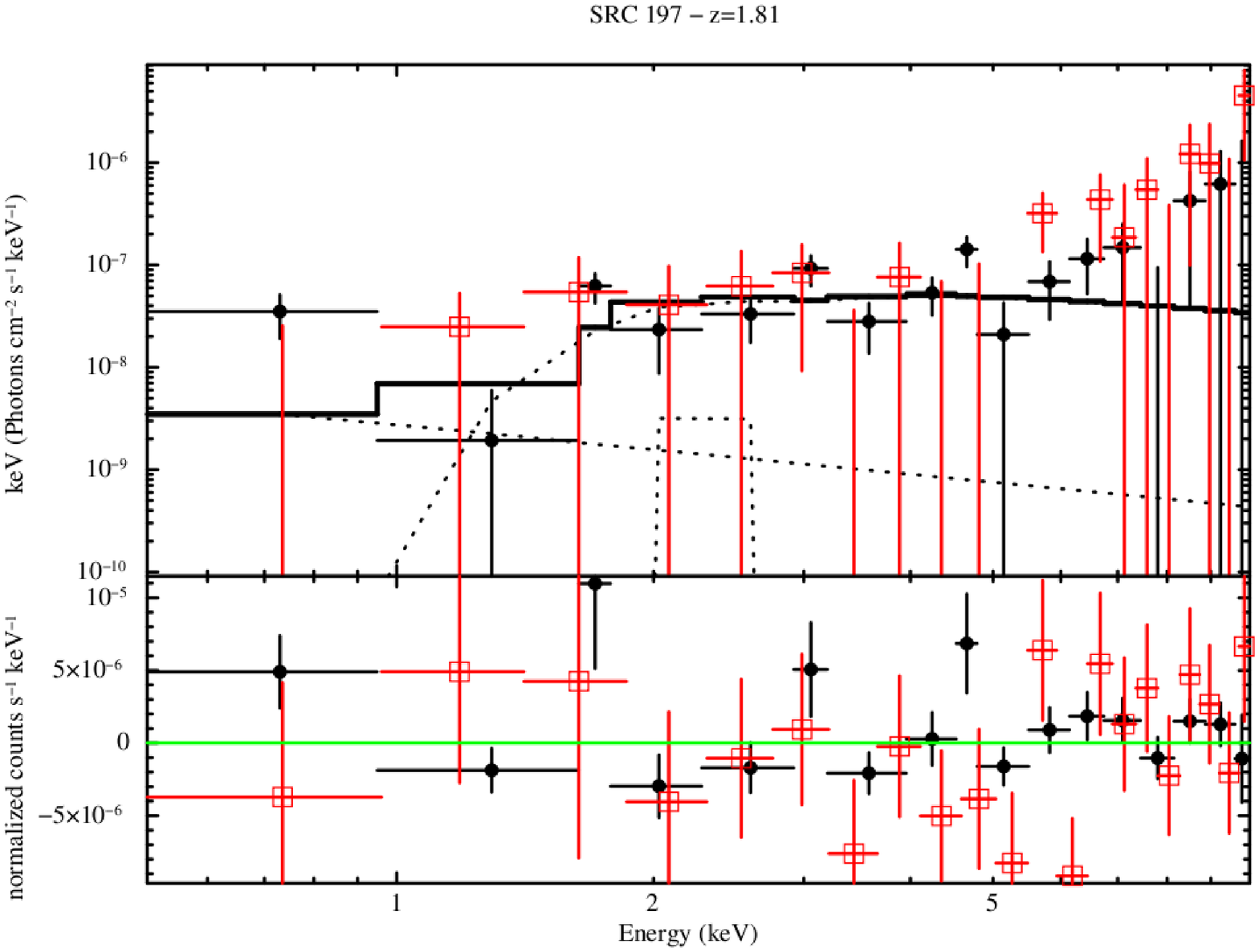}    &    \includegraphics[angle=0,width=0.3\textwidth]{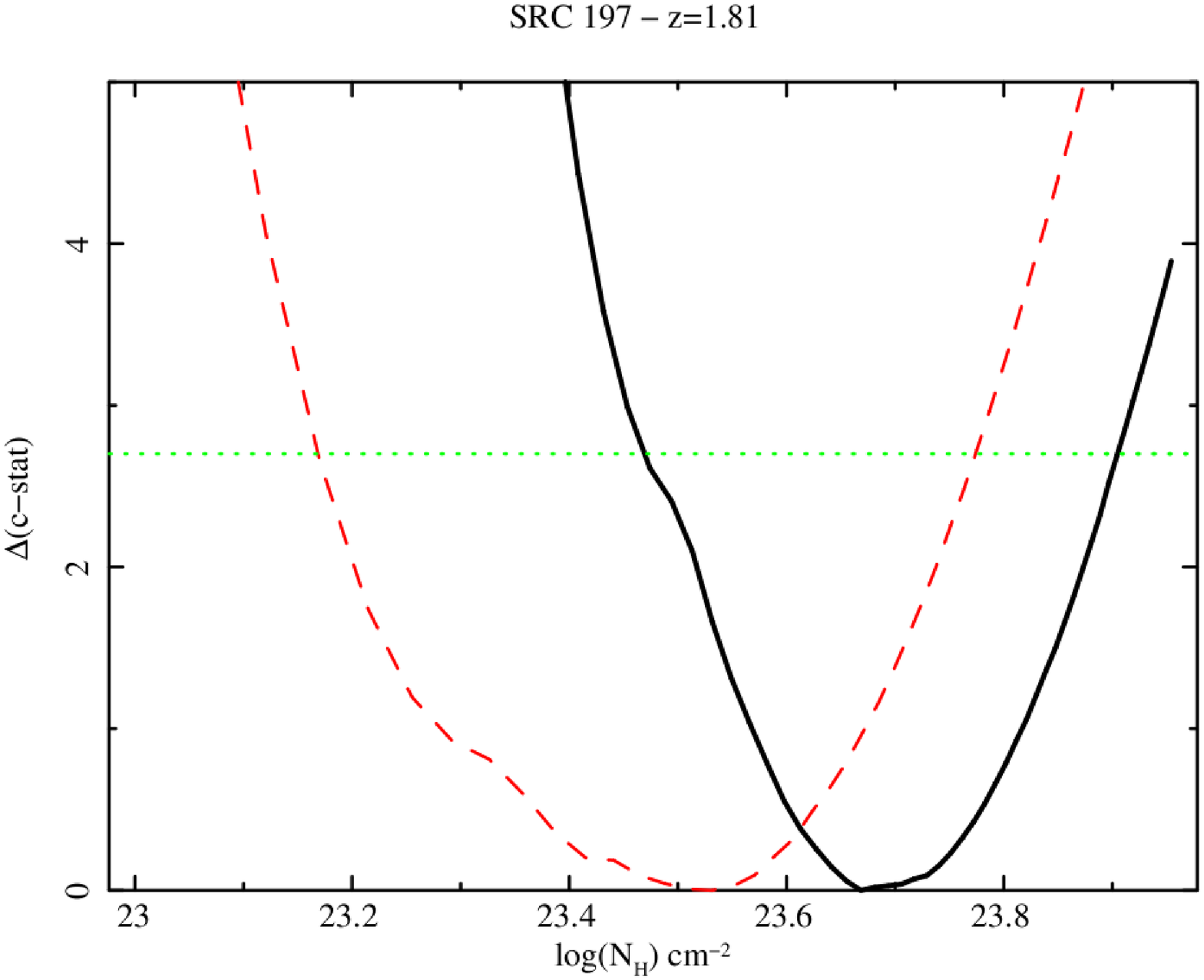}    &    \includegraphics[angle=0,width=0.3\textwidth]{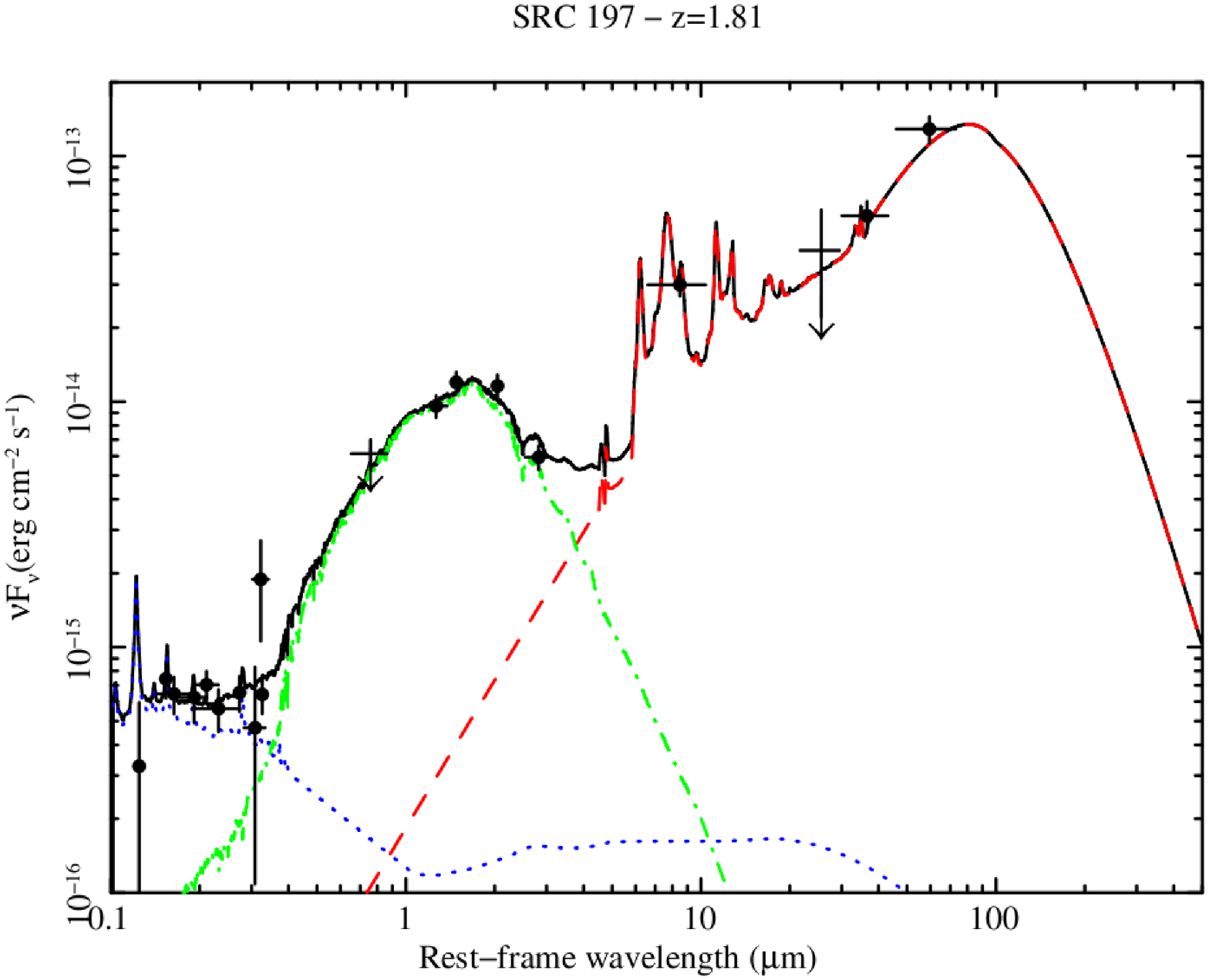}\\
    \includegraphics[angle=0,width=0.3\textwidth]{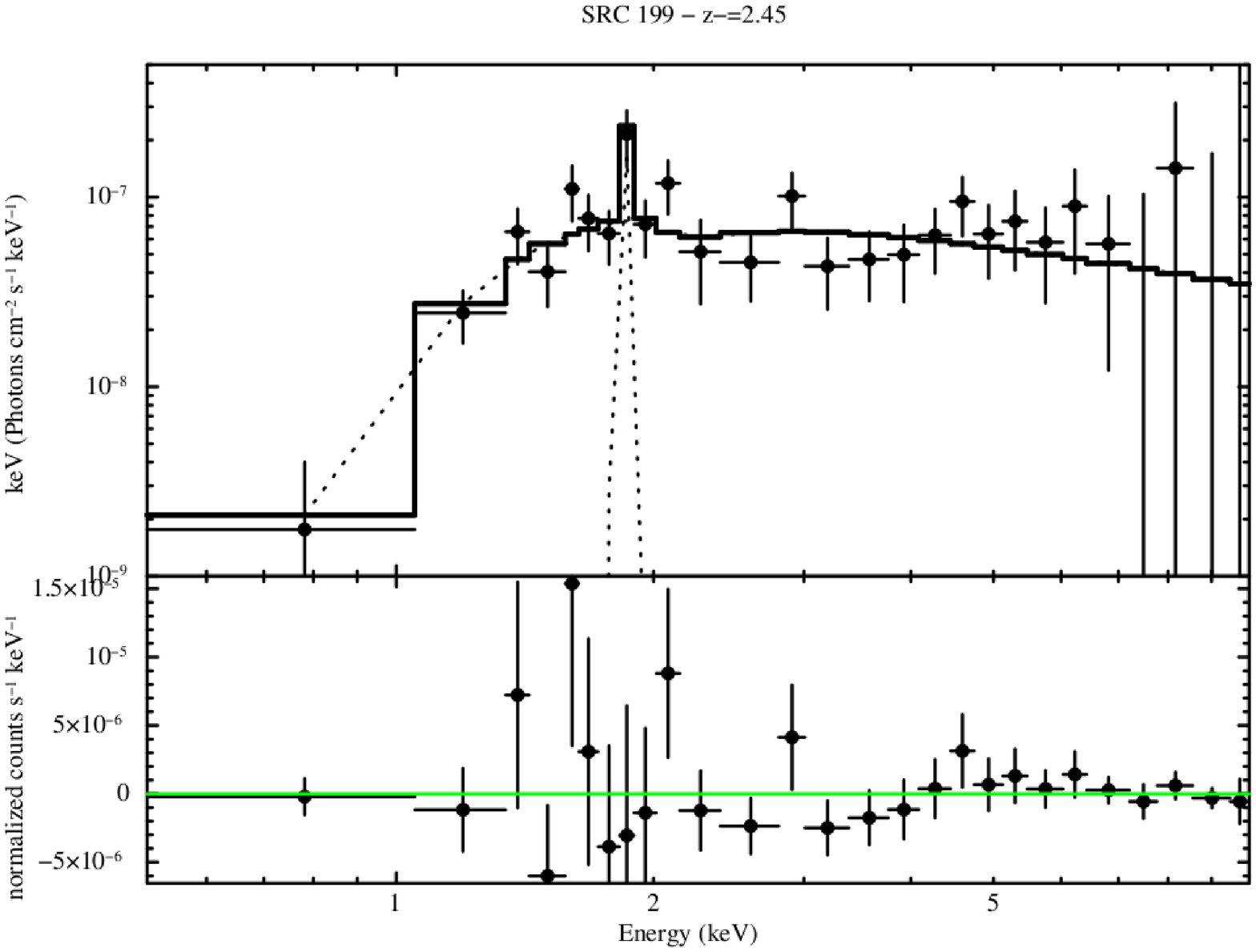} &    \includegraphics[angle=0,width=0.3\textwidth]{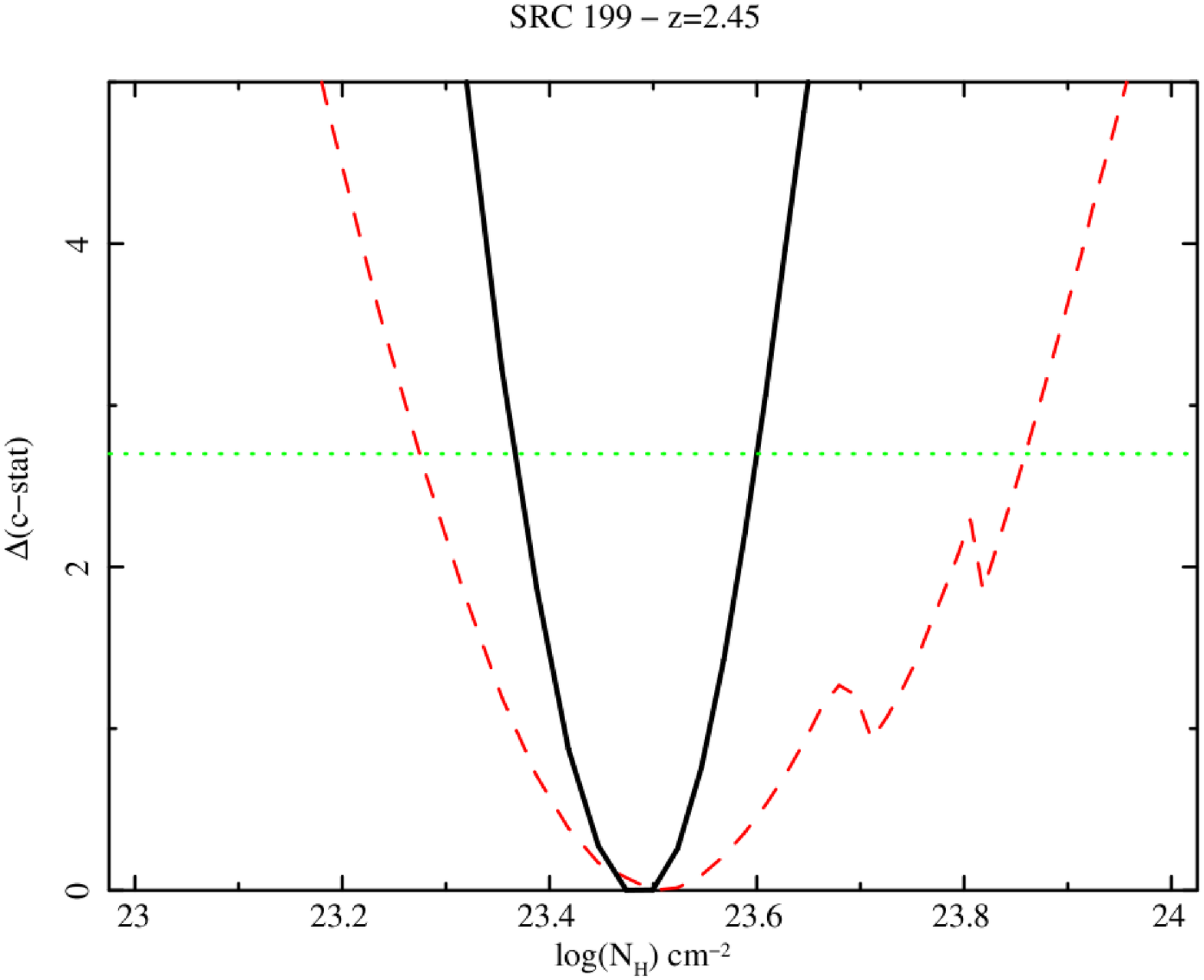} &    \includegraphics[angle=0,width=0.3\textwidth]{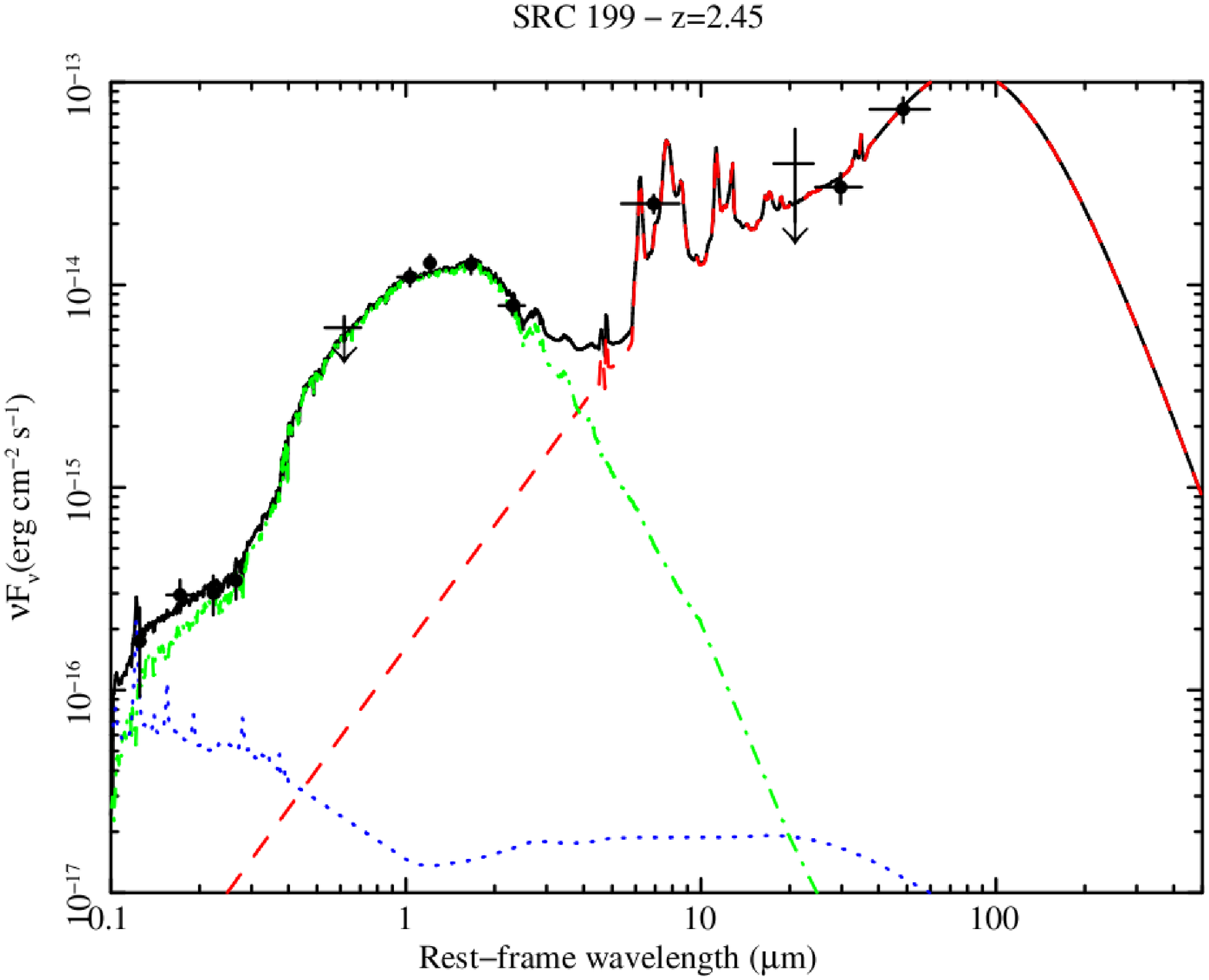}\\
\end{array}
$$
\caption{{\it Left:} X-ray spectral fits (unfolded model and
  residuals) for the X-ray detected DOGs in our sample. Filled circles
  and empty squares correspond to {\it Chandra} and {\it XMM-Newton}
  data, respectively. {\it Middle:} Comparison between the confidence
  contours for the column density values computed for the 4 Ms data
  (dashed line) and 6 Ms data (solid line). The horizontal dotted
  line indicates the 90\% confidence level. {\it Right:} SED fits for
  our X-ray DOGs sample. Solid line: full model; dotted line: AGN
  contribution; dash-dotted line: stellar contribution; and dashed line:
  starburst contribution. Our Compton-thick candidates are sources 95,
  230, and 309.}
\end{figure*}

\begin{figure*}[!htbp] 
\ContinuedFloat
\centering
$$
\begin{array}{ccc}
    \includegraphics[angle=0,width=0.3\textwidth]{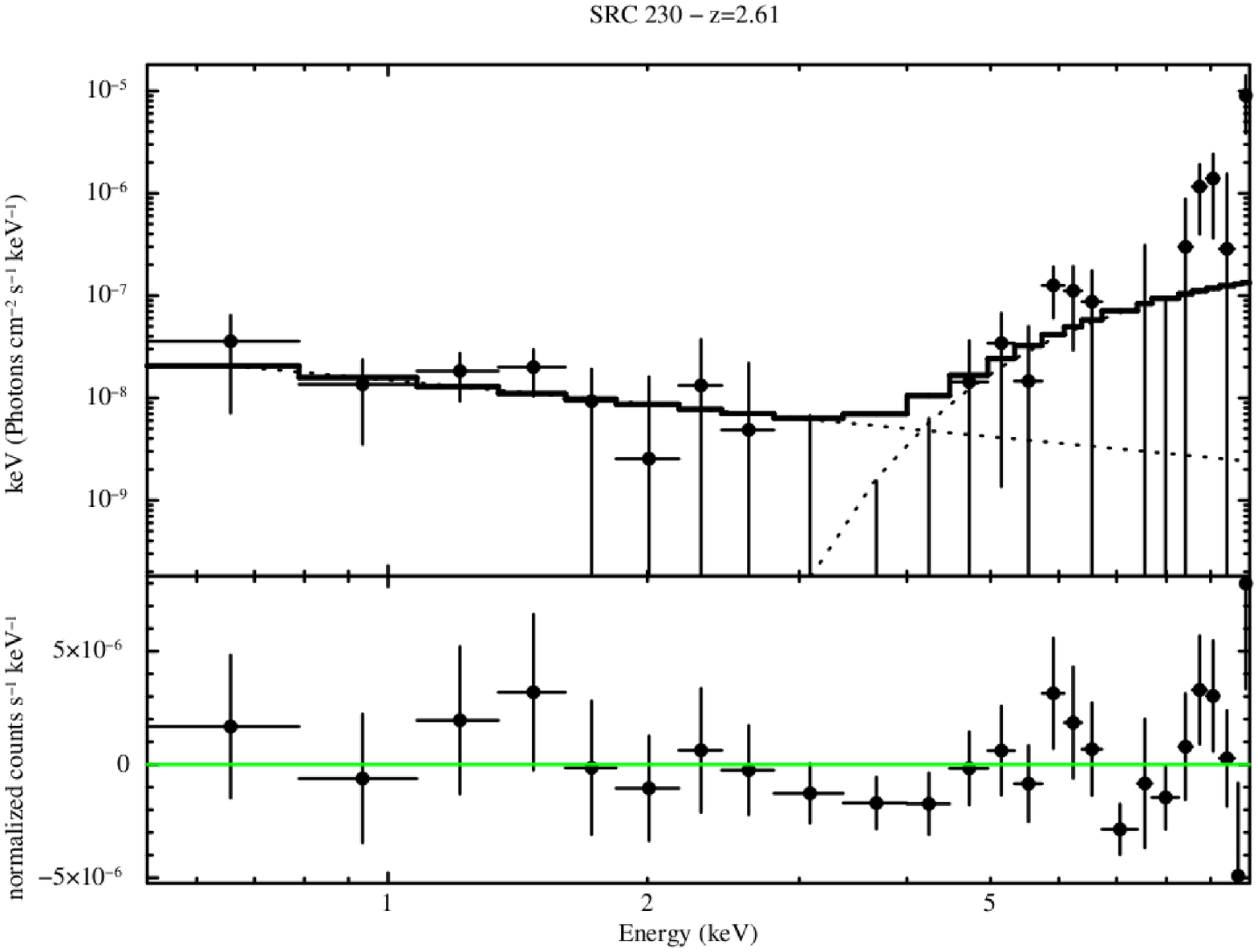} &    \includegraphics[angle=0,width=0.3\textwidth]{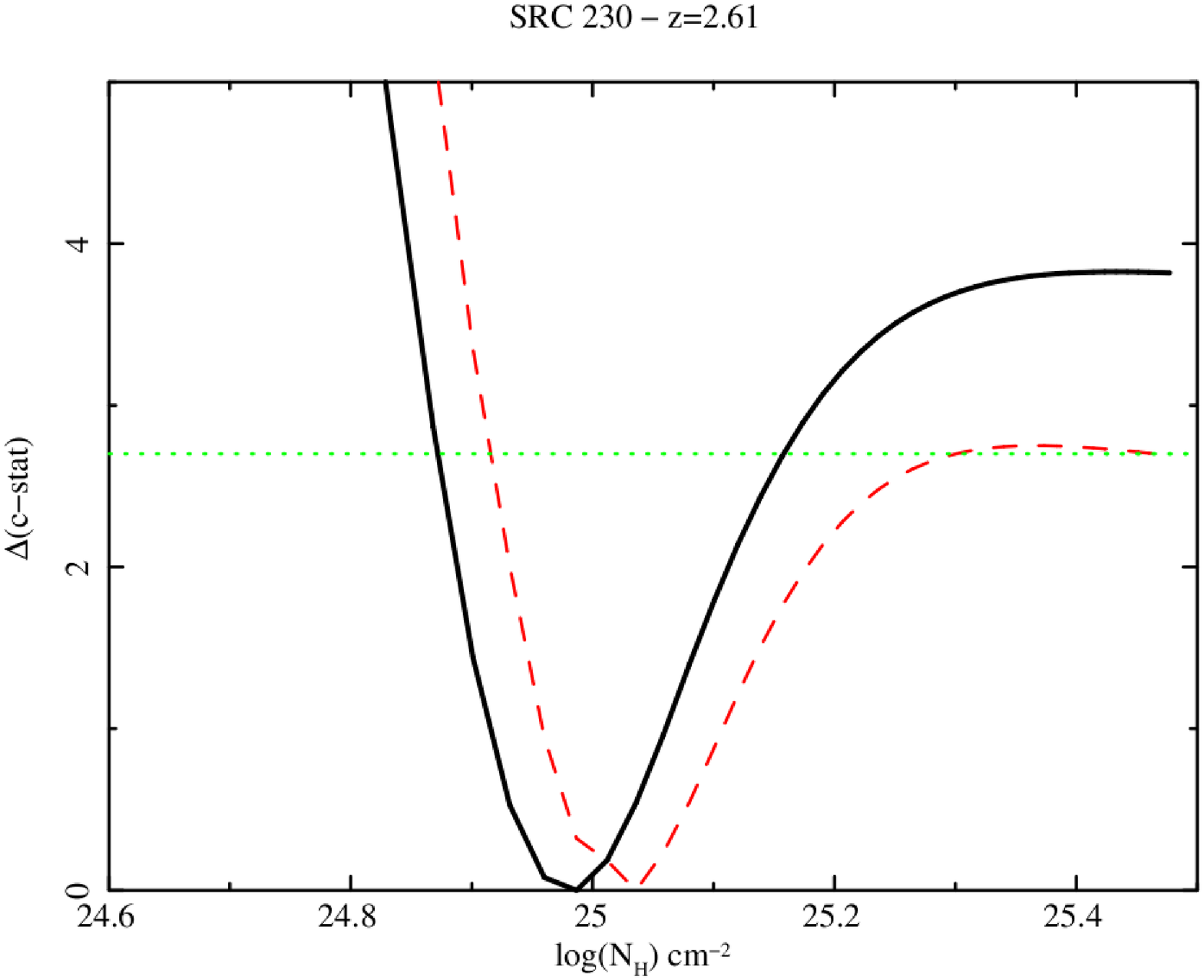} &    \includegraphics[angle=0,width=0.3\textwidth]{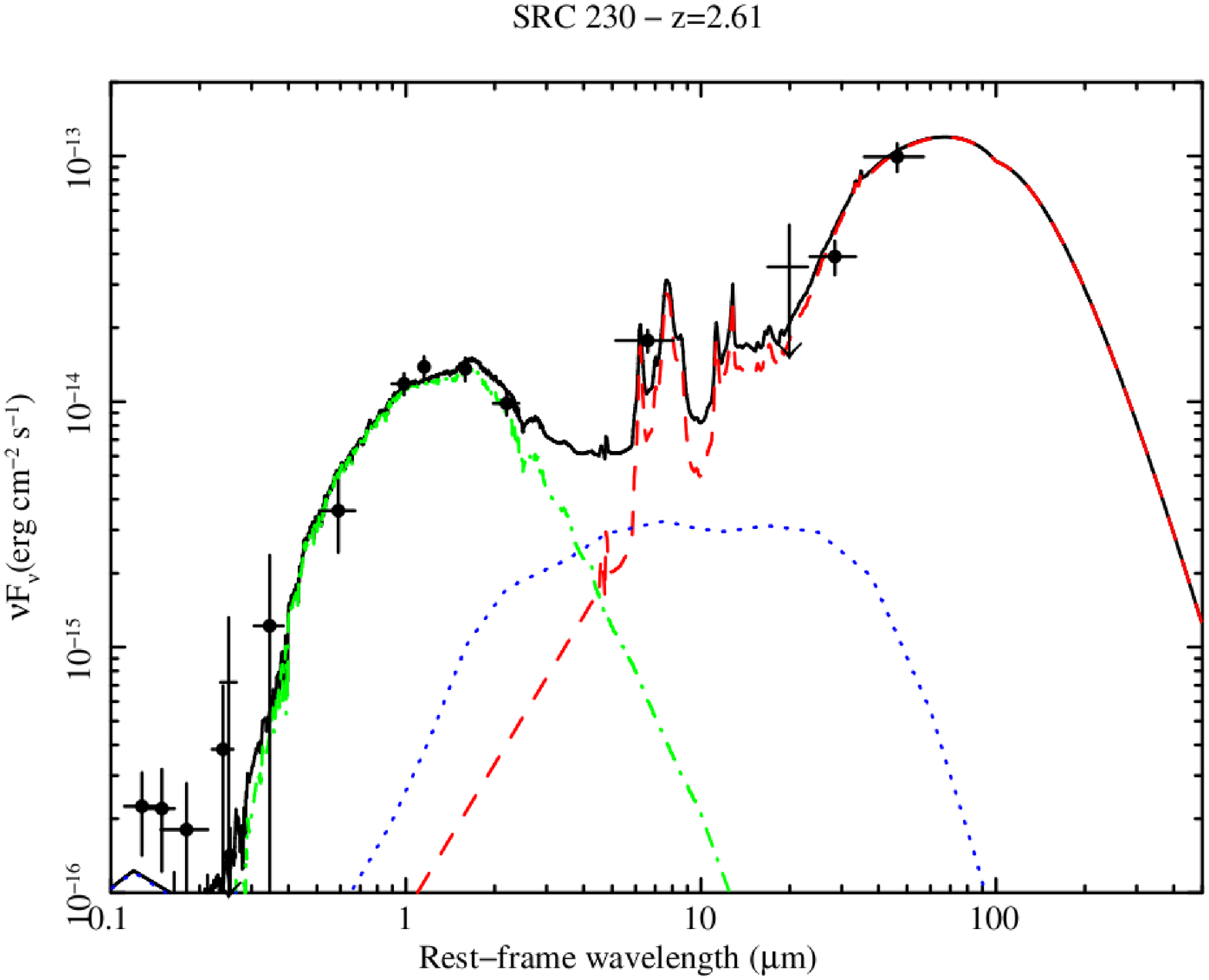}\\
    \includegraphics[angle=0,width=0.3\textwidth]{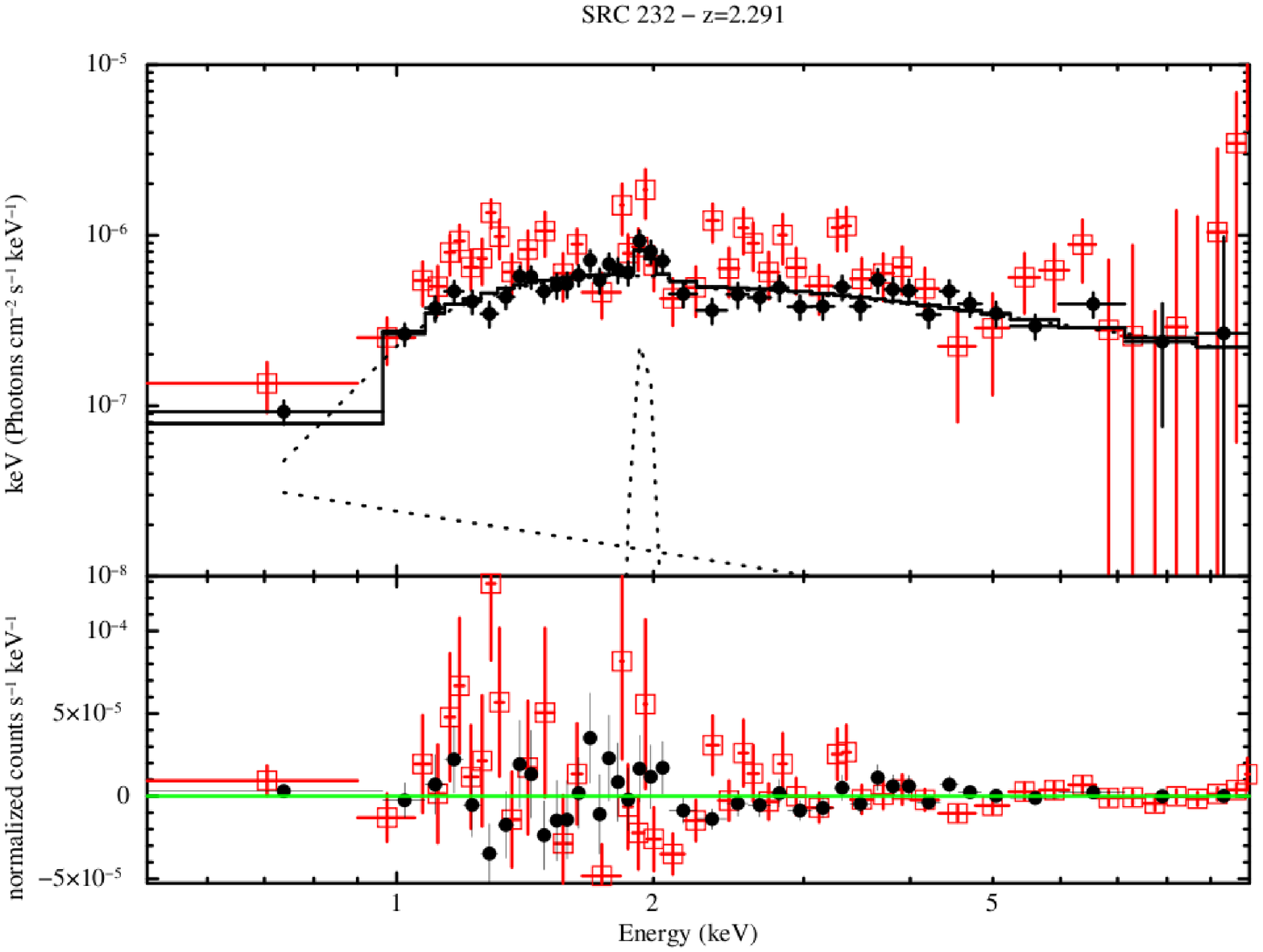} &    \includegraphics[angle=0,width=0.3\textwidth]{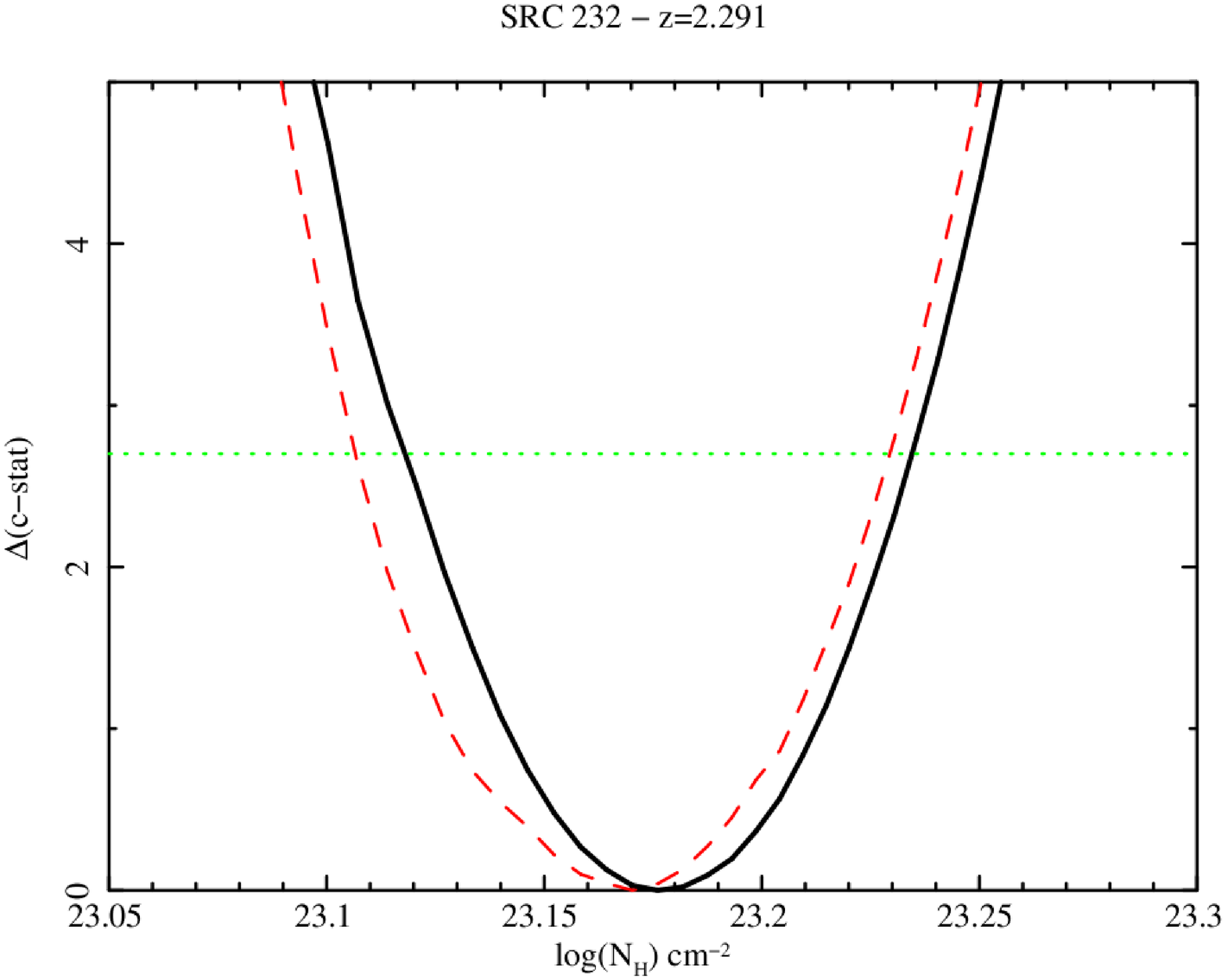} &    \includegraphics[angle=0,width=0.3\textwidth]{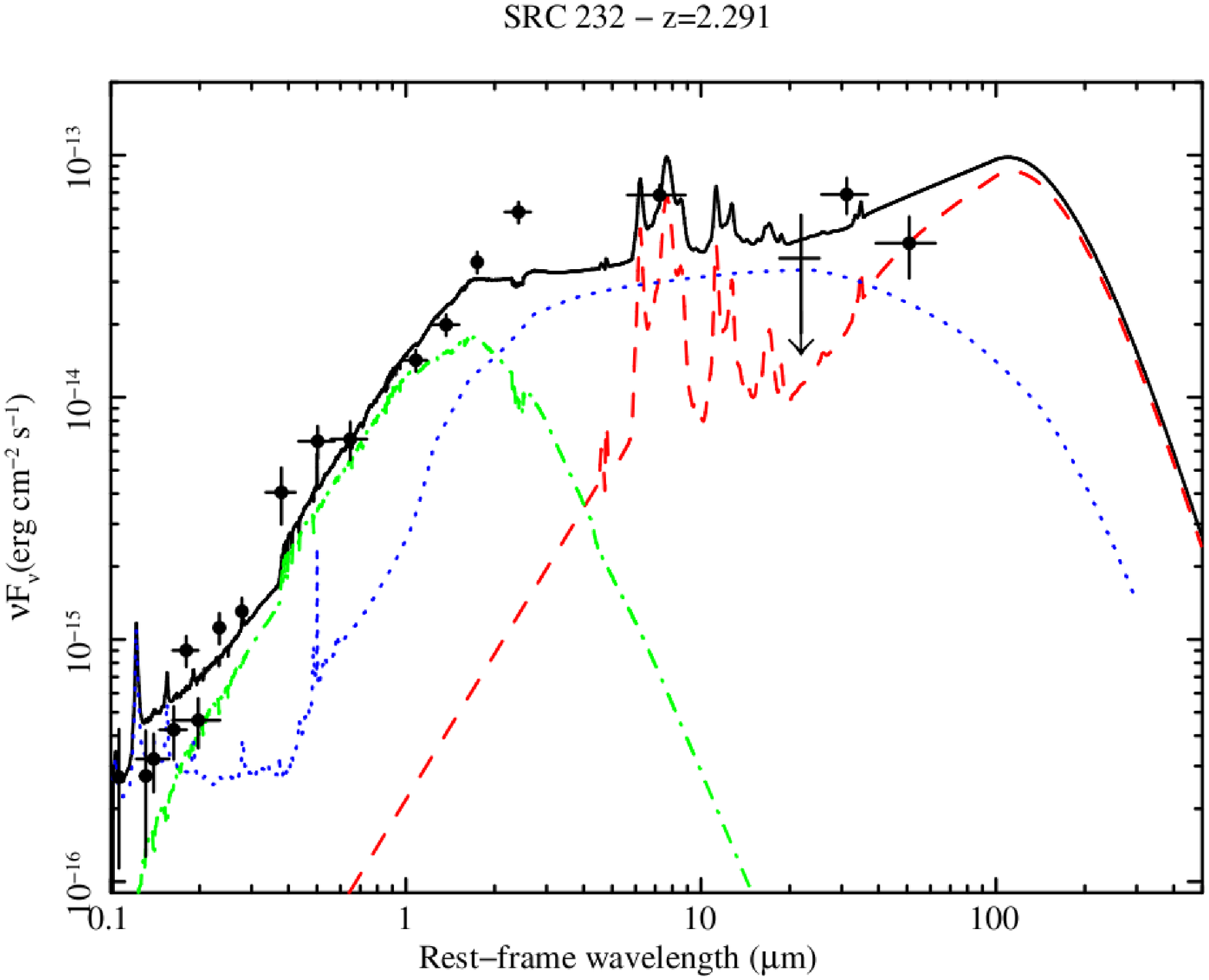}\\ 
    \includegraphics[angle=0,width=0.3\textwidth]{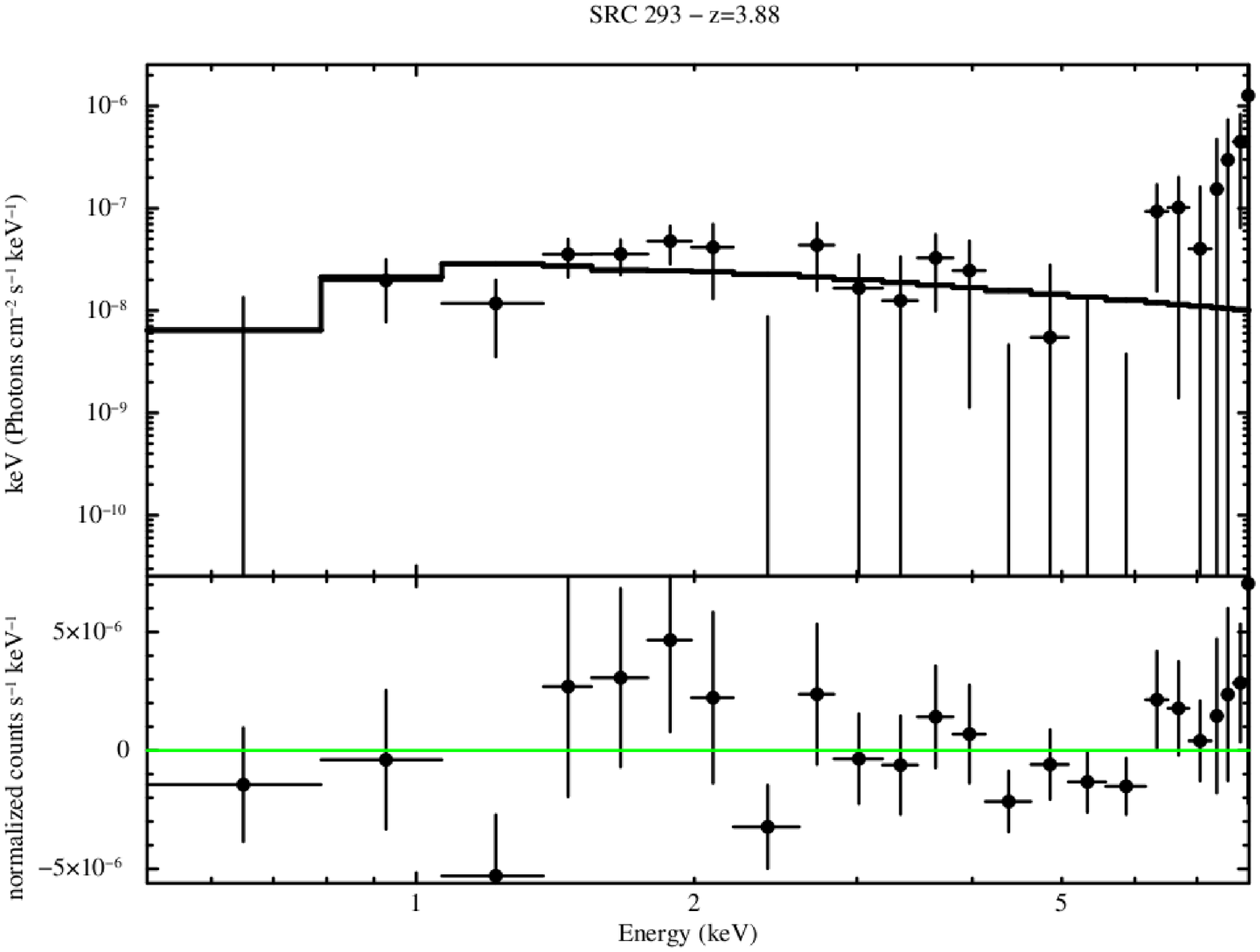} &    \includegraphics[angle=0,width=0.3\textwidth]{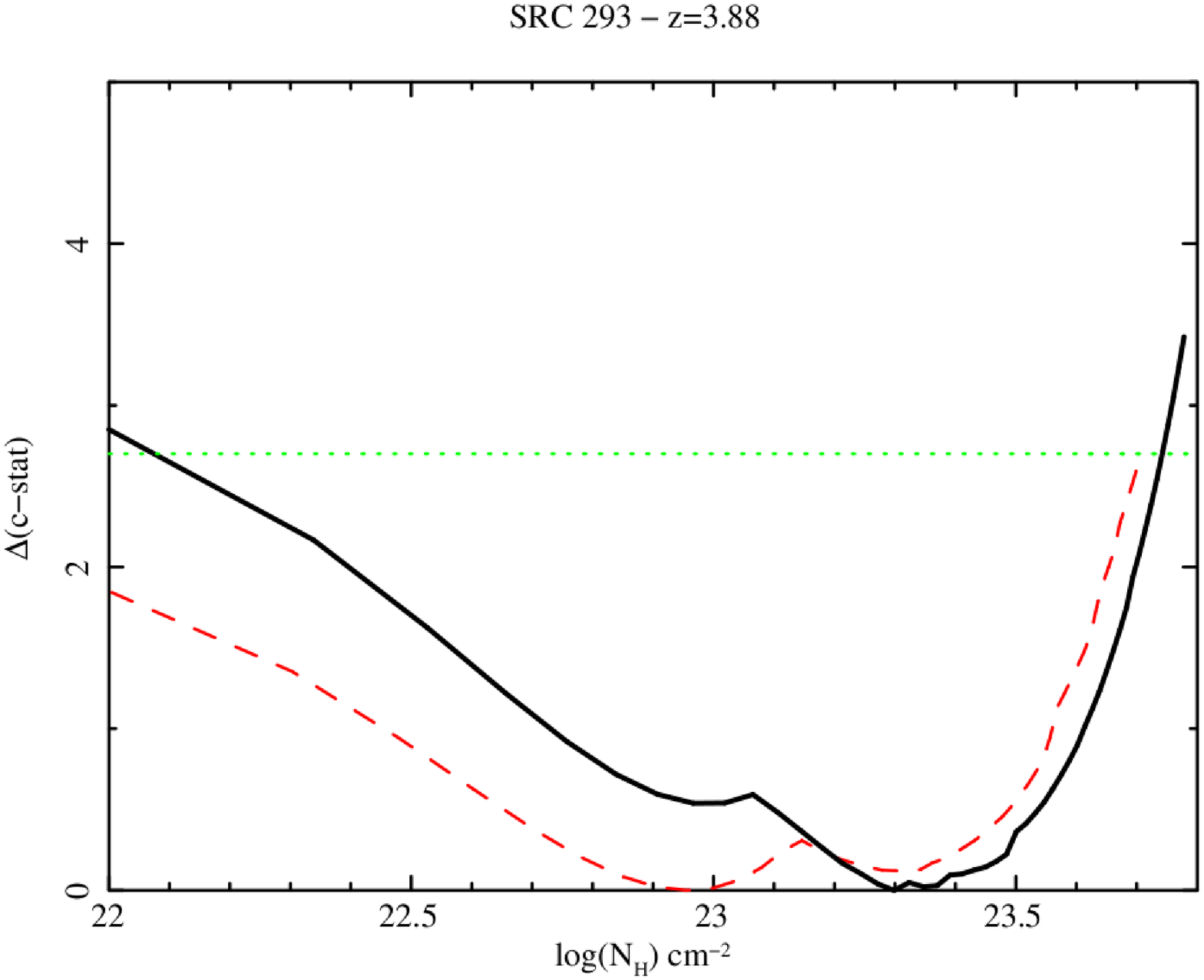} &    \includegraphics[angle=0,width=0.3\textwidth]{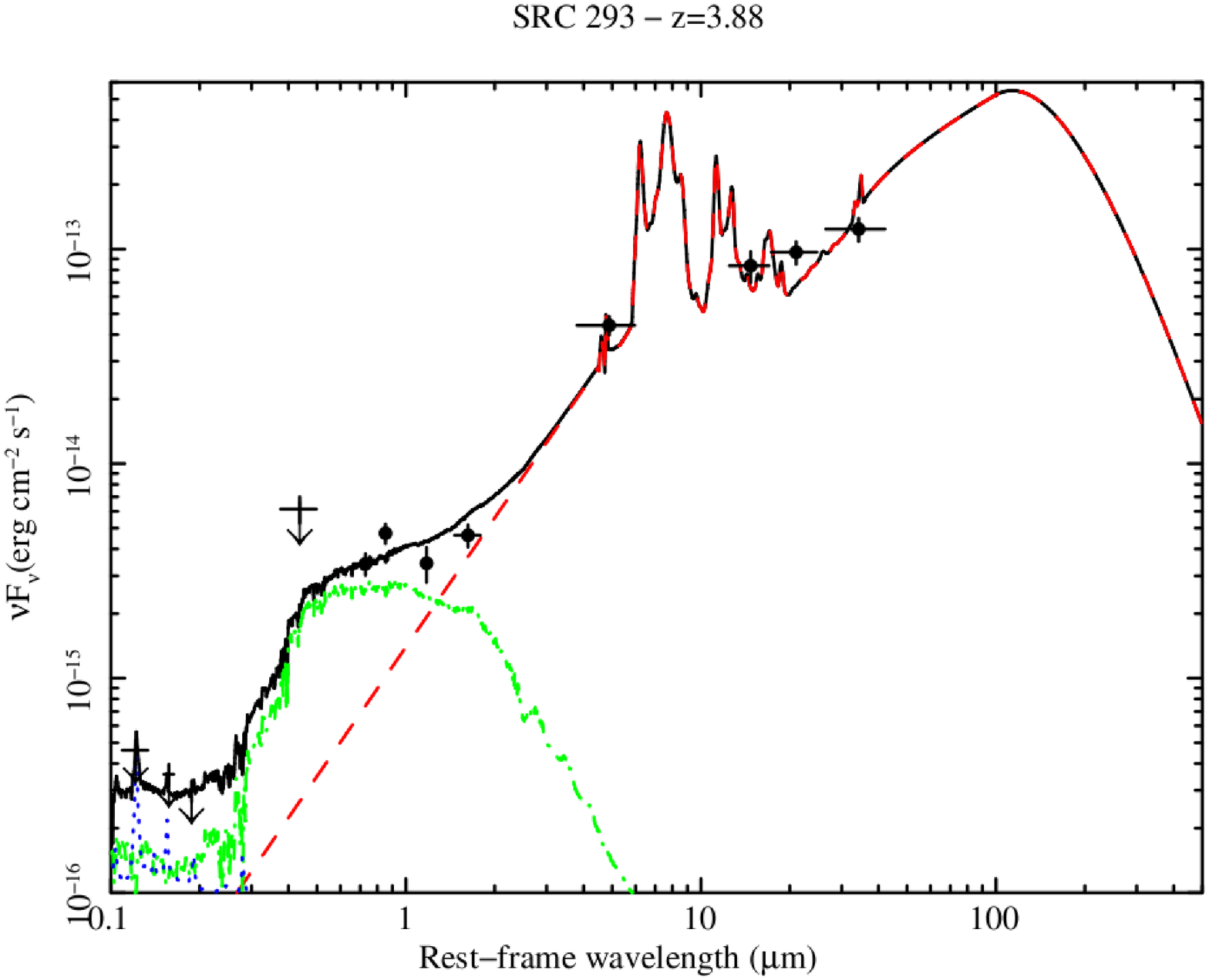}\\
    \includegraphics[angle=0,width=0.3\textwidth]{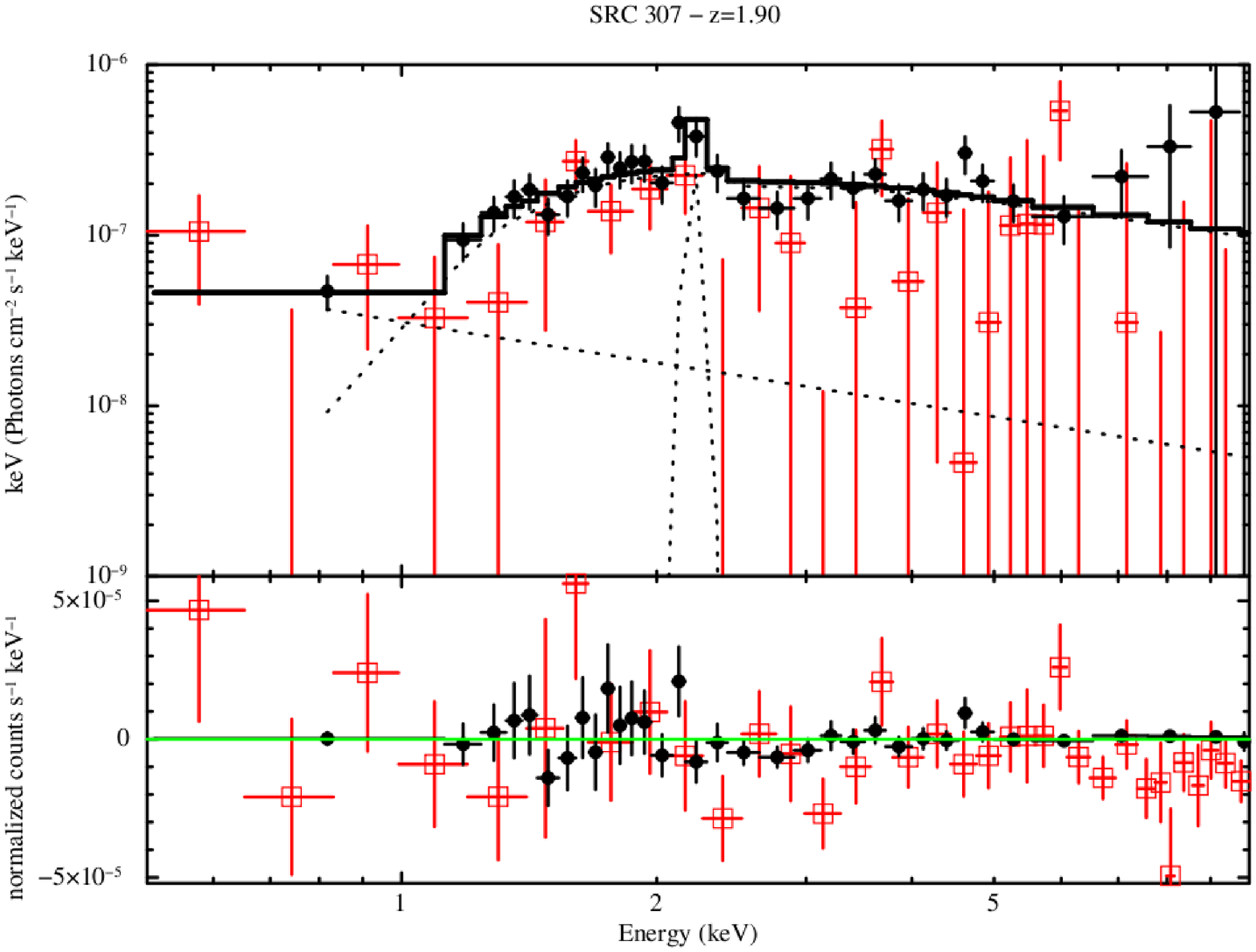}    &    \includegraphics[angle=0,width=0.3\textwidth]{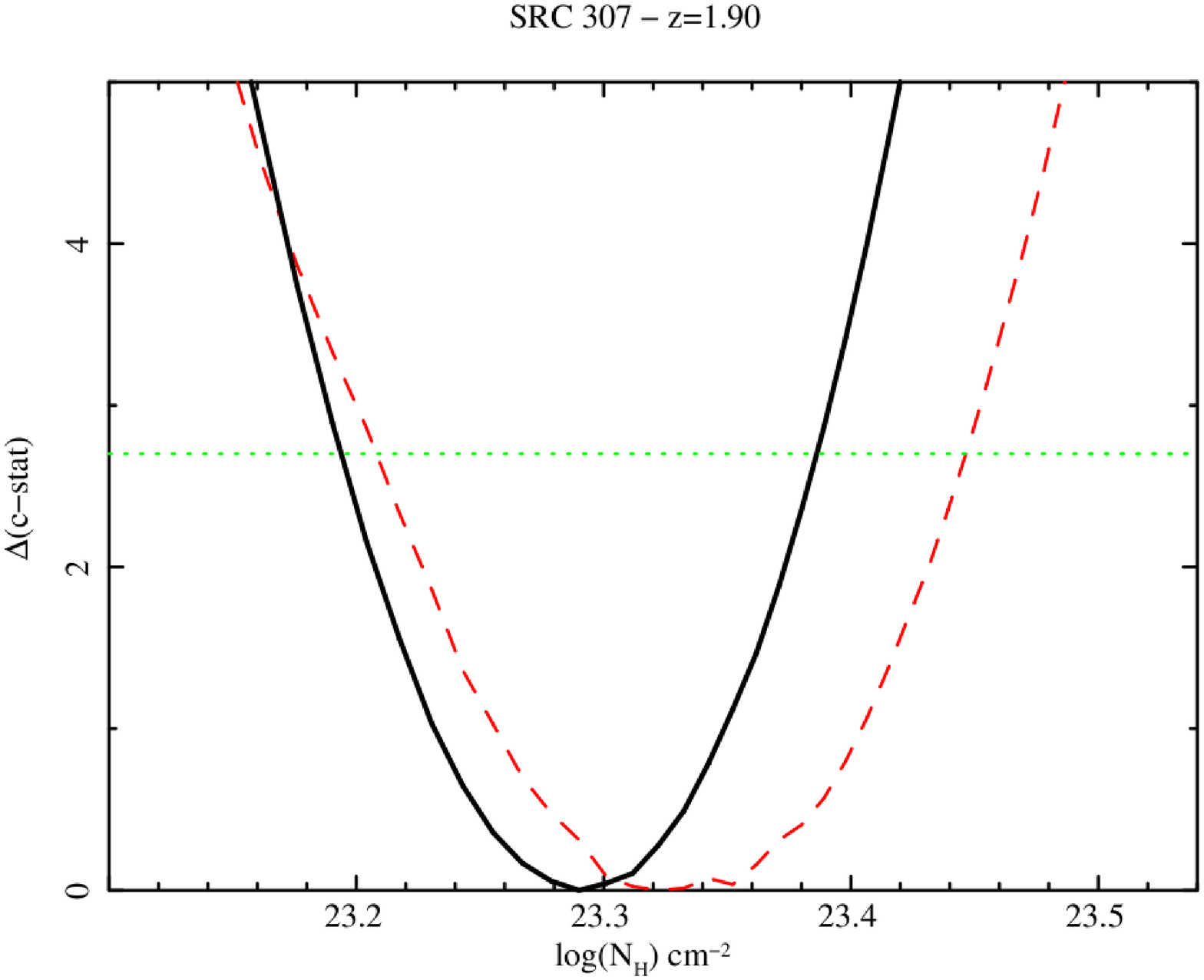}    & \includegraphics[angle=0,width=0.3\textwidth]{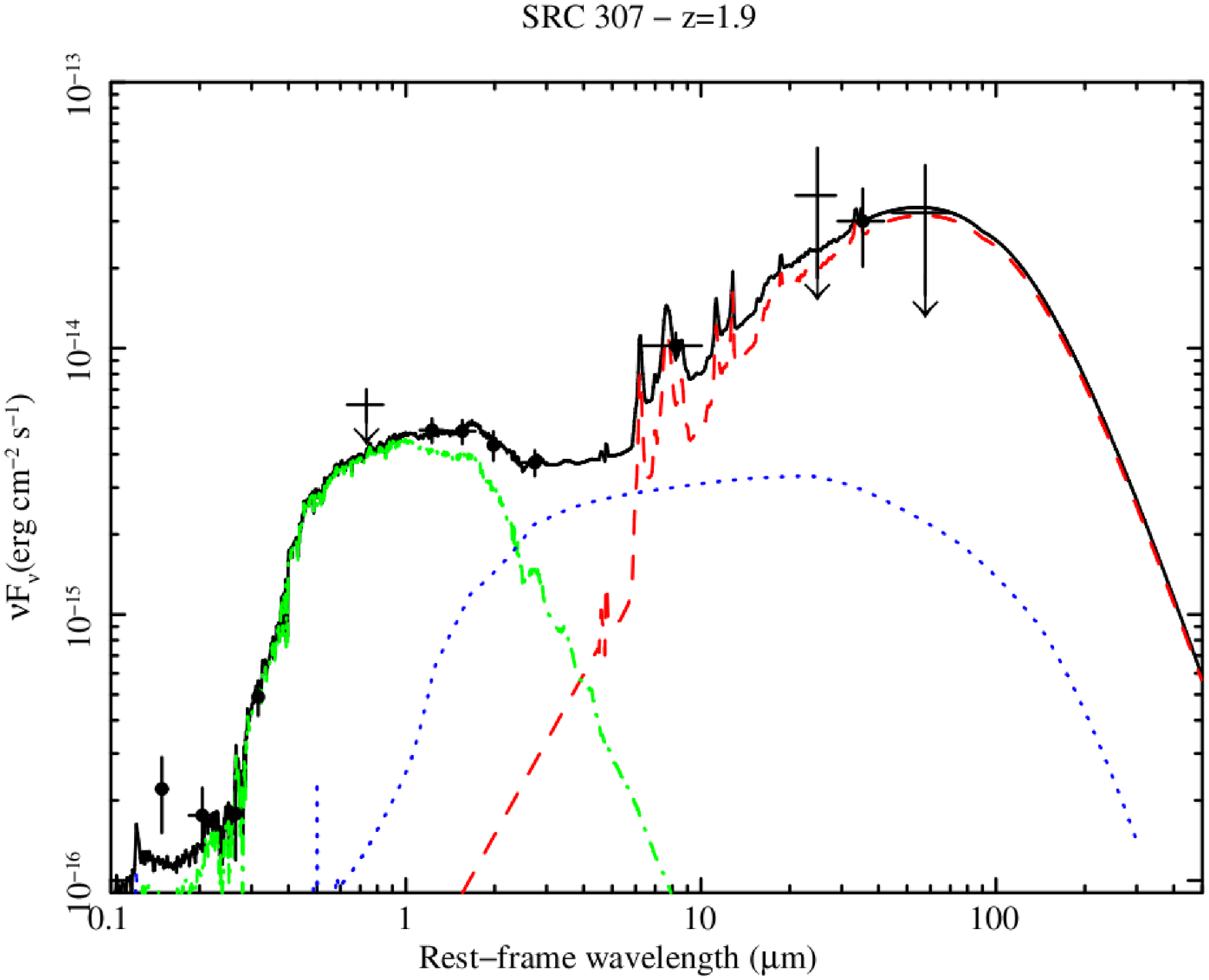}\\
    \includegraphics[angle=0,width=0.3\textwidth]{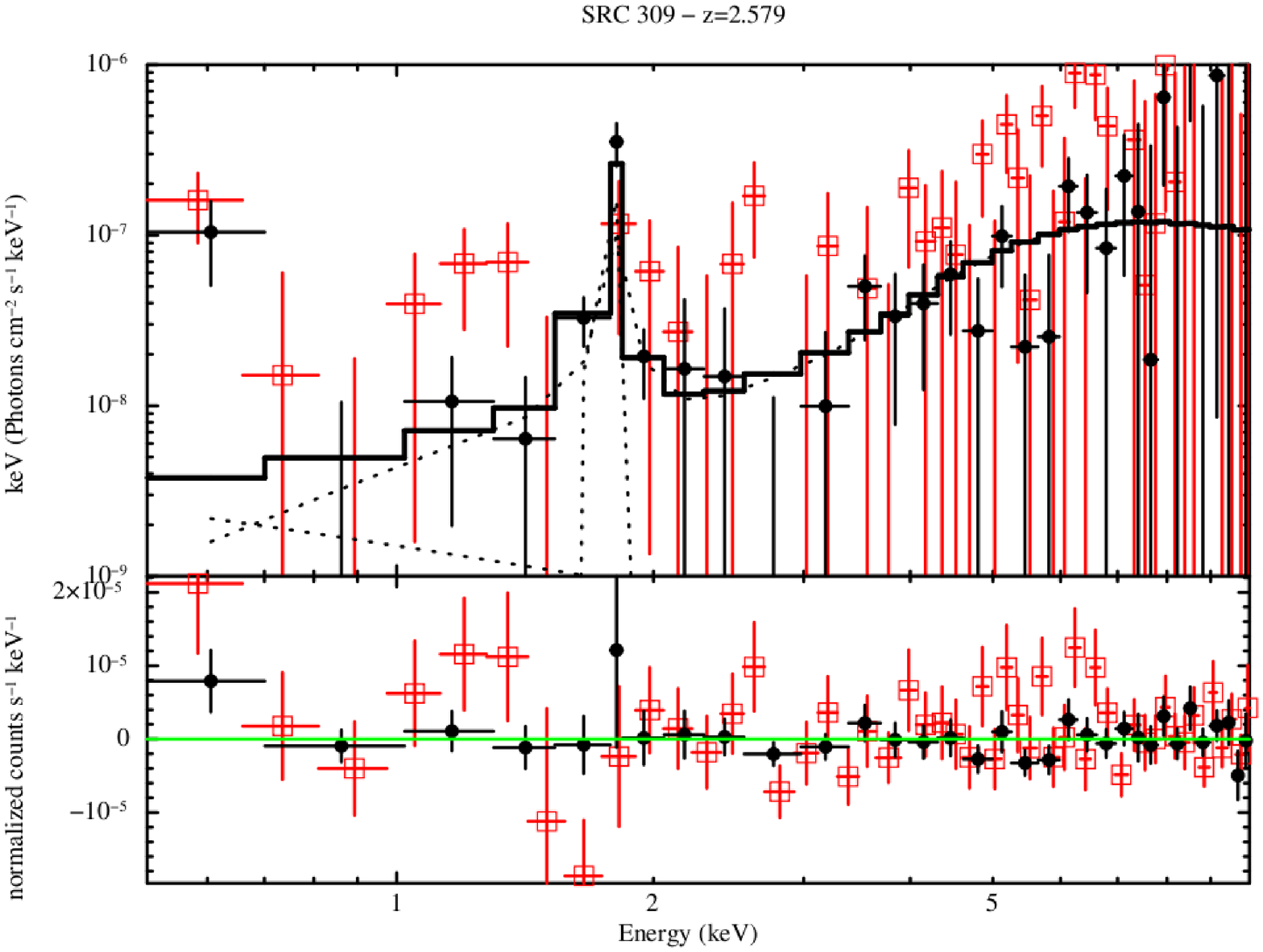}    &    \includegraphics[angle=0,width=0.3\textwidth]{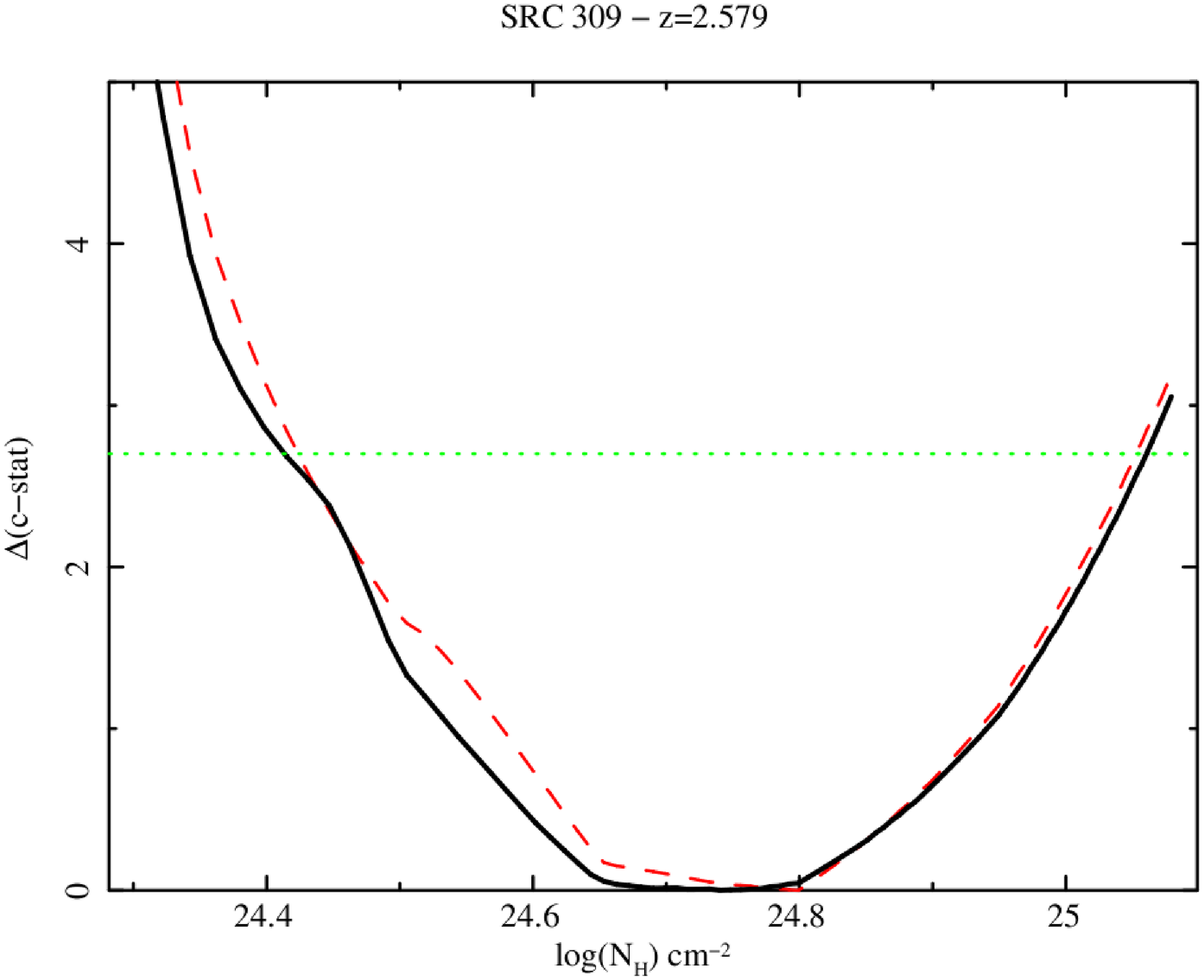}    & \includegraphics[angle=0,width=0.3\textwidth]{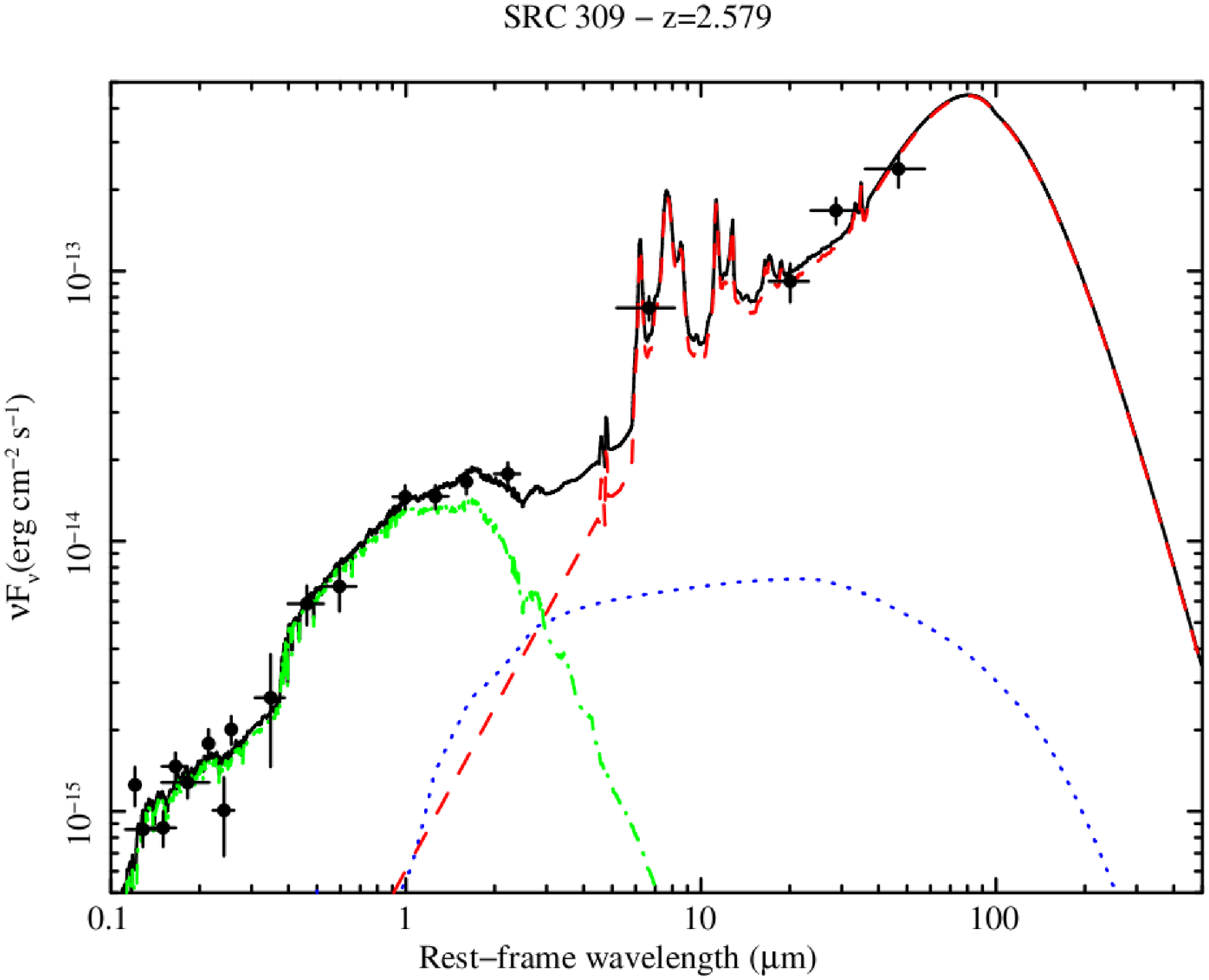}\\
\end{array}
$$
  \caption{Continued.}
\end{figure*}
\begin{figure*}[!htbp] 
\ContinuedFloat
\centering
$$
\begin{array}{ccc}
    \includegraphics[angle=0,width=0.3\textwidth]{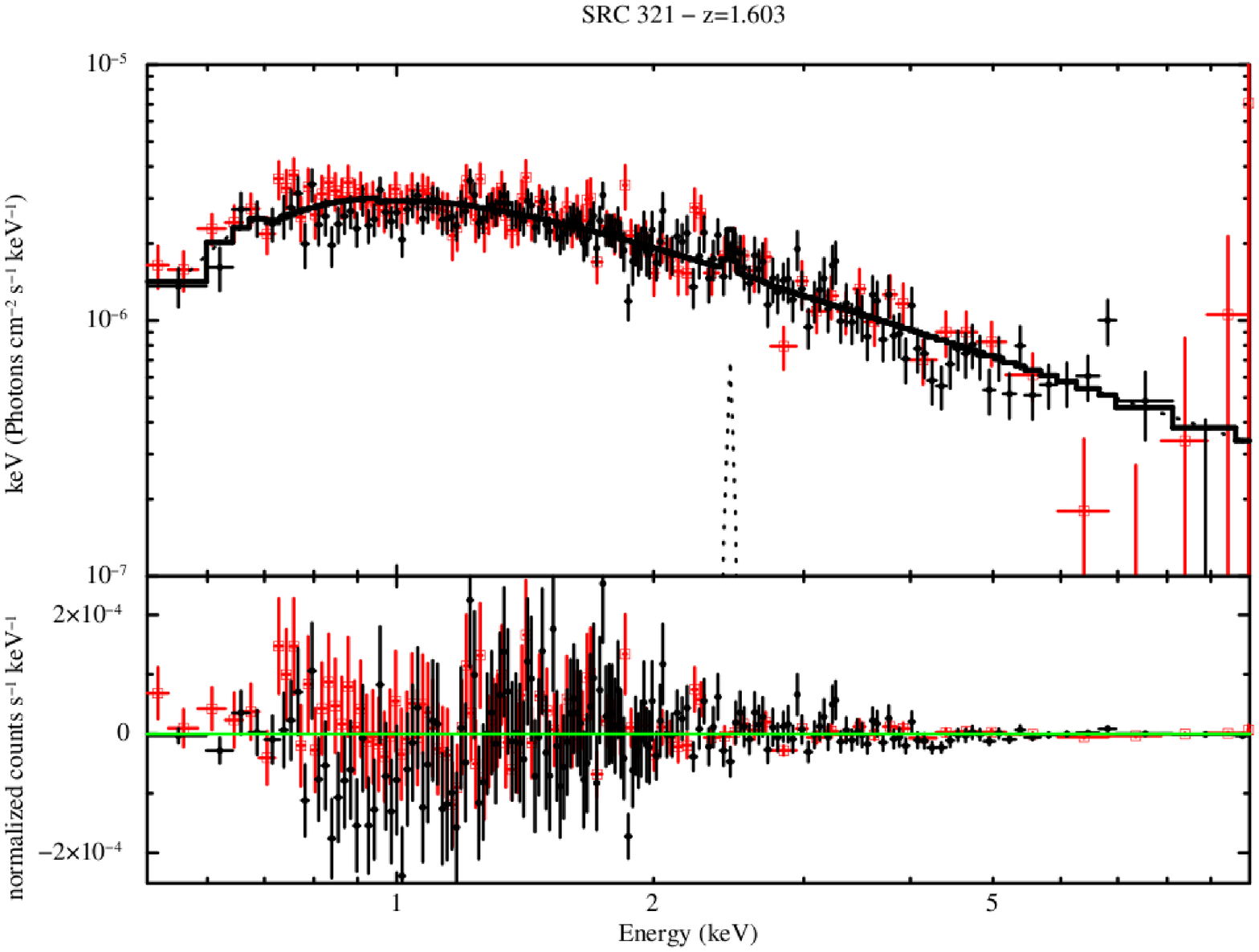}    &    \includegraphics[angle=0,width=0.3\textwidth]{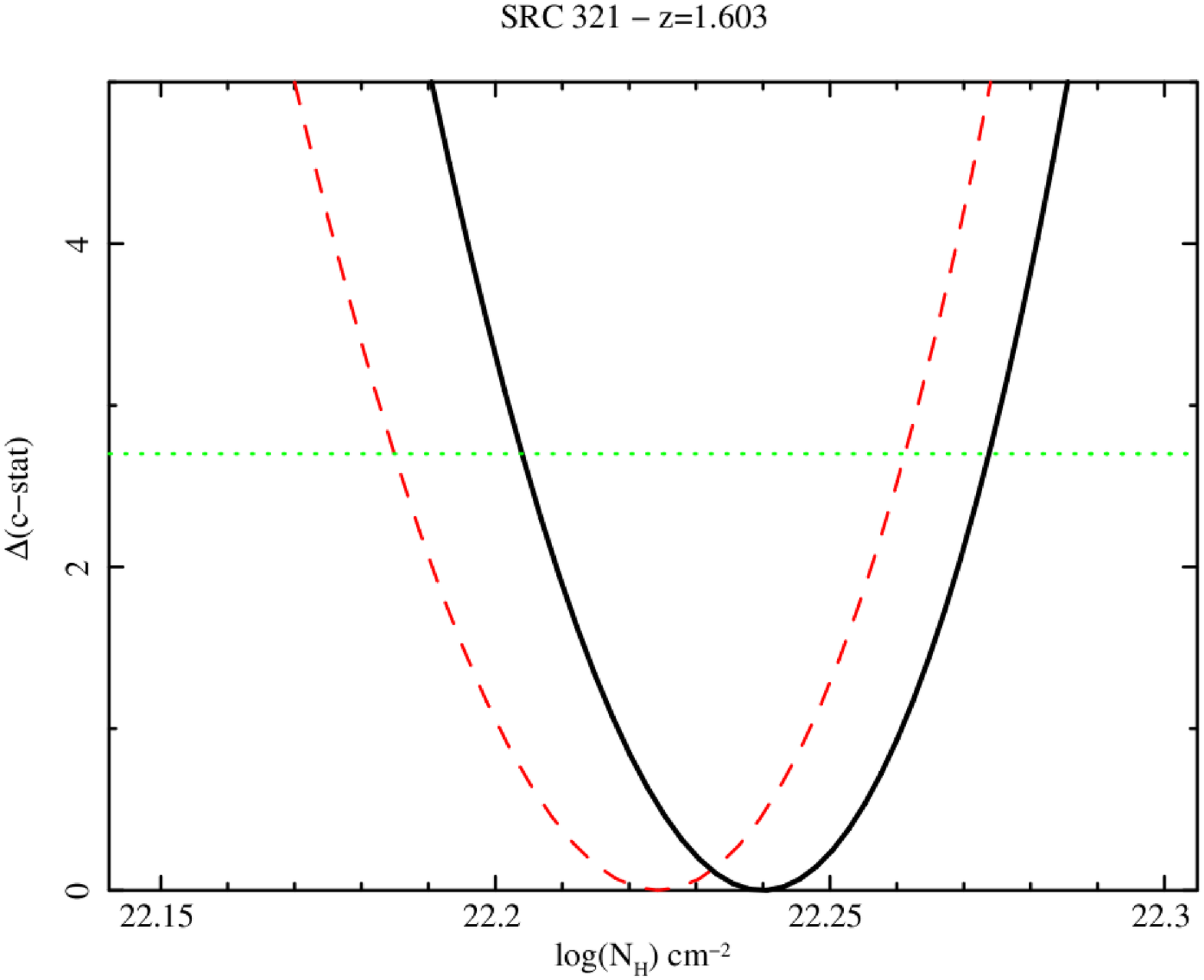}    & \includegraphics[angle=0,width=0.3\textwidth]{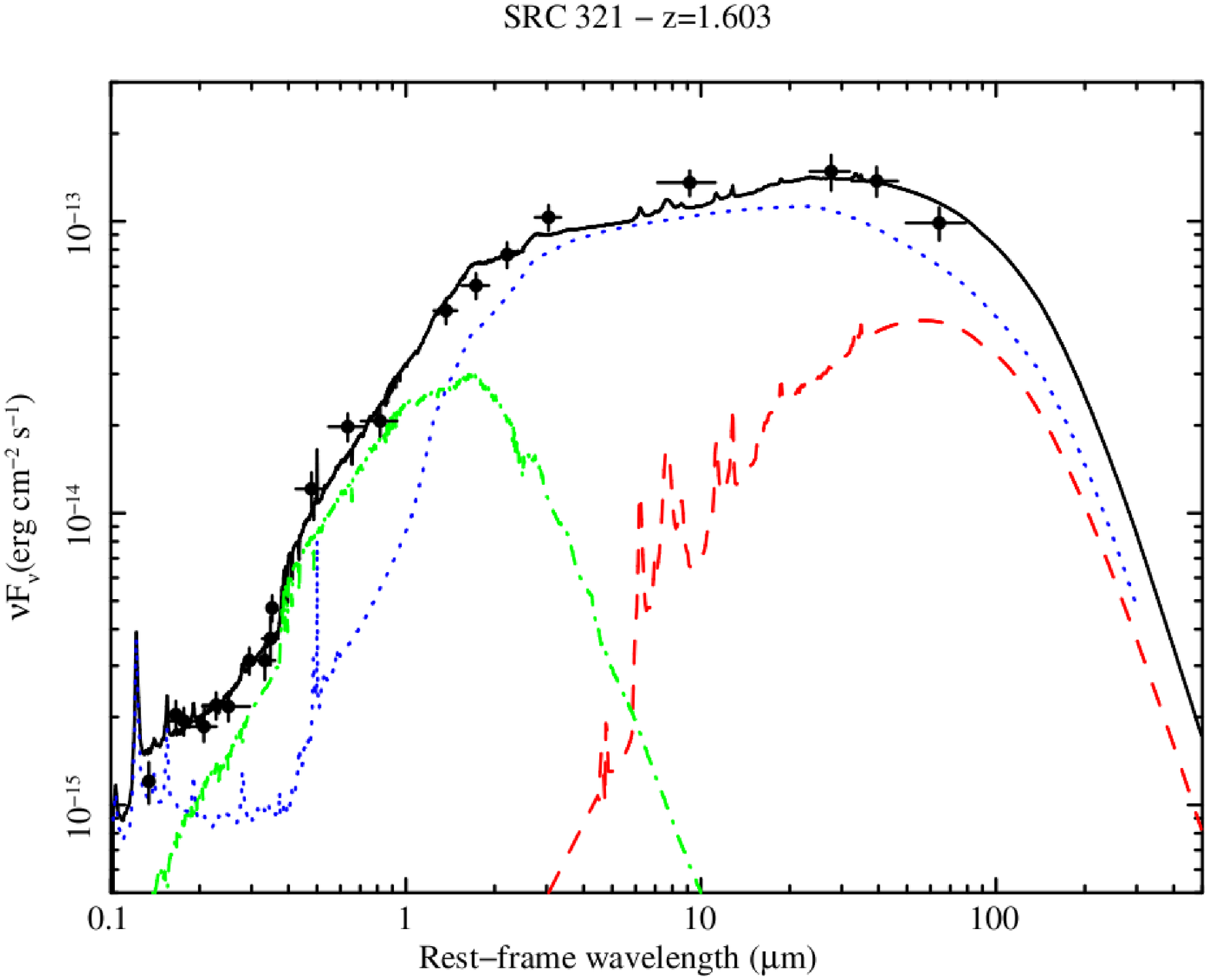}\\
    \includegraphics[angle=0,width=0.3\textwidth]{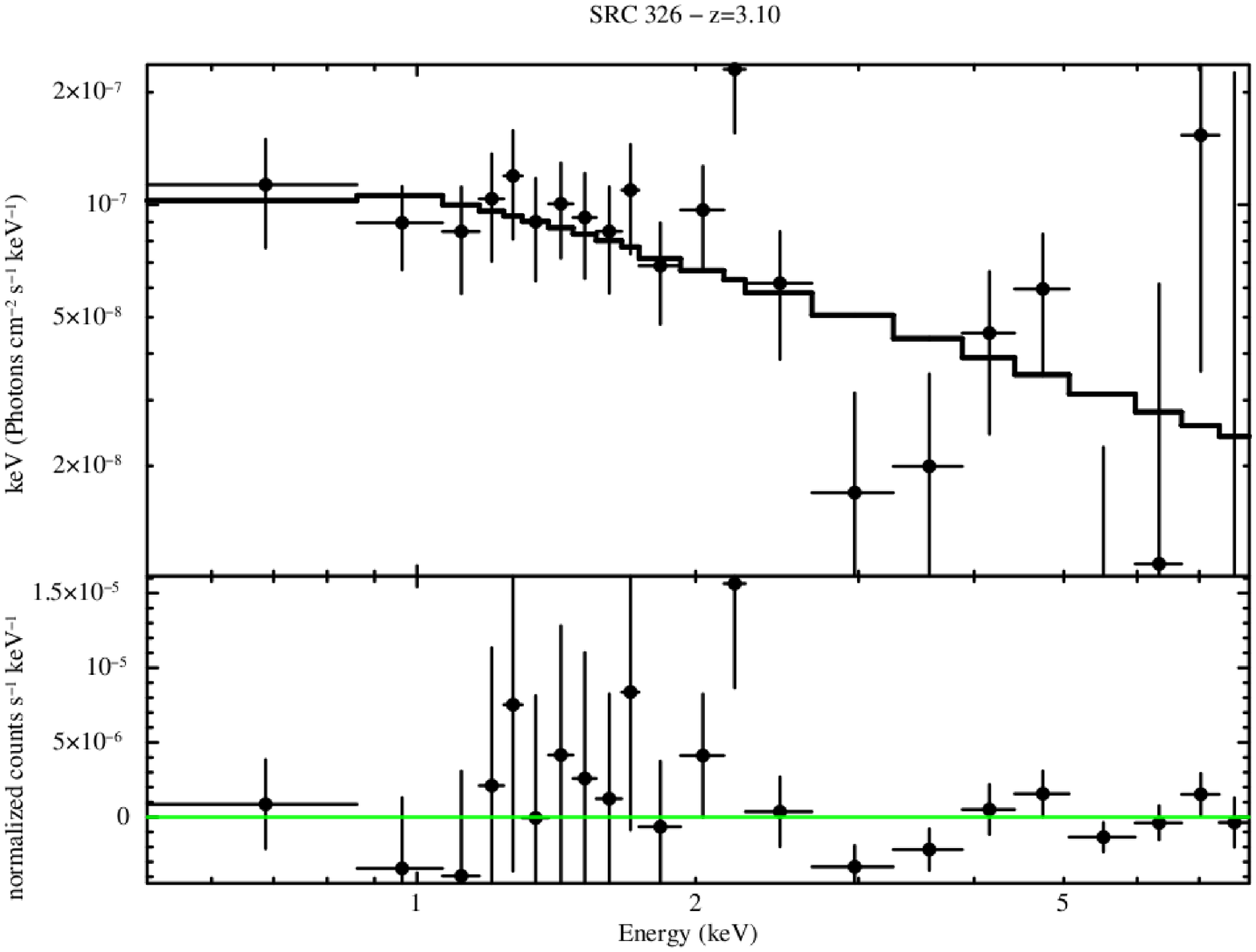}    &    \includegraphics[angle=0,width=0.3\textwidth]{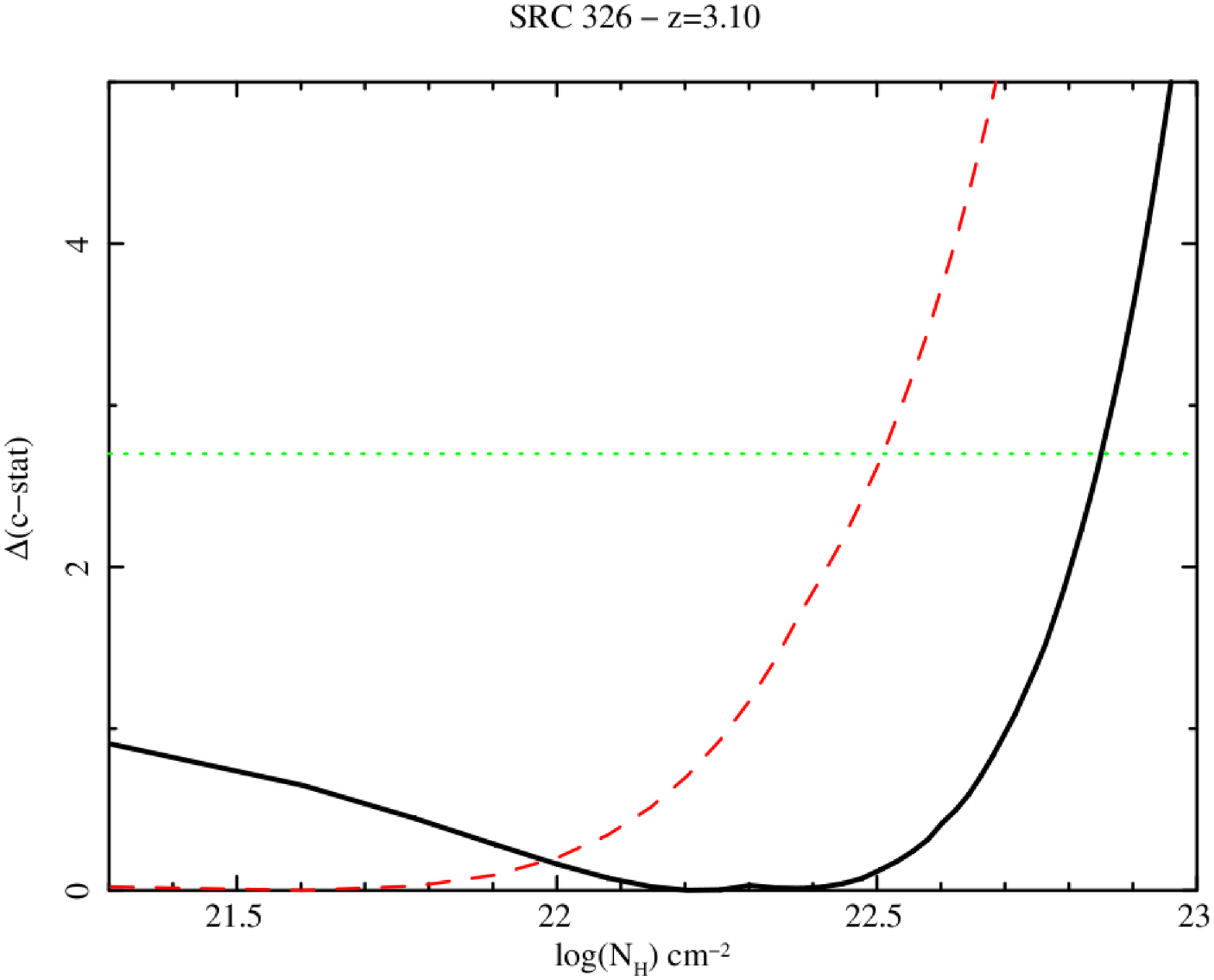}    & \includegraphics[angle=0,width=0.3\textwidth]{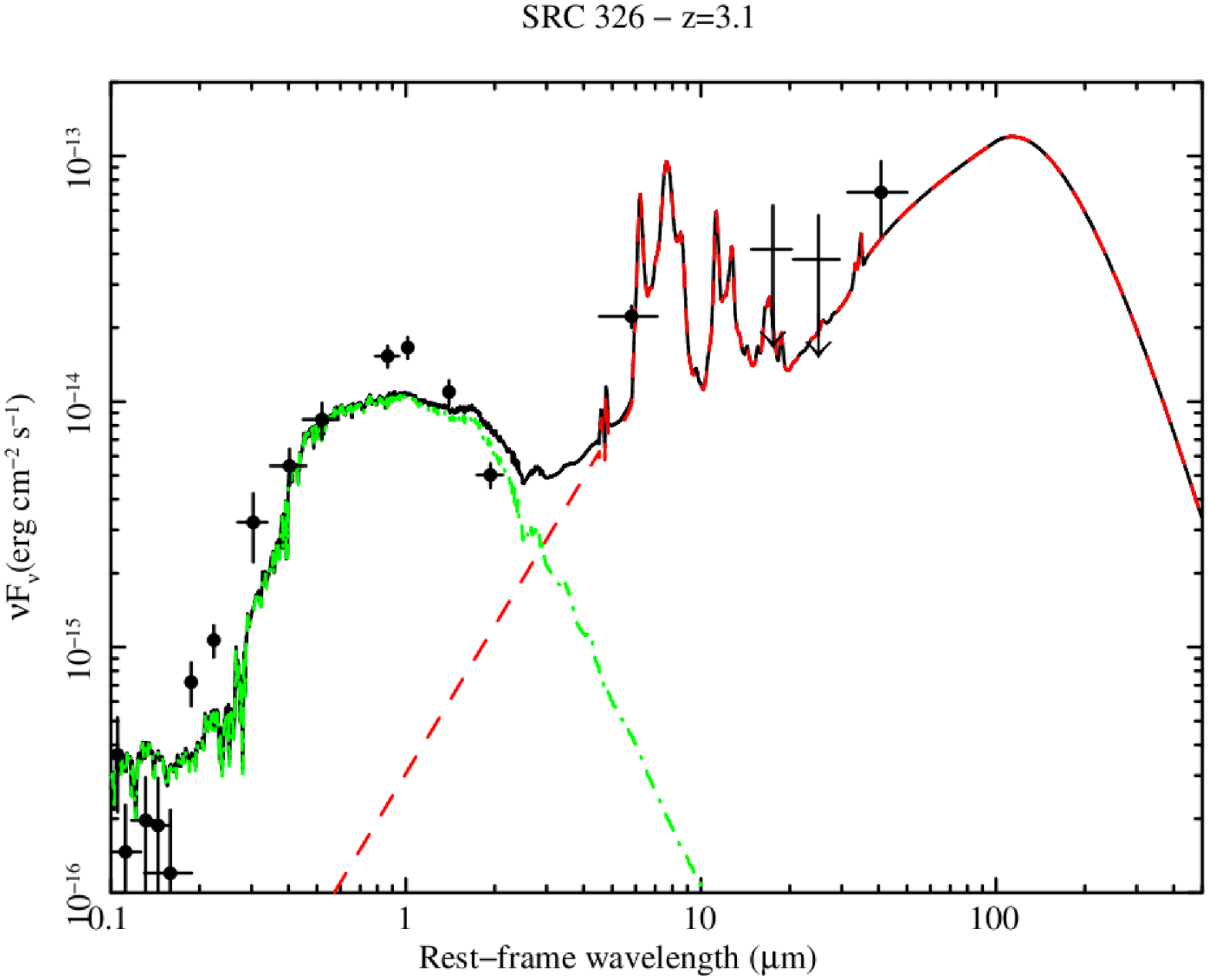}\\
   \includegraphics[angle=0,width=0.3\textwidth]{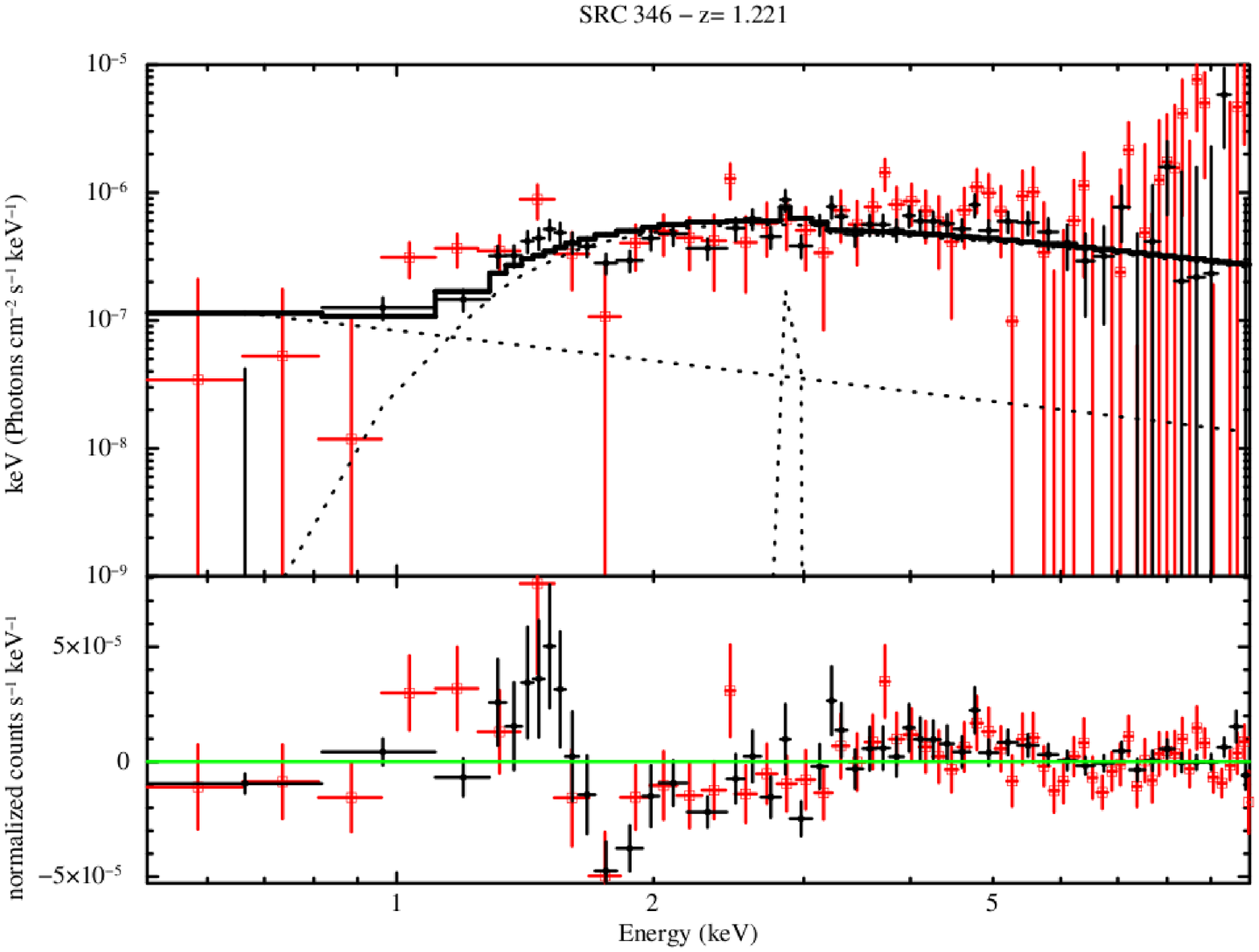}   &    \includegraphics[angle=0,width=0.3\textwidth]{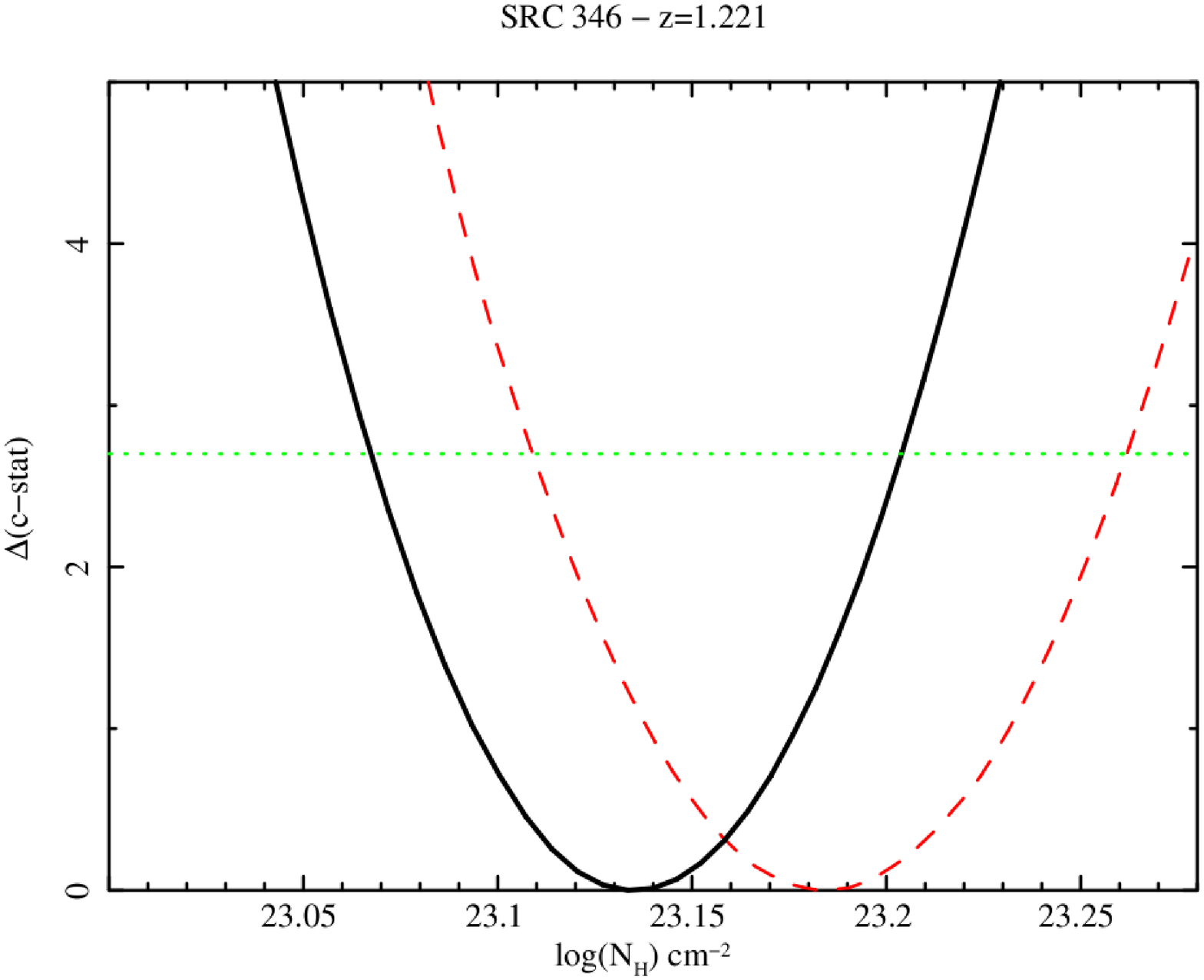}   & \includegraphics[angle=0,width=0.3\textwidth]{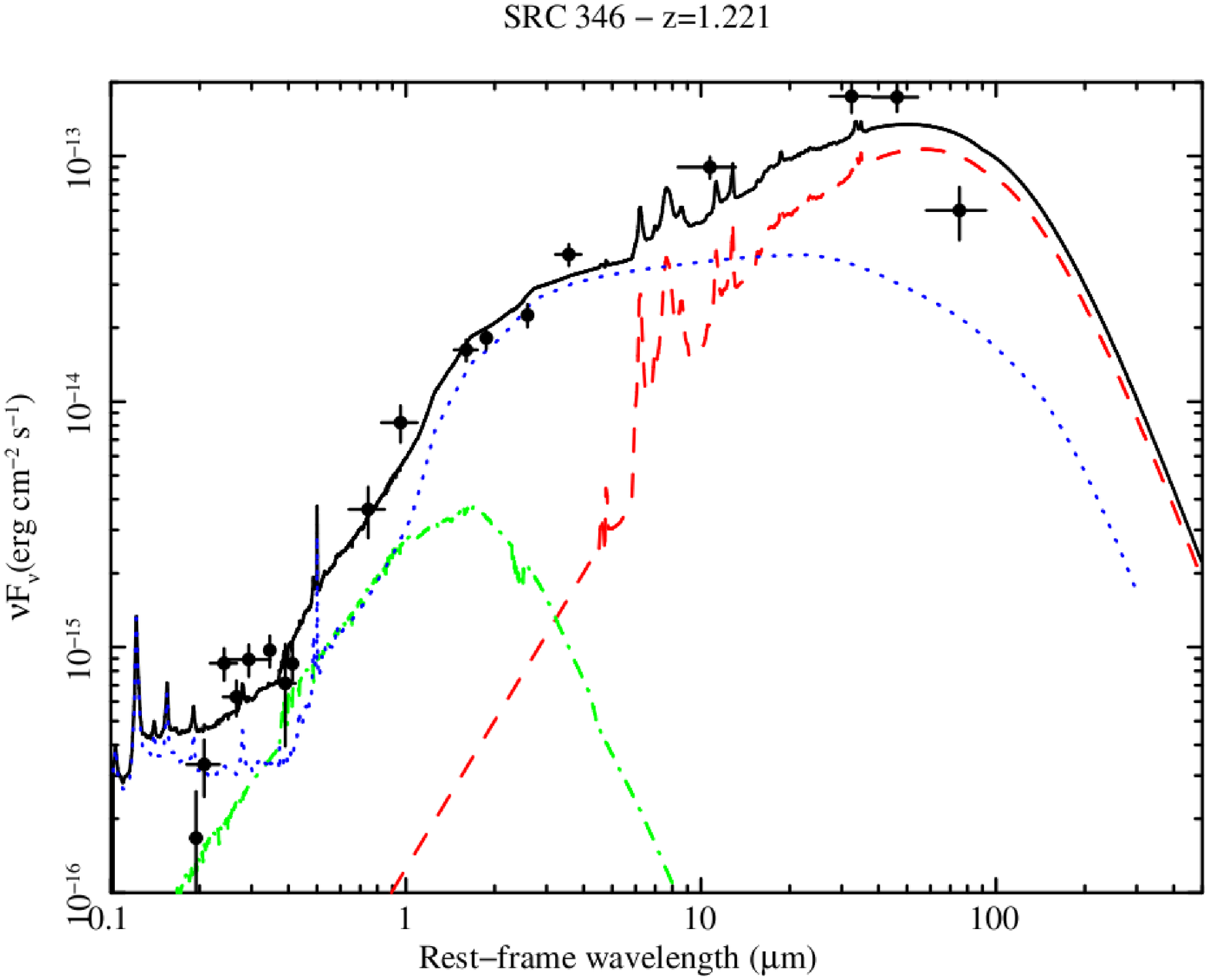}\\
    \includegraphics[angle=0,width=0.3\textwidth]{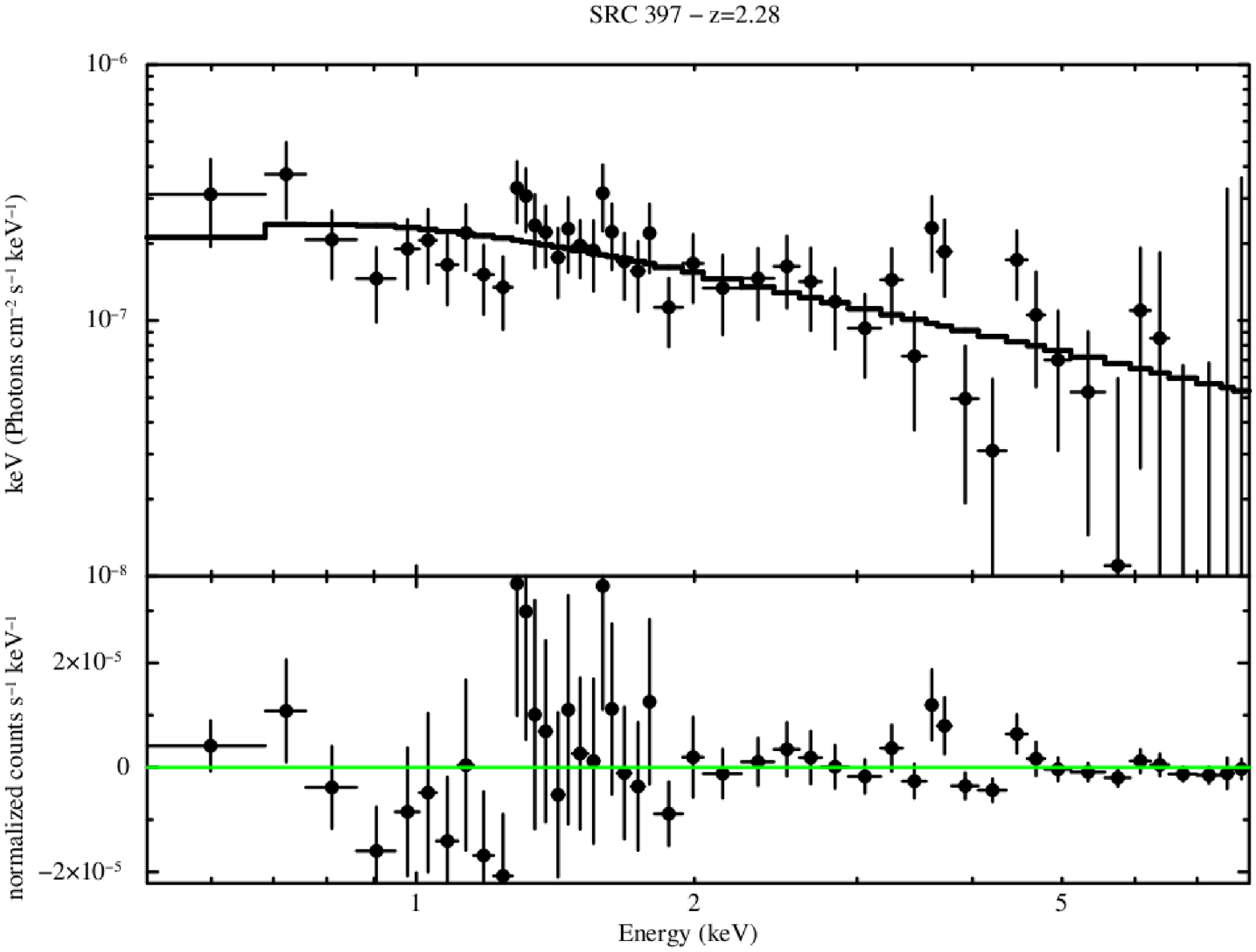}    &    \includegraphics[angle=0,width=0.3\textwidth]{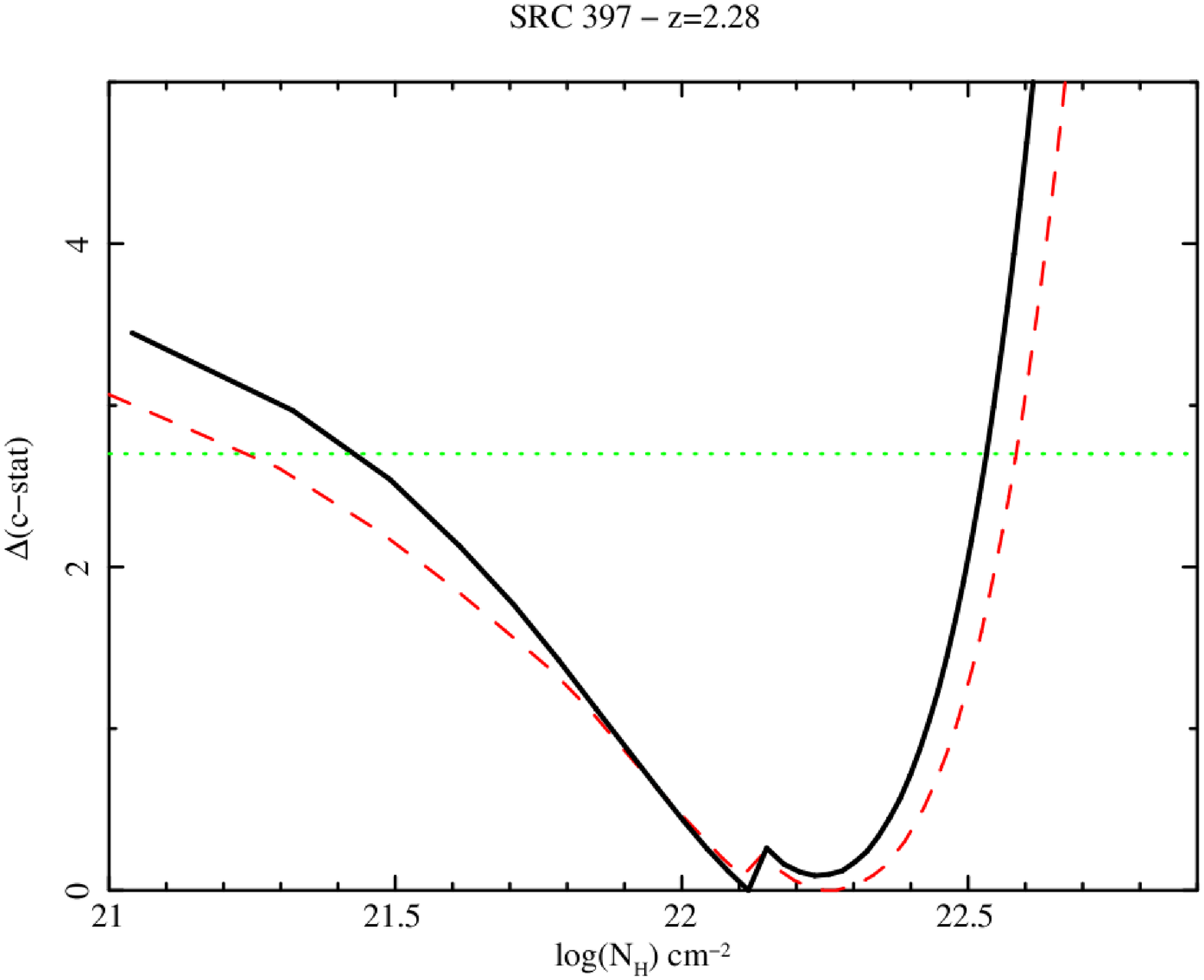}    & \includegraphics[angle=0,width=0.3\textwidth]{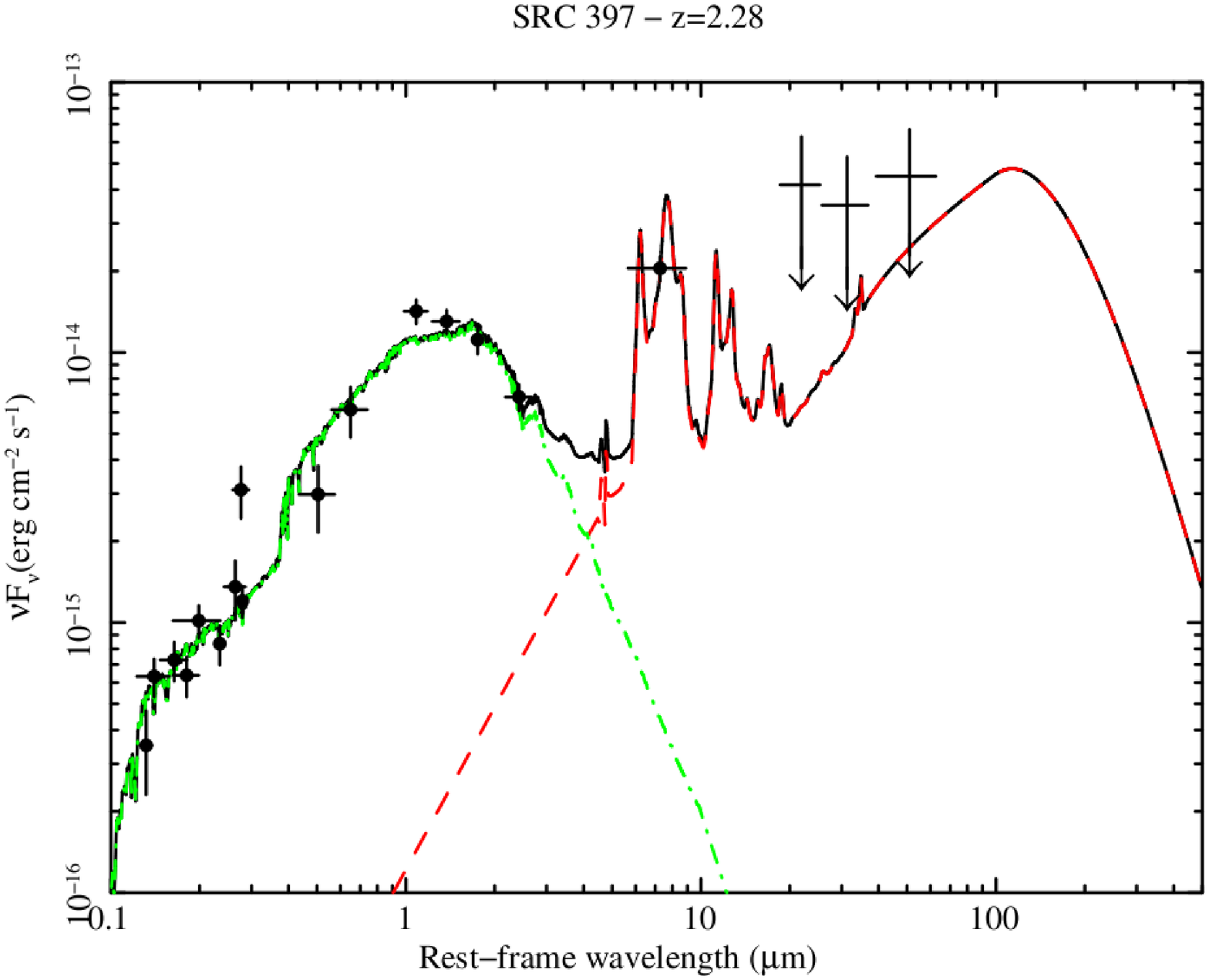}\\ 
\end{array}
$$
  \caption{Continued.}
\label{ctsxrayfit}
\end{figure*}

\section{Spectral energy distributions}

To obtain the AGN contribution to the IR emission in our sample, we
took advantage of the multi-wavelength coverage of the CDF-S region to
construct and fit the spectral energy distributions (SEDs) of our
sources. There are enough multi-wavelength data for all of our 14
X-ray detected DOGs to obtain reliable SED fits, but there are only
enough data for 34 of the X-ray undetected DOGs to carry out the same
analysis.

We used the SEd Analysis with BAyesian Statistics
(SEABASs\footnote{\url{http://astro.dur.ac.uk/~erovilos/SEABASs/}};
\citealt{rovilos14}) fitting code that combines SED templates and fit
them using a maximum likelihood method with the posibility of
including priors. We used three sets of templates to account for the
stellar, starburst, and AGN contributions. The stellar templates are
those from the library of synthetic templates of
\citet{bruzual03}. For the starburst templates we used two sets of
libraries: those presented in \citet{chary01}, and
\citet{mullaney11}. For the AGN contribution, we used the library of
\citet{silva04}. In a small number of sources, the AGN contribution
did not seem to be estimated properly, so we also included the AGN
templates described in \citet{polletta07}.

The IR luminosities obtained from the SED fitting are listed in Table
\ref{sedsfits} and Table~\ref{sedsfitsnox}, for the X-ray detected and
the X-ray undetected DOGs, respectively. The L$_{IR}$ derived from the
SED fitting are not always over the 10$^{12}$ L$\odot$ value usually
reported for DOGs, but they are over or very close to this value in
the vast majority of cases. The resulting SEDs for the X-ray detected
DOGs, showing each contribution separately, are plotted in the right
panel of Fig,~\ref{ctsxrayfit}. We do not find any significant AGN
contribution to the mid-IR emission in the two cases of sources 326
and 397, although the presence of an AGN is confirmed by their X-ray
properties. Moreover, in four additional cases (sources 170, 197, 199,
and 293), the AGN emission does not contribute significantly to the
mid-IR luminosity. In the case of X-ray undetected DOGs, we do not
find any AGN contribution in 16 cases (47\% of the X-ray undetected
DOGs with reliable SED fits). It is important to note that for 70\% of
the X-ray detected DOGs and 85\% of the undetected DOGs with available
SED fits, the derived L$_{12\mu m}$ for the starburst component is
larger than for the AGN component.

\begin{table*}[!htbp]
\caption{Results from the SED fits for the X-ray detected DOGs.} 
\label{sedsfits}
\centering
$$
\begin{tabular}{ccccccccc}
\hline\hline
LID & z & $\rm log L_{IR}(8-1000\mu m)$   & $\rm log L_{12\mu m}$   & $\rm log L^{SB}_{12\mu m}$  &  $\rm log L^{AGN}_{12\mu m}$ & $\rm log L^{stellar}_{12\mu m}$ & log Stellar mass & SFR  \\
 &   & $\rm L_{\odot}$ &\lunits & \lunits &  \lunits & \lunits & $\rm M_{\odot}$ & $\rm M_{\odot}/yr$   \\
(1) & (2) & (3) & (4) & (5) & (6) & (7) & (8) & (9)\\
 \hline
95 & 5.22 & 13.45 & 46.28 & 46.08 & 45.85 & 43.93 & 12.30 & 4029 \\
117 & 1.51 & 11.47 & 44.38 & 44.04 & 44.12 & 41.96 & 9.87 & 37 \\
170 & 1.22 & 11.51 & 44.09 & 44.08 & 42.16 & 41.96 & 10.91 & 56 \\
197 & 1.81 & 12.11 & 44.82 & 44.82 & 42.59 & 42.42 & 11.11 & 220 \\
199 & 2.45 & 12.38 & 45.09 & 45.08 & 41.97 & 42.75 & 11.34 & 410 \\
230 & 2.61 & 12.50 & 44.96 & 44.87 & 44.23 & 42.82 & 11.51 & 525 \\
232 & 2.291 & 12.39 & 45.34 & 44.91 & 45.13 & 42.85 & 10.76 & 272 \\
293 & 3.88 & 13.55 & 46.26 & 46.26 & 42.74 & 42.30 & 10.95 & 6164 \\
307 & 1.90 & 11.71 & 44.50 & 44.36 & 43.93 & 41.87 & 10.52 & 79 \\
309 & 2.579 & 13.01 & 45.74 & 45.71 & 44.59 & 42.82 & 11.33 & 1722 \\
321 & 1.603 & 12.22 & 45.33 & 44.34 & 45.28 & 42.69 & 10.93 & 75 \\
326 & 3.10 & 12.66 & 45.37 & 45.37 & - & 42.70 & 11.35 & 790 \\
346 & 1.221 & 11.87 & 44.78 & 44.41 & 44.54 & 41.49 & 9.21 & 90 \\
397 & 2.28 & 11.94 & 44.65 & 44.65 & - & 42.65 & 11.17 & 150 \\
\hline \hline
\end{tabular}
$$
\begin{list}{}{}
\item The columns are: (1) \chandra ID from the \citet{luo10} catalogue. (2)  Redshift. (3) Total IR luminosity (8-1000 $\mu m$). (4) Total luminosiy at 12 $\mu$m. (5) Luminosity of the starburst component at 12 $\mu$m. (6) Luminosity of the AGN component at 12 $\mu$m. (7) Luminosity of the stellar component at 12 $\mu$m. (8) Stellar mass. (9) Star formation rate. 
\end{list}
\end{table*}

\begin{table*}[!htbp]
\caption{Results from the SED fits for X-ray undetected DOGs with SED fitting} 
\label{sedsfitsnox}
\centering
$$
\begin{tabular}{cccccccccc}
\hline\hline
RA & DEC & z & $\rm log L_{IR}(8-1000\mu m)$   & $\rm log L_{12\mu m}$  & $\rm log L^{SB}_{12\mu m}$  &  $\rm log L^{AGN}_{12\mu m}$ & $\rm log L^{stellar}_{12\mu m}$  & log Stellar mass & SFR  \\
 & &  & $\rm L_{\odot}$ &\lunits & \lunits &  \lunits & \lunits & $\rm M_{\odot}$ & $\rm M_{\odot}/yr$   \\
(1) & (1) & (2) & (3) & (4) & (5) & (6) & (7) & (8) & (9)\\
 \hline
53.0022 & -27.7390 & 1.78 & 12.67 & 44.77 & 44.57 & 44.32 & 42.45 & 11.04 & 788 \\
53.0050 & -27.7768 & 1.90 & 12.94 & 45.01 & 44.52 & 44.83 & 42.94 & 11.63 & 1412 \\
53.0203 & -27.7798 & 1.04 & 11.52 & 44.28 & 44.28 & - & 42.39 & 11.35 & 58 \\
53.0354 & -27.6901 & 1.74 & 12.46 & 45.22 & 45.21 & 43.26 & 42.70 & 11.39 & 502 \\
53.0365 & -27.8875 & 1.95 & 12.59 & 44.93 & 44.13 & 44.86 & 42.49 & 11.18 & 569 \\
53.0501 & -27.8331 & 1.76 & 12.24 & 45.00 & 44.99 & 42.85 & 42.36 & 11.05 & 303 \\
53.0602 & -27.8761 & 2.17 & 12.15 & 44.91 & 44.90 & - & 42.61 & 11.30 & 249 \\
53.0733 & -27.7642 & 1.81 & 11.93 & 44.69 & 44.68 & 43.07 & 42.43 & 11.08 & 149 \\
53.0772 & -27.8595 & 2.88 & 12.92 & 45.68 & 45.68 & - & 43.12 & 11.77 & 1473 \\
53.0860 & -27.7095 & 1.97 & 12.07 & 44.83 & 44.82 & 42.90 & 42.67 & 10.58 & 206 \\
53.0898 & -27.9399 & 1.51 & 12.66 & 45.42 & 45.42 & - & 42.83 & 11.42 & 812 \\
53.0965 & -27.8518 & 1.65 & 11.65 & 44.41 & 44.40 & 42.67 & 42.20 & 10.89 & 78 \\
53.0965 & -27.6725 & 3.04 & 12.60 & 45.36 & 45.35 & 43.58 & 42.80 & 11.45 & 695 \\
53.1215 & -27.8214 & 2.70 & 12.40 & 45.16 & 45.15 & 42.96 & 42.54 & 11.23 & 442 \\
53.1385 & -27.6719 & 1.77 & 12.09 & 44.85 & 44.85 & - & 42.61 & 11.29 & 217 \\
53.1552 & -27.9485 & 2.19 & 12.46 & 44.47 & 44.05 & 44.26 & 42.26 & 10.61 & 473 \\
53.1578 & -27.7041 & 2.46 & 12.75 & 45.50 & 45.50 & - & 43.07 & 11.72 & 982 \\
53.1598 & -27.8502 & 2.03 & 12.14 & 44.90 & 44.90 & - & 42.45 & 11.11 & 244 \\
53.1614 & -27.6515 & 1.75 & 11.91 & 44.67 & 44.67 & - & 42.56 & 11.25 & 144 \\
53.1636 & -27.8907 & 2.87 & 12.81 & 45.74 & 45.42 & 45.46 & 42.86 & 11.45 & 817 \\
53.1677 & -27.8304 & 1.57 & 12.15 & 44.91 & 44.91 & - & 42.53 & 11.21 & 251 \\
53.1736 & -27.7227 & 1.93 & 11.96 & 44.72 & 44.71 & - & 42.48 & 11.17 & 160 \\
53.1790 & -27.6836 & 1.87 & 12.34 & 45.10 & 45.09 & 43.20 & 42.64 & 11.29 & 380 \\
53.1868 & -27.8316 & 2.10 & 11.99 & 44.75 & 44.74 & 42.86 & 42.50 & 11.16 & 170 \\
53.1926 & -27.8921 & 1.44 & 11.99 & 44.75 & 44.75 & - & 42.24 & 11.10 & 174 \\
53.1949 & -27.9490 & 1.72 & 11.84 & 44.60 & 44.59 & 43.15 & 42.20 & 10.11 & 120 \\
53.1983 & -27.7479 & 1.34 & 11.87 & 44.64 & 44.63 & - & 42.44 & 10.34 & 131 \\
53.2131 & -27.6718 & 1.93 & 11.84 & 44.60 & 44.58 & 43.24 & 42.50 & 10.73 & 119 \\
53.2132 & -27.6613 & 1.41 & 12.05 & 44.80 & 44.80 & - & 42.43 & 11.20 & 196 \\
53.2170 & -27.7635 & 1.59 & 11.98 & 44.73 & 44.73 & - & 42.66 & 11.34 & 168 \\
53.2232 & -27.7195 & 1.01 & 11.77 & 43.69 & 43.68 & - & 42.10 & 11.06 & 101 \\
53.2234 & -27.7151 & 1.78 & 12.50 & 44.52 & 44.09 & 44.32 & 42.38 & 10.61 & 518 \\
53.2456 & -27.9310 & 1.72 & 12.21 & 44.97 & 44.97 & - & 42.54 & 11.23 & 291 \\
53.2611 & -27.8583 & 2.14 & 11.87 & 44.64 & 44.62 & 43.27 & 42.30 & 10.95 & 130 \\
\hline \hline
\end{tabular}
$$
\begin{list}{}{}
\item The columns are: (1) Optical coordinates (J200). (2)  Redshift. (3) Total IR luminosity (8-1000 $\mu m$) (4) Total luminosiy at 12$\mu$m. (5) Luminosity of the Starburst component at 12$\mu$m. (6) Luminosity of the AGN component at 12$\mu$m. (7) Luminosity of the stellar component at 12$\mu$m. (8) Stellar masss. (9) Star formation rate. 
\end{list}
\end{table*}

\section{Discussion} 

\subsection{Relation between 12 $\mu m$ and 2-10 keV luminosities}

A strong correlation has been found between mid-IR and X-ray emission
in AGN (\citealt{krabbe01}, \citealt{gandhi09}, \citealt{levenson09},
\citealt{mateos15}, \citealt{stern15}). This was expected since
absorbed AGN emission would be re-processed and re-emitted in the
IR. This correlation has been proposed as a possible selection
technique for CT AGN. CT AGN, because of the strong supression of the
X-ray continuum, should fall below this correlation if we plot the
observed X-ray luminosity against their IR emission. However, it is
difficult to isolate the AGN IR emission due to contamination by the
galaxy starlight and the star formation IR emission.

\begin{figure}[!htbp]
\centering
 \includegraphics[angle=0,width=9.5cm]{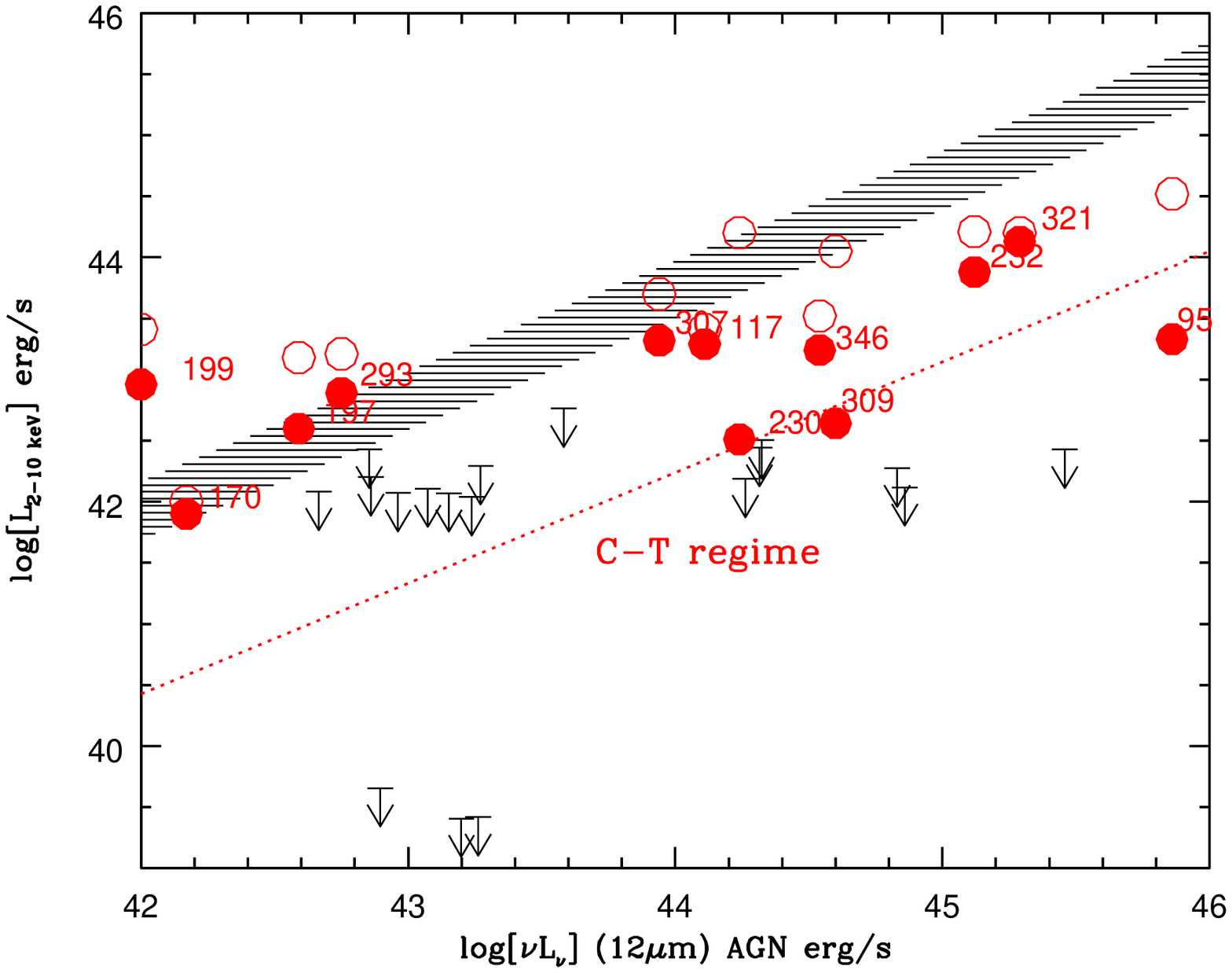}\\
 \caption{$\nu$L$_{\nu 12\mu m}$ for the AGN component vs.
   L$_{\rm{2-10 keV}}$. Shaded region corresponds to
   \citealt{gandhi09} relation. Filled and open circles correspond to
   observed and absorption-corrected L$_{X}$ luminosities,for the X-ray detected DOGs, respectively. Arrows correspond to
     X-ray luminosity upper limits for the X-ray undetected DOGs.}
\label{corr}
\end{figure}

We used the estimated AGN luminosity at 12 $\mu m$ from the SED
fitting and the 2-10 keV observed luminosities from the X-ray spectral
fits to see whether our sources follow the observed correlation (see
Fig.~\ref{corr}). The shaded region in Fig.~\ref{corr} corresponds to
the relation presented in \citet{gandhi09}. We also plotted the
intrinsic 2-10 keV luminosity (open circles), i.e. corrected by the
measured absorption. Although all our DOGs do not follow the
correlation well, the three candidate CT AGN are clearly located
towards the expected CT region, i.e. the region below the line that
corresponds to a factor of 30 lower X-ray luminosity, as is typical in
many CT AGN. It is important to point out that the relation presented
in \citet{gandhi09} is based on high-spatial resolution data, whereas
ours is derived from SED fitting, which could explain the deviations
from the correlation in \citet{gandhi09} in our case.

Nine of the X-ray undetected DOGs lie in the CT region, however, for
three of these (the three less luminous ones), the extremely low
luminosities accompanied by rather high redshifts ($\langle$z$\rangle$
$\sim$ 1.8) do not strongly support a CT classification.

\subsection{Star formation rates}

We took advantage of the SED decomposition we performed to study the
star formation properties of our sample. In star-forming galaxies, the
star formation rate (SFR) and stellar mass follow a relation called
the main sequence (MS)
\citep{daddi07a,elbaz07,whitaker12,behroozi13,speagle14}, which
depends on the stellar mass and evolves with redshift. Outliers from
this relation, such as local ULIRGs and some submm-selected galaxies
(SMGs), are undergoing intense starbursts episodes that are probably
driven by major mergers.

\begin{figure}[!htbp]
\centering
 \includegraphics[angle=-90,width=9cm]{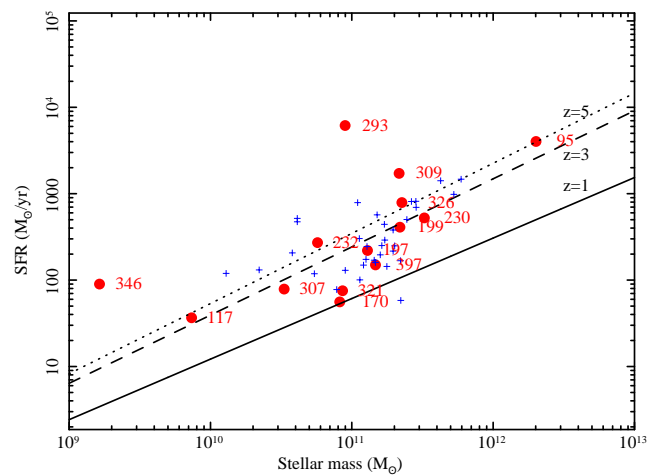}\\
 \caption{Stellar mass vs. SFR for the X-ray detected DOGS
     (filled circles), and X-ray undetected DOGs (crosses). The
   solid, dashed, and dotted lines correspond to the MS relation
   presented in \citet{speagle14} for z = 1, 3, and 5, respectively.}
\label{sfr}
\end{figure}

The stellar mass is an output parameter of our SED fitting. To obtain
the SFR, we converted the IR luminosity (8 - 1000 $\mu m$) of the
starburst component to SFR using the relation in \citet{kennicutt98},
which assumes Salpeter initial mass function. The resulting values are
plotted in Fig.~\ref{sfr} along with the MS for star-forming galaxies
at z=1, 3, and 5 from \citet{speagle14}. Our DOGs occupy a wide region
of the plot, as has also been found for the {\it Herschel} detected
DOGs presented in \citet{calanog13}, but we do not find significant
differences between the CT candidates and the rest of the
sample. Although our sample is too small, the most prominent
characteristic of our three CT candidates is to be among the DOGs with
the highest stellar masses. The average stellar mass and standard
deviation for the X-ray detected DOGs are $\langle M\star \rangle
=2.6\times10^{11}M\odot$ and $\sigma=4.9\times10^{11}M\odot$; whereas
they are $\langle M\star \rangle =1.8\times10^{11}M\odot$ and
$\sigma=1.3\times10^{11}M\odot$ for the X-ray undetected
DOGs. Considering our three CT X-ray detected DOGs, and the six CT
candidates among the X-ray undetected DOGs, the fraction of CT AGN
with $\langle M\star \rangle >2\times10^{11}M\odot$ is $\sim$ 33\%,
whereas it is only $\sim$ 12\% at lower masses.

\subsection{Previous X-ray spectral fits}

The X-ray spectral fits were carried out in a different way in
\citet{georg11}, so a direct comparison between the results is not
possible. Eight sources from CDF-S in \citet{georg11} were classified
as possibly highly absorbed or CT AGN according to their flat photon
index ($< 1$) at high energies. All of these sources are highly absorbed
(N$_{H}$ $>$ 10$^{23}$ cm$^{-2}$) according to our spectral analysis,
including two of our three CT candidates. Source 230, one of our CT
candidates, was not flagged as possible CT in \citet{georg11} because
of the extremely small number of counts in its X-ray spectrum. For the
remaining sources in \citet{georg11} not considered as highly
absorbed, we obtain column densities of the order of or lower than
10$^{22}$ cm$^{-2}$.

We find only three possible CT candidates. The argument in favour of a big
percentage of CT AGN among X-ray undetected DOGs comes mainly from the
flat photon index in their stacked spectra. However, for 6 out of our
14 sources, we also find a very flat ($<$ 1.4) photon index if left
free to vary, but we find only 3 CT candidates when computing the
actual column densities. So it is possible, as suggested in
\citet{georg11}, that the flat photon index in the stacked/averaged
spectra could be the result of a mixture of a low fraction of CT
sources combined with highly (but Compton-thin) absorbed sources. To
test this hypothesis, we simulated a sample of AGN composed, similarly
to the sample studied here of 20\% CT AGN, 44\% highly absorbed AGN,
22\% moderately absorbed AGN, and 14\% unabsorbed AGN. We then jointly
fitted these simulated spectra, obtaining a photon index $\sim$ 1.1.

\citet{brightman14} presented X-ray spectral analysis for the 4 Ms
      {\it Chandra} spectra in the CDF-S. Our spectral fitting results
      are in agreement with those presented in \citet{brightman14} in
      most cases. In a few cases, there are small differences in the
      obtained column densities because of different photometric
      redshifts were used or because we used a more complex spectral
      model thanks to the availability of better spectral data. As for
      our three CT candidates, only source 309 is classified as such
      in \citet{brightman14}. For the remaining two CT candidates in
      our sample (sources 95 and 230) and because of the limited
      number of counts, they only fitted the photon index for source
      230 (obtaining a flat value $\sim$ 0.8) and all parameters were
      fixed in the case of source 95. In both cases, the column
      density was fixed to 10$^{20}$ cm$^{-2}$, i.e. they were
      classified as unabsorbed AGN. In any case, the location of these
      two sources in the L$_{\rm 2-10\,keV}$/L$_{12\mu m}$ plot
      supports their CT nature. In the case of source 170,
      \citet{brightman14} found it to be highly absorbed (almost CT),
      and we only found a low upper limit for the amount of
      absorption. Given the extremely low number of counts, a flat
      reflection dominated spectrum cannot be rejected in this
      case. Besides, the large uncertainty in its computed X-ray
      luminosity makes it consistent with the CT region.

\subsection{Compton-thick/highly absorbed AGN fraction}

Out of the 14 DOGs studied here, 9 sources have $N_H> 10^{23}$
\cunits. If we compare the results in \citet{georg11}, which uses data
from the CDF-N and CDF-S, with the results in \citet{lanzuisi09}, it
appears that despite the fact that the SWIRE sample is much brighter
(its median flux is $\sim1\times10^{-13}$ \funits), the percentage of
highly absorbed sources ($>10^{23}$ \cunits) is not very different
compared to our sample. However, when using the deeper data from this
work (in which we only used data from the CDF-S), we find a higher
percentage of highly absorbed sources ($\sim$ 64\%) and a much higher
percentage of CT sources ($\sim$ 20\%).

In \citet{fiore09} they suggested that a higher percentage of CT
sources could be hidden among the X-ray undetected DOGs given the very
flat photon index obtained from stacked images. We also find that the
spectra of our sources are flat but, after the spectral analysis, we
find that the flat photon index comes from moderate to high absorption
instead of from a high percentage of CT sources. To search for more CT
candidates in our sample, we also attempted a join fit considering
only the sources with spectroscopic redshifts available (only four
sources), but the resulting best-fit model included only moderate
amounts of absorption.

Our results could be consistent with the CT fraction estimated in
\citet{georgakakis10} ($\sim$ 30\%) as far as the X-ray detections are
concerned. If we take into account the possibility that source 170 is
a reflection-dominated CT AGN, as we mentioned in the previous
section, our fraction of CT sources among X-ray detected DOGs would
reach $\sim$ 30\% as well.

\begin{figure}[!htbp]
\centering
 \includegraphics[angle=0,width=9cm]{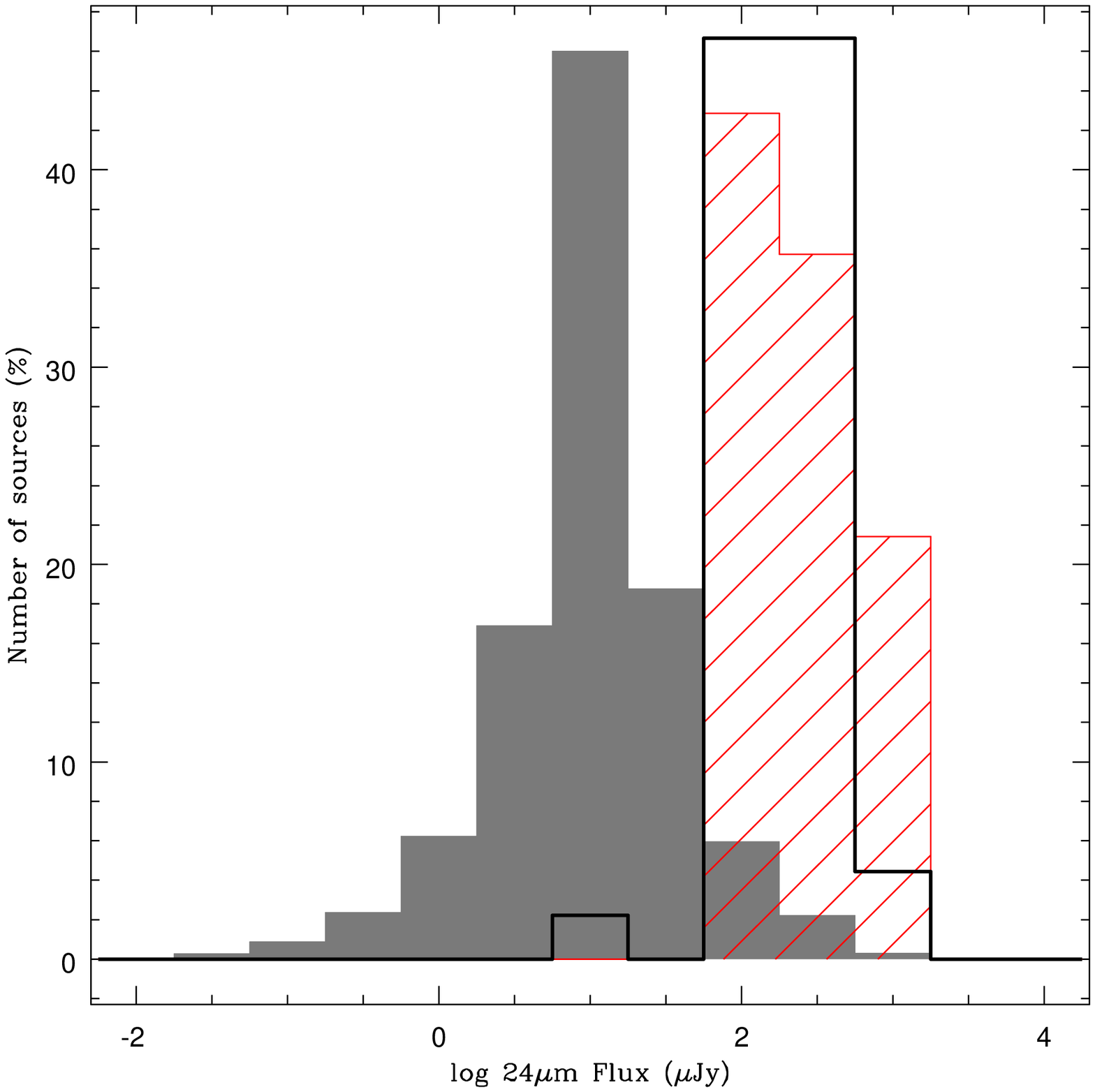}\\
 \caption{Distribution of 24 $\mu m$ fluxes for all the detected
   sources in the CDF-S (filled histogram),  X-ray detected DOGs
   (line-shaded histogram), and X-ray undetected DOGs (empty histogram).}
\label{24flux}
\end{figure}

Regarding the full DOGs sample, if we only consider the most secure CT
AGN (the three candidates among the X-ray detected and the six
candidates among the X-ray undetected DOGs), we estimate a fraction of
13\% CT AGN among the whole DOGs population. If we assume that all the
undetected DOGs without SED fitting available are CT candidates, this
fraction would increase to 44\%. However, strong star formation is
also expected to produce a high $f_{24\mu m}$/f$_R$ ratio by
increasing the 24 $\mu m$ flux. As the AGN contribution becames
stronger, it dominates the 24 $\mu m$ flux so that the fraction of AGN
among DOGs increases for higher fluxes
\citep{rigu15,donley12,fiore09,dey08}. We examined the 24 $\mu m$
fluxes of the X-ray detected and undetected DOGs. Both samples display
very high 24 $\mu m$ fluxes, although the X-ray detected DOGs show
marginally higher fluxes (see Fig.~\ref{24flux}). By applying a K-S
test, we find that the probability that the two samples are drawn from
the same population is high (p$\sim$0.5). Nevertheless, we find that
47\% of the X-ray undetected DOGs with SED fits do not need any AGN
contribution to their IR emission. Therefore, a significant percentage
of the remaining X-ray undetected DOGs could be still be powered by
star formation rather than by an AGN.

\section{Conclusions}

We have explored the properties of the X-ray detected and undetected
DOGs in the CDF-S. This is an extension of the work presented in
\citet{georg11}, but we had the advantages of not only the
availability of deeper (6 Ms) {\it Chandra} observations of the
sources presented in \citet{georg11}, but also the addition of deep (3
Ms) {\it XMM-Newton} observations in the CDF-S. We also present a more
accurate estimate of the AGN contribution to the IR luminosity by SED
fitting from the UV to the far-IR. Out of the 70 DOGs composing the
full sample, 14 are X-ray detected. For the remaining 56 DOGs, there
were enough multi-wavelength data available for 34 of these DOGs to
perform a similar SED-fitting based analysis as for the X-ray detected
sources.\\

From the X-ray spectral analysis, we find that most (9/14) of the
X-ray detected DOGs in our sample show high obscuration (N$_{H}$
$>10^{23}$ \cunits). However, only three (maybe four) of our X-ray
detected DOGs could be CT AGN, so we estimate a CT fraction of 20-30
\% among X-ray detected DOGs. Many of our DOGs (6/14) show flat photon
indices ($\Gamma$ $\lesssim$ 1.4), but only three display CT
absorption. Therefore, caution must be exercised when estimating the
fraction of CT AGN from the photon index of stacked/averaged spectra.\\

Considering the full DOGs sample, the fraction of CT AGN among the
whole DOGs population could range from 13 to 44\%. X-ray detected DOGS
seem to have marginally higher 24 $\mu m$ fluxes and CT X-ray detected
DOGs seem be hosted in galaxies with higher stellar masses than X-ray
undetected DOGs, but a bigger sample is necessary to confirm these
results.

\begin{acknowledgements}   
  We thank the anonymous referee for his/her thorough review and
  suggestions, which significantly contributed to improve the quality
  of this paper.  We acknowledge the use of {\it Spitzer} data
  provided by the {\it Spitzer} Science Center.  The {\it Chandra}
  data were taken from the Chandra Data Archive at the Chandra X-ray
  Center. Based on observations obtained with {\it XMM-Newton}, an ESA
  science mission with instruments and contributions directly funded
  by ESA Member States and NASA. AC acknowledges funding from the
  PROTEAS project within GSRT’s KRIPIS action, funded by Greece and
  the European Regional Development Fund of the European Union under
  the O.P. Competitiveness and Entrepreneurship, NSRF 2007-2013 and
  the Regional Operational Program of Attica.
\end{acknowledgements}
\bibliographystyle{aa}
\bibliography{corral_cdfs_dogs}

\end{document}